\documentclass[useAMS,usenatbib]{mn2e}
\usepackage{psfrag,graphicx}
\usepackage{txfonts}
\usepackage{natbib}
\usepackage{ulem}
\usepackage{color}
\usepackage{xspace}
\usepackage{enumitem}

\def \msol{M$_{\odot}$\xspace}
\def \kmps{\,km\,s$^{-1}$\xspace}
\def \cmcube{\,cm$^{-3}$\xspace}
\def \nhtot{$n_{\rm H}$\xspace}
\def \tff{$t_{\rm ff}$\xspace}

\title[How chemistry influences cloud, stars, and the IMF]{How chemistry influences cloud structure, star formation, and the IMF}
\author[S. Hocuk, S. Cazaux, M. Spaans and P. Caselli]{S. Hocuk$^{1}$\thanks{E-mail: seyit@mpe.mpg.de}, S. Cazaux$^{2}$, M. Spaans$^{2}$
 and P. Caselli$^{1}$\\
$^{1}$Max-Planck-Instit\"{u}t f\"{u}r extraterrestrische Physik, Giessenbachstrasse 1, 85748 Garching, Germany \\
$^{2}$Kapteyn Astronomical Institute, University of Groningen, P. O. Box 800, 9700 AV Groningen, Netherlands}

\begin{document}

\date{Submitted ?; Accepted ?}
\maketitle

\label{firstpage}
\begin{abstract}
{In the earliest phases of star-forming clouds, stable molecular species, such as CO, are important coolants in the gas phase. Depletion of these molecules on dust surfaces affects the thermal balance of molecular clouds and with that their whole evolution. For the first time, we study the effect of grain surface chemistry (GSC) on star formation and its impact on the initial mass function (IMF). We follow a contracting translucent cloud in which we treat the gas-grain chemical interplay in detail, including the process of freeze-out. We perform 3D hydrodynamical simulations under three different conditions, a pure gas-phase model, a freeze-out model, and a complete chemistry model. The models display different thermal evolution during cloud collapse as also indicated in \citealt{2014MNRAS.438L..56H}, but to a lesser degree because of a different dust temperature treatment, which is more accurate for cloud cores. The equation of state (EOS) of the gas becomes softer with CO freeze-out and the results show that at the onset of star formation, the cloud retains its evolution history such that the number of formed stars differ (by 7\%) between the three models. While the stellar mass distribution results in a different IMF when we consider pure freeze-out, with the complete treatment of the GSC, the divergence from a pure gas-phase model is minimal. We find that the impact of freeze-out is balanced by the non-thermal processes; chemical and photodesorption. We also find an average filament width of 0.12 pc ($\pm$0.03 pc), and speculate that this may be a result from the changes in the EOS caused by the gas-dust thermal coupling. We conclude that GSC plays a big role in the chemical composition of molecular clouds and that surface processes are needed to accurately interpret observations, however, that GSC does not have a significant impact as far as star formation and the IMF is concerned.}
\end{abstract}

%OPTIONAL: changes for the abstract in order to shorten down to 255 words instead of 311 now.
% (1) remove: first sentence '{In the earliest...'
% (2) start with: '{Depletion of gas-phase molecules...'
% (3) remove the sentence: 'The models display ... for cloud cores'

\section{Introduction}
\label{sec:introduction}
Substantial depletion of gas-phase species has been observed in various star-forming regions such as low-mass starless cores and high-mass star-forming clouds \citep[e.g.,][]{2002ApJ...569..815T, 2004A&A...416..191T, 2011ApJ...738...11H, 2012MNRAS.423.2342F, 2013ApJ...775L...2L, 2014A&A...570A..65G}. Depletion occurs when species are drained from the gas-phase and deposited on grain surfaces. Observational data indicate that depletion factors can go up to 80 and beyond, where the depletion factor is defined as the expected abundance over the observed abundance of a gas-phase species, while gas temperatures of up to 25\,K can exist in high-mass pre-stellar cores \citep{2012MNRAS.423.2342F}. This leaves the warmer gas to coexist with the colder dust when there is sufficient time for depletion to become significant \citep{2001ApJ...557..736G}. The importance of freeze-out on the (thermo)dynamics of an evolving interstellar cloud may eventually impact star formation altogether.

At number densities exceeding \nhtot$\gtrsim 3\times10^4$ \cmcube, where \nhtot is the total hydrogen nuclei number density, freeze-out is not expected to play a significant role on the thermal balance since gas and grain will be tightly coupled and their temperatures will approach each other. Heat transfer through collisions with dust rather than line emission will then dominate the cooling of a cloud. However, below this critical density, and with sufficient UV extinction for molecules to survive, depletion can perturb the thermal balance by inhibiting cooling, especially by freezing out CO. This will increase the temperature of the gas as shown in \cite{2014MNRAS.438L..56H}, hereafter Paper\,I. In regions such as cold dense clumps and cores with dust temperatures below 20\,K, CO molecules will be locked up in ice. These CO molecules cannot thermally evaporate back into the gas phase and the freeze-out timescale is sufficiently short \citep[$\lesssim10^4$\,yr,][]{2015A&A...577A.102V} to deplete all molecules within a cloud lifetime ($10^6$\,yr). Other molecules, like water and methanol, await a similar fate. In models and numerical simulations that consider freeze-out, when not including other means of desorption, usually near 100\% depletion is seen in cold ($\sim$10\,K) environments, much greater than what is observed. Unless a strong heat source like a nearby protostar reveals the rich organic chemistry on dust mantles \citep{2003ApJ...593L..51C}, frozen molecules will remain solid and, therefore, unable to contribute to cooling through ro- vibrational line emissions.

Non-thermal desorption, however, can liberate the frozen species that are thermally locked. Photodesorption is often considered as an effective physical mechanism to release surface molecules \citep[e.g.,][]{2011ApJ...739L..36F, 2013A&A...560A..73G, 2015ApJ...803...52K} to the gas phase, thereby explaining how unexpected high amount of molecules can exist in the gas phase \citep[e.g.,][]{2012A&A...541L..12B, 2012ApJ...759L..37C, 2014ApJ...795L...2V}. Molecules can be photodesorbed by a direct process or an indirect process through exciting neighboring molecules \citep[e.g.,][]{2013ApJ...779..120B, doi:10.1021/acs.jpca.5b02611}. Photons suffer, however, from attenuation deeper into the cloud because of the high extinction naturally arising towards dense regions. This is also generally the place where the ices are known to exist. Cosmic rays (CRs), on the other hand, can penetrate deeply and excite H$_2$ molecules, which then fluoresce producing UV photons locally. This is a secondary form of photodesorption, one that is only marginally affected by the column density. Even though the flux of secondary UV photons is much lower than the general background UV field, CR-induced UV (CRUV) can dominate the UV photon production at high visual extinctions ($A_V>10$\,mag). The process that is dubbed CR desorption, which is not included in this work, typically indicates the desorption following the dust heating upon direct collision with a CR. This can proceed via `whole-grain heating' \citep{1985A&A...144..147L, 1993MNRAS.261...83H} or through explosive desorption due to `impulsive spot heating' \citep{2015ApJ...805...59I}.

Another way to transfer species from the solid phase into the gas phase is through chemical desorption \citep{2007A&A...467.1103G, 2010A&A...522A..74C, 2013NatSR...3E1338D}, also known as reactive desorption \citep[e.g.,][]{2013ApJ...769...34V}. This non-thermal desorption is not affected by the depth of the cloud, but is rather the result of rich and reactive chemistry on surfaces. The exothermicity of the reaction, the binding energy of the species, and the degrees of freedom of the formed products is what determines the likelihood of this process. It should be noted that `chemically driven desorption' \citep{2000MNRAS.314..273T} is a more specific mechanism that occurs through an energetic reaction on grain surfaces, such as the Eley-Rideal (ER) mechanism of H$_2$ formation, kicking out a molecule next to it. Chemical desorption has been shown to be an important desorption process in cold dense cores \citep{2015A&A...576A..49H, 2015MNRAS.447.4004R, 2015A&A...576A..91N}.

All of these processes cause the molecules to experience phase-transitions depending on the conditions of the environment and the importance of each process. It will affect the cooling properties of the gas. Moreover, some species, like CO, are optically thick at cloud densities of $\geq 10^4$\cmcube such that a decrease in the abundance will result in the reduction of self-shielding, counterbalancing each other to a certain degree. So, the question becomes: Does freeze-out have a strong enough impact on the thermal balance to affect star formation and if so, can other (non-thermal) processes counter act the depleting nature of freeze-out? Paper\,I showed that freeze-out does impact the thermal balance of an evolving cloud enough to change the equation of state (EOS). The role of the gas-dust interplay on the final stellar masses should therefore be carefully studied, and, if significant, need to be properly addressed in the theories of star and filament formation \citep[see also][]{2015arXiv151006544S}.

The stellar initial mass function (IMF) in the Milky Way is often observed to follow a universal, Salpeter-like \citep{1955ApJ...121..161S} power-law at high masses ($M\gtrsim0.5 M_{\odot}$), see for example a review by \cite{2010ARA&A..48..339B}. Although the low-mass end of the IMF is still a topic of debate \citep{2011ApJ...735L..13V,2014ApJ...784..162P}, it is generally thought that gravity causes the power-law behavior at the high-mass end \citep[e.g.,][]{2013ApJ...766L..17S}, while more recently the idea of preferential attachment has been proposed \citep{2015arXiv151105200K} that is similar to the competitive accretion theory \citep{1997MNRAS.285..201B}. In extragalactic regions, variations in the IMF are more frequently reported \citep[see][and references therein]{2013pss5.book..115K}. Aside from the variation in the distribution of \textit{stellar} masses, a link is often made to the fragmentation of a cloud into clumps and cores and the \textit{core} mass distribution \citep{2010A&A...518L.106K, 2010A&A...518L.102A}. In order to fully dissect this problem, one has to explore the earlier stages of cloud evolution. Numerical simulations have been able to reproduce the log-normal shape of the IMF with the power-law tail. However, dust chemistry has never been considered as an important factor in shaping the IMF, even though it was thought to be able to have an impact on the thermodynamics of a gas cloud at \nhtot $\leq 10^{4.5}$\,\cmcube \citep{2001ApJ...557..736G}, thus, also affecting the cloud fragmentation. \cite{2005MNRAS.359..211L} suggested that a departure from an isothermal EOS might be enough to set the characteristic mass of the IMF, while also affecting filament sizes as the isothermal case provides support for a cylinder. \cite{2012ApJ...760L..28P} showed that a sub-isothermal EOS already enhances filamentary structures. In this study we investigate the impact of GSC on star-forming clouds by considering different levels of complexity of the GSC and show numerically how strong a collapsing cloud, thereby the IMF, can be influenced by different GSC processes.

This work is a follow-up on an earlier study on the effects of freeze-out during cloud evolution. In Paper\,I, we showed that freeze-out had a significant effect on the thermal balance of a diffuse cloud evolving into a molecular cloud, leading to a fluctuation of the EOS around the critical value of unity. We hypothesized that this could lead to an altered fragmentation epoch with the end result of different star formation efficiencies and different initial stellar masses at the moment of star formation. In this paper, we initiate our simulations from a translucent cloud stage and continue upon this project with higher resolution simulations. In addition to the previous work, we added an extra element to our models by also testing against a complete grain surface chemistry model as was recently presented in \cite{2015A&A...576A..49H}. The complete chemistry model has the aforementioned non-thermal desorption processes that can counter the effect of freeze-out. One major difference is that in this work we adopt a different, but more accurate treatment to compute the dust temperature.

For the impact on star formation, we focus on the process of freeze-out, because it has likely the most dramatic effect on the thermodynamics. This has consequences on the pressure and the density evolution of the cloud. Since molecule freezing is being affected by many processes, the compete gas-grain chemical interplay must be considered in order to treat the freeze-out process most accurately. Next to freeze-out, other important influences on the thermodynamics by dust grains will arise by forming increasingly more complex molecules and with that the many line transitions they allow, like methanol once they are released into the gas phase, but detailed line cooling from complex molecules is currently beyond the scope of this work.

Our paper is divided into five sections. In Sect.\,\ref{sec:numericalmethod}, we explain our models, outline our code, and show the initial conditions. In Sect.\,\ref{sec:analyticalmethod}, we describe the chemistry and our chemical network, with further details on this given in appendix\,\ref{app:AA}. We address the thermal balance, providing all the heating and cooling processes used in this work, where the equations are given in appendix\,\ref{app:BB}. We also elaborate on the adopted dust temperature calculation method. In Sect.\,\ref{sec:results}, we present our results on the chemical and thermal evolution and the stellar mass distribution. Finally, in Sect.\,\ref{sec:conclusion} we summarize our conclusions and discuss the caveats.

\section{Numerical method}
\label{sec:numericalmethod}
\subsection{The code}
The code that is used to perform the numerical simulations in this work is the adaptive-mesh refinement (AMR) hydrodynamical code {\sc flash} 4.0 \citep{2000ApJS..131..273F, Dubey2009512}. Hydrodynamic equations were solved with the directionally split piecewise-parabolic method \cite[PPM;][]{1984JCoPh..54..174C}, which is well-suited to handle gas dynamics and contact discontinuities, such as shocks. The Poisson equations were solved with the Multigrid solver in which gravity is coupled to the Euler equations through the momentum and energy equations. We employed the multi-species unit that is able to follow each species with its own properties in order to track multiple fluids. To prevent overshoots in the mass fractions as a result of the PPM advection, we applied the consistent multi-fluid advection scheme \citep{1999A&A...342..179P}.

Our work combines the physical processes on a wide range of scales associated with the field of star formation, cloud dynamics, thermodynamics, and chemistry. We used time-dependent heating and cooling rates to solve the thermal energy equation at each mesh point (grid cell) and perform rate equations for computations of the chemistry in gas and solid phases. The cloud dynamics advances through the hydrodynamic calculations of our code. Magnetic fields were not included in this work. We presume that our cloud is supercritical and thus not supported by magnetic fields.

For our simulations we consider and include processes such as gravity, turbulence (non-driven), shocks, external radiation field, polytropic EOS, multiple species, sink particles, thermodynamics, and chemistry. Descriptions for the well-tested physics modules that are not default to the {\sc flash} code can be found in \cite{2010A&A...522A..24H} and Paper\,I. Column densities are calculated from the UV flux at each cell by tracing 14 equally weighted rays with long characteristics \citep[see][]{2015A&A...576A..49H}. We use the sink particle routine that is described in \cite{2011A&A...536A..41H}, which adopts the methods from \cite{2004ApJ...611..399K} and \cite{2010ApJ...713..269F}. The {\sc flash} code has also an integrated sink particle module that is similar to our own. The key physics modules important for this work, surface chemistry and thermodynamics, are explained in detail in Sect.\,\ref{sec:analyticalmethod}, appendices \ref{app:AA} and \ref{app:BB}, and, concerning the equations and formulations of the GSC, also in \cite{2015A&A...576A..49H}.

We point out that there are several differences in the numerical prescription as compared to the work done in Paper\,I. The main differences are, the version of the chemistry database with newer reaction rates and coefficients (see Sect.\,\ref{sec:chemistry} and Tables \ref{tab:appendixG2} and \ref{tab:appendix1}), extending non-thermal desorption to multiple species (see Sect.\,\ref{sec:nonthermal}), and the adopted expression for the dust temperature (see Sect.\,\ref{sec:tdust}). The changes were driven by most recent experimental and observational findings. These changes are next to the increased spatial and temporal resolution of the simulation and the addition of sink particles in order to follow star formation.

\subsection{Three models}
\label{sec:models}
To address the impact of freeze-out and non-thermal desorption to star formation we devise three models. We run a simulation with each model in order to better understand the underlying physical impact. Each model is identical to the other in every other aspect than the ones mentioned below. We identify our models in the following way:

\begin{description}\itemsep5pt
\item[\textit{Pure gas}:] For this model we exclude any surface species from our calculations. This does not mean that there is no dust, it means that the chemistry is occurring in the gas phase only. Dust is considered in the thermal balance with the gas-grain coupling and we do calculate the dust temperature. The initial species abundances slightly differs from the models with surface species (see Sect.\,\ref{sec:species}).

\item[\textit{Gas+freeze-out}:] For this model we include surface species alongside the gas species, but disallow any reactions or photoprocesses on surfaces. This is to exclude all forms of non-thermal desorption. Only accretion of gas-phase species and thermal evaporation are permitted. This includes the H$_2$ formation through the ER process.

\item[\textit{Full chemistry}:] For this model we include all species and the complete set of reactions. The simulation is much more computationally demanding, while the chemical time-step is also generally much smaller. This does not affect the hydrodynamical time-step, which may otherwise be a factor in the formation of sink particles as this partially depends on time resolution.
\end{description}

\subsection{Initial conditions}
\begin{figure}
\includegraphics[scale=0.1]{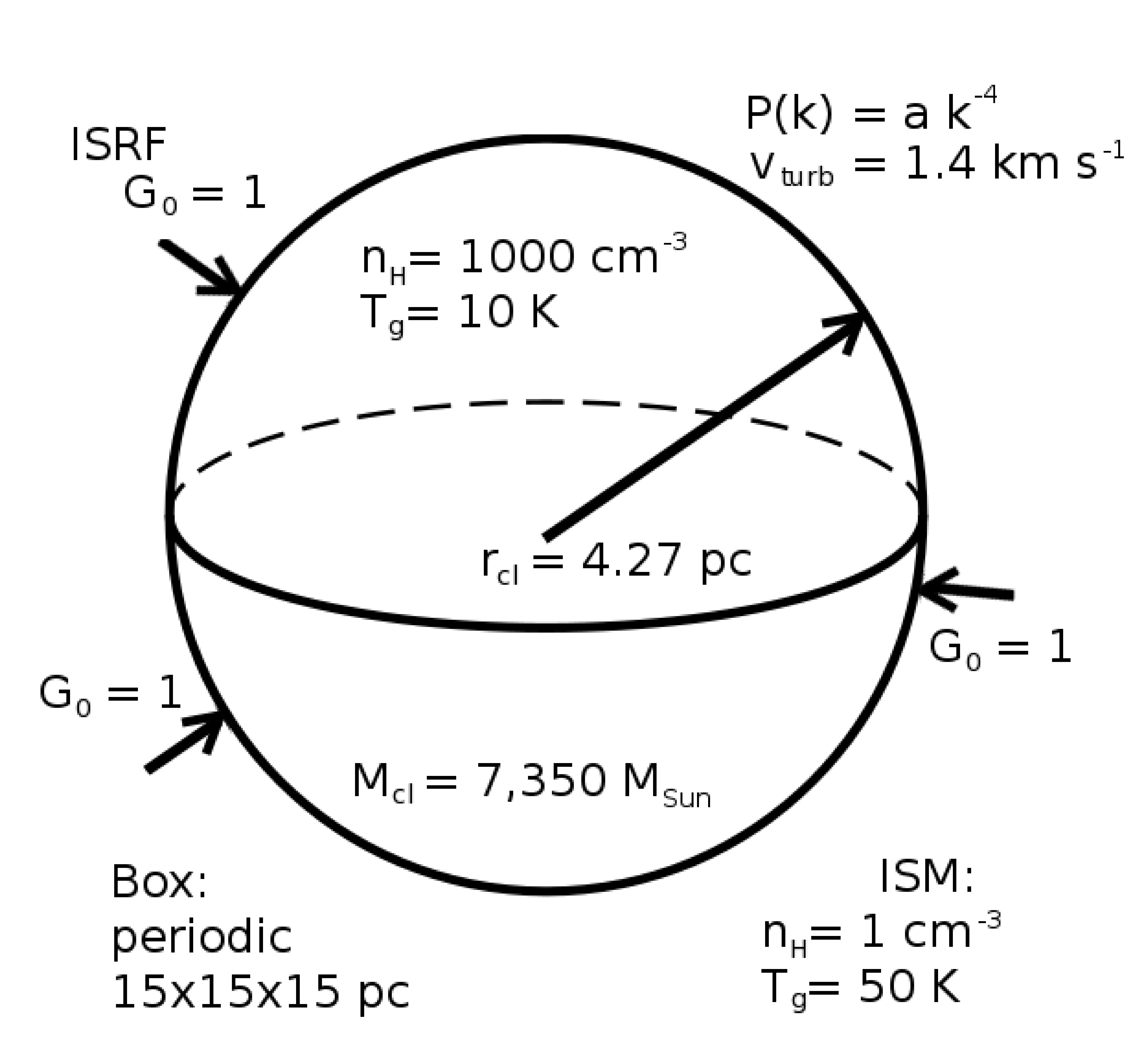}
\caption{A graphical representation of the initial setup.}
\label{fig:init}
\end{figure}

We simulate gravitationally bound spherical clouds that are in a stage of collapse. Each cloud is initially in a translucent stage with uniform conditions inside the cloud. A translucent cloud is a molecular cloud with visual extinction in the 1-5\,mag and density in the 10$^2$-10$^4$\,\cmcube range in which both photoprocesses and gas-phase reactions play an important role in the chemistry \citep[e.g.,][]{2015MNRAS.452.3969C}. The cloud edge is similar (density, temperature, pressure, and abundances) to the diffuse atomic medium and has different initial conditions compared to the cloud interior. We allow for a smooth transition from the cloud edge to the uniform inner parts. The transition region covers the range from the outer edge to 30\% of the cloud radius, which corresponds to a visual extinction ($A_V$) of unity. The initial number density of the model clouds is \nhtot = $10^{3}$\cmcube. The interstellar environment as well as the cloud edge have a density of \nhtot = 1\cmcube. The model clouds have an effective radius of 4.27\,pc and a total mass of 7.35$\times10^{3}$ \msol. The total mass in the rest of the simulation domain is negligible, and $<1\%$, relative to the cloud mass. A graphical representation of the initial conditions is given in Fig.\,\ref{fig:init}.
 \\

The translucent clouds are established with turbulent initial conditions similar to the interstellar clouds residing in the Milky Way. The turbulent velocity width is $\rm \Delta v_{\rm turb}$ = FWHM = 1.4\kmps and has a power spectrum of $\rm P(k) \propto k^{-4}$, following the empirical laws for compressible fluids \citep{1981MNRAS.194..809L, 1999ApJ...522L.141M, 2004ApJ...615L..45H}. The decaying turbulence, with a virial parameter $\alpha_{\rm vir} \simeq 0.3$, is not strong enough to support gravitational collapse. The turbulence will, however, create a clumpy and filamentary density structure of the cloud in a timescale much shorter than a free-fall time ($t_{\rm ff}$) or a cooling time ($t_{\rm cool}$).

A background UV field radiation flux of $G_0=1$ in terms of the Habing field \citep{1968BAN....19..421H} is assumed pertaining to the average interstellar medium (ISM) conditions of our Milky Way. Due to the photoelectric heating, this yields a temperature gradient from 10\,K inside the cloud to 50\,K just at the edge of the cloud \citep{2005A&A...436..397M}, while the dust temperature increases from 6 to 18\,K inside out. The isothermal sound speed of the cloud in this region ranges between $\rm c_{s}=0.20$ and 0.45\kmps. The temperature will not stay fixed as the thermal balance is solved time-dependently by considering the most important heating and cooling processes, see Sect.\,\ref{sec:thermodynamics}. The temperature in the diffuse ISM (\nhtot $\leq$ 1 \cmcube), for example, can go up from several hundred to a few thousand Kelvin due to photoelectric heating and shocks.

All hydrogen inside the three translucent clouds is set to molecular form while it is atomic at the cloud edge. A mixture exists in the transition region. We obtained these abundances by running a 1D model until convergence was reached, which is explained in more detail in Sect.\,\ref{sec:species}. Atomic species are also more ionized at the cloud edge and neutral inside. See Tables \ref{tab:species} and \ref{tab:species2} for the edge and center abundances. In total, we include 41 different gas-phase and surface-bound `ice' species in our models.

We placed our clouds in cubic boxes of size 15\,pc$^3$, with periodic boundary conditions. We refined our grid according to a strict Jeans length ($\lambda_{\rm J}$) criterion, with a resolution of $n_{\rm Jeans}\geq\,16$ cells/$\lambda_{\rm J}$. Given a maximum resolution of 1024$^3$ cells and utilizing the AMR, the spatial resolution yields 15\,pc/1024 $\simeq$ 0.015\,pc, which equals to a smallest size scale of slightly below 3000 AU.

\section{Analytical method}
\label{sec:analyticalmethod}
\subsection{Chemistry}
\label{sec:chemistry}
We perform time-dependent rate equations that include both gas-phase and surface reactions. Ices and gas-phase species are described in the same manner, in number densities, to make the interaction easier. The inter-exchange between the two phases proceed through the adsorption and the desorption processes, or through the ER mechanism. Chemical reactions involving dust grains are divided into six categories

\begin{description}[style=multiline,leftmargin=0.8cm]%,font=\normalfont]
 \item[(\ref{sec:accretion})] accretion (adsorption) of gas phase species on to grains,
 \item[(\ref{sec:evaporation})] evaporation (thermal desorption) of ices,
 \item[(\ref{sec:twobody})] two-body surface reactions,
 \item[(\ref{sec:crs})] CR processes,
 \item[(\ref{sec:uvreactions})] photoprocesses that include photodissociation, \\
 photodesorption, and CRUV photoprocesses,
 \item[(\ref{tab:appendix3})] chemisorption of H including the ER formation of H$_2$.
\end{description}
The headings within the parentheses point to the specific sections in the appendix where these processes are described in more detail. 
The Tables \ref{tab:appendixG1}, \ref{tab:appendix1}, and \ref{tab:appendix3} in the appendix display the considered surface reactions in this work.

Gas-phase reactions are taken from the Kinetic Database for Astrochemistry version 2014 \citep[\textit{KiDA},][]{2015ApJS..217...20W}. This version has over 1000 more reactions as compared to the previous version \citep{2012ApJS..199...21W}, which in our case only amounts to 19 additional gas-phase reactions because we do not utilize the entire network (see Sect.\,\ref{sec:species}). In addition, the newer version also has many different reaction coefficients. Almost all the new coefficients are for photoprocesses particularly important for this study. In summary, the photoprocesses in the new version are more efficient at low columns, but more quickly attenuated towards higher opacities. We refer the reader to the \textit{KiDA} database for the equations and the corresponding parameters. The gas-phase reactions in our network include bimolecular reactions, charge-exchange reactions, radiative associations, associative detachment, electronic recombination and attachment, ionization or dissociation of neutral species by UV photons, and ionization or dissociation of species by direct collision with CR particles or by secondary UV photons following H$_2$ excitation. Tables \ref{tab:appendixG1} and \ref{tab:appendixG2} in the appendix show all the gas-phase reactions considered in this work.

Chemical reactions are solved using a fast and stable semi-implicit scheme following \cite{2008A&A...490..521S}. This is an improved scheme over the first-order backwards differencing (BDF) method developed by \cite{1997NewA....2..209A}. The manner in which this solver is applied can also be found in \cite{2015A&A...576A..49H}. The present grain surface chemistry rate equations are not modified at low fluxes to account for the discrete nature of grain particles \citep{1998ApJ...495..309C, 2008A&A...491..239G}.

\subsection{Chemical network}
\label{sec:species}
We consider in our simulations a total of 41 species. Only species with the elements H, C, and O are considered in this work. Tables\,\ref{tab:species}\,and\,\ref{tab:species2} list the species and their initial abundances, whereas Table\,\ref{tab:species2} is not applicable for the model \textit{pure gas}.
\begin{table}
\caption{List of gas-phase species and their initial abundances ($n_{x_i}$/\nhtot)}
\begin{tabular}{lcccc}
\hline
\hline
                      &                    & {Pure gas species} & {With ice species} \\
Species               & Cloud edge         & Cloud center       & Cloud center        \\
\hline
H                     & 1.0                & 4.7$\times10^{-5}$ & 4.1$\times10^{-5}$ \\
H$^{-}$               & 0                  & 0                  & 0                  \\
H$^{+}$               & 0                  & 0                  & 0                  \\
H$_{2}$               & 2.2$\times10^{-5}$ & 0.5                & 0.5                \\
H$_2^{+}$             & 0                  & 0                  & 0                  \\
H$_3^{+}$             & 0                  & 0                  & 0                  \\
O                     & 2.9$\times10^{-4}$ & 1.6$\times10^{-4}$ & 8.3$\times10^{-7}$ \\
O$^{-}$               & 0                  & 0                  & 0                  \\
O$^{+}$               & 0                  & 0                  & 0                  \\
O$_{2}$               & 0                  & 1.9$\times10^{-6}$ & 1.9$\times10^{-8}$ \\
C                     & 0                  & 0                  & 5.0$\times10^{-5}$ \\
C$^{-}$               & 0                  & 0                  & 0                  \\
C$^{+}$               & 1.3$\times10^{-4}$ & 0                  & 0                  \\
OH                    & 0                  & 1.5$\times10^{-7}$ & 3.2$\times10^{-7}$ \\
OH$^{+}$              & 0                  & 0                  & 0                  \\
CO                    & 0                  & 1.3$\times10^{-4}$ & 3.0$\times10^{-5}$ \\
CO$_{2}$              & 0                  & 0                  & 0                  \\
H$_{2}$O              & 0                  & 5.2$\times10^{-7}$ & 1.3$\times10^{-7}$ \\
H$_{2}$O$^+$          & 0                  & 0                  & 0                  \\
H$_{3}$O$^+$          & 0                  & 0                  & 0                  \\
HCO                   & 0                  & 0                  & 0                  \\
HCO$^+$               & 0                  & 0                  & 0                  \\
H$_{2}$CO             & 0                  & 0                  & 0                  \\
CH$_{3}$O             & 0                  & 0                  & 0                  \\
CH$_{3}$OH            & 0                  & 0                  & 0                  \\
e$^{-}$               & 1.3$\times10^{-4}$ & 0                  & 0                  \\
\hline
\end{tabular}
\label{tab:species}
\end{table}
\begin{table}
\caption{List of surface species and their initial abundances ($n_{x_i}$/\nhtot)}
\begin{tabular}{lclc}
\hline
\hline
Species               & Center abundance  & Species               & Center abundance \\
\hline
$\bot$ H              & 0                  & $\bot$ CO$_{2}$       & 0                  \\
$\bot$ H$_{\rm c}$    & 0                  & $\bot$ H$_{2}$O       & 2.1$\times10^{-4}$ \\
$\bot$ H$_{2}$        & 0                  & $\bot$ HO$_{2}$       & 0                  \\
$\bot$ O              & 0                  & $\bot$ H$_{2}$O$_{2}$ & 0                  \\
$\bot$ O$_{2}$        & 0                  & $\bot$ HCO            & 0                  \\
$\bot$ O$_{3}$        & 0                  & $\bot$ H$_{2}$CO      & 0                  \\
$\bot$ OH             & 0                  & $\bot$ CH$_{3}$O      & 0                  \\
$\bot$ CO             & 5.0$\times10^{-5}$ & $\bot$ CH$_{3}$OH     & 0                  \\
\hline
\end{tabular}
\label{tab:species2} \\
Notes. The symbol $\bot$ denotes a surface/ice species. $\rm \bot H_c$ is the chemically adsorbed counterpart of $\rm \bot H$.
\end{table}
Of these, 25 are gas-phase and 16 are surface species. Together they comprise 280 reactions. The complete set of reactions is listed in appendix\,\ref{app:CC}.

The choice of initial values can sometimes have an effect on the final results. The presence or absence of a species at the start, either on dust or in the gas-phase, may influence the thermodynamical evolution. This is something we noticed during our trial runs. Therefore, it is appropriate to start with more realistic initial abundances according to the conditions of the model cloud. This creates a small difference in the beginning between the models. We obtained our initial abundances by performing two 1D simulations of a diffuse cloud, with and without surface species. We used atomic initial conditions, except for hydrogen which was in molecular form, and typical Galactic elemental abundances as given in Table \ref{tab:species} for the cloud edge. We ended these runs when the diffuse clouds evolved into the starting conditions of this work. We neglected all the abundances below a value of $n_{x_i}/n_{\rm H} = 10^{-9}$.

For all three models, we adopt solar metallicity and a dust-to-gas mass ratio of 0.01, a common value for local galaxies \citep{2014Natur.505..186F}. A lower ratio or metallicity is expected to result in a reduced star-formation rate, even in the same galaxy \citep{2015MNRAS.451..103Y}. The grain-size distribution is that of \cite{1977ApJ...217..425M}, hereafter MRN.

\subsection{Non-thermal desorption}
\label{sec:nonthermal}
We consider non-thermal desorption from two different channels, i.e., desorption through UV photons and through exothermic reactions. We plan to include desorption by direct CR impact in the future. The photodesorption is included for the species CO, H$_2$O, H$_2$CO, CH$_3$OH which is induced either by primary UV photons or by secondary UV photons from CRs. The photodesorption rate is calculated as
\begin{eqnarray}
R_{\rm phdes} &=& n_{x_i} f_{\rm ss} k_{\rm phdes} ~~~\rm cm^{-3} s^{-1},
\\ \nonumber
k_{\rm phdes} &=& \frac {\chi F_{\rm Draine}} {4 n_{\rm surf} N_{\rm lay}} Y_i \simeq 2.16\times10^{-8} \chi Y_i ~~~\rm s^{-1},
\label{eq:phdes1}
\end{eqnarray}
where $R_{\rm phdes}$ is the photodesorption rate and $k_{\rm phdes}$ is the rate coefficient following \cite{2012A&A...537A.138C}. In these equations $\chi$ is the UV field strength \citep{1978ApJS...36..595D}, whereas the photon flux produced by this field per unit area is $F_{\rm Draine} = 1.921\times10^8$ $\rm cm^{-2} ~s^{-1}$ \citep{2009A&A...501..383W}. Furthermore, $n_{\rm surf} = 1.11\times10^{15}$ $\rm cm^{-2}$ is the surface density of available absorption sites per unit grain area assuming $3$\,\AA{} separation between sites, $N_{\rm lay} = 2$ is the assumed number of ice layers that photons can penetrate for photodesorption \citep{2006JChPh.124f4715A, 2010JChPh.132r4510A, 2010A&A...522A.108M}, $n_{x_i}$ is the number density of species $x_i$, $f_{\rm ss}$ is the self-shielding factor, which is obtained from \cite{1996ApJ...468..269D} for H$_2$ and from \cite{2009A&A...503..323V} for CO, and $Y_i$ is the photodesorption yield per photon. The photodesorption yields are given in Table \ref{tab:yields}.
\begin{table}
\caption{Photodesorption yields}
\begin{tabular}{lrc}
\hline
\hline
Species  & Yield & Reference \\
\hline
CO	 & $1\times10^{-2}$      & {\cite{2011ApJ...739L..36F}, and refs. therein} \\ %older Oberg prediction was 2.7e-3 also adopted by German
H$_2$O	 & $f(x,T)\times10^{-3}$ & {\cite{2009ApJ...693.1209O}} \\
H$_2$CO	 & $1\times10^{-3}$      & {\cite{2013A&A...560A..73G}} \\ %assumption by guzman by relating it to CO
CH$_3$OH & $\leq 2\times10^{-3}$ & {\cite{2009A&A...504..891O}} \\ %actually 2.1e-3 and for pure methanol ice, guzman used 2e-4
\hline
\end{tabular}
\label{tab:yields} \\
Notes. Here, $f(x,T) = (1.3 + 0.032\times T ) (1 - e^{-x/\ell(T)} )$. See the reference.
\end{table}

It is advisable to note that the photodesorption yield for methanol should be taken as an upper limit. Recent work by Mu{\~n}oz Caro et al. (in preparation) suggests that the actual yields are much lower, by about a factor 10. It seems that methanol preferably dissociates rather than to desorb. On a similar note, CO photodesorption yields are found to be increasing with decreasing temperature \citep{2010A&A...522A.108M}. These are not considered here.

The CRUV photon flux is computed by following once again \cite{2012A&A...537A.138C},
\begin{eqnarray}
R_{\rm crdes} &=& n_{x_i} k_{\rm crdes} ~~~\rm cm^{-3} s^{-1},
\\ \nonumber
k_{\rm crdes} &=& \frac {N_{\rm CU}} {4 n_{\rm surf} N_{\rm lay}} Y_i \times \left(\frac{\zeta_{\rm CR}}{5\times10^{-17} \rm s^{-1}}\right) \simeq 225.2 ~\zeta_{\rm CR} Y_i ~~~\rm s^{-1},
\label{eq:phdes2}
\end{eqnarray}
where $N_{\rm CU} = 10,000$ is the flux of locally generated CRUV photons in cm$^{-2}$ s$^{-1}$ for molecular clouds \citep{2004A&A...415..203S} and $\zeta_{\rm CR}$ is the CR ionization rate per H$_2$ molecule. For the $\zeta_{\rm CR}$ in this work we take $5\times10^{-17} \rm s^{-1}$ \citep{2000A&A...358L..79V, 2006PNAS..10312269D, 2009A&A...501..619P}.

Chemical desorption is included for all the reactions that are exothermic. We use a probabilistic approach for the reaction products to desorb into the gas phase or to stay on the surface. Our simulations are run with the desorption probabilities that have been recently obtained experimentally by \cite{2013NatSR...3E1338D} and \cite{2016A&A...585A..24M}, with the assumption that we have a mixed graphite/silicate dust. Table \ref{tab:chemdes} provides the desorption probabilities.
\begin{table}
\caption{Chemical desorption probabilities}
\begin{tabular}{lclccc}
\hline
\hline
& & & \multicolumn{3}{c}{Probability for products to desorb} \\
\multicolumn{3}{c}{Surface reaction} & & Bare  & Ice \\
\hline
$\bot$H + $\bot$H		& $\rightarrow$ & H$_2$       & & 0.900 & 0.800 \\
$\bot$H + $\bot$O		& $\rightarrow$ & OH          & & 0.500 & 0.250 \\
$\bot$H + $\bot$OH		& $\rightarrow$ & H$_2$O      & & 0.600 & 0.300 \\
$\bot$H + $\bot$O$_3$		& $\rightarrow$ & OH + O$_2$  & & 0.080 & 0.016 \\
$\bot$H + $\bot$H$_2$O$_2$	& $\rightarrow$ & OH + H$_2$O & & 0.025 & 0.025 \\
$\bot$H + $\bot$CO		& $\rightarrow$ & HCO         & & 0.100 & 0.050 \\
$\bot$H + $\bot$HCO		& $\rightarrow$ & H$_2$CO     & & 0.080 & 0.016 \\
$\bot$H + $\bot$H$_2$CO		& $\rightarrow$ & CH$_3$O     & & 0.040 & 0.008 \\
$\bot$H + $\bot$CH$_3$O		& $\rightarrow$ & CH$_3$OH    & & 0.080 & 0.016 \\
$\bot$O + $\bot$O		& $\rightarrow$ & O$_2$       & & 0.600 & 0.050 \\
\hline
\end{tabular}
\label{tab:chemdes}
\end{table}
For ice, if a desorption probability could not be derived from the experiments because the probability was too low, we assume that the chemical desorption is one-fifth of the desorption probability derived on bare surfaces. More precise measurements are being performed. We use the methods from \cite{2015A&A...576A..49H} to calculate the chemical desorption rate, which is also explained in Sect.\,\ref{sec:twobody} with additional details.

One important point is that when considering both photodesorption and chemical desorption, there is the possibility that the desorption yields provided for photons does already partially include chemical desorption, since photons can also dissociate molecules only to see them immediately reform. The photodesorption yields cannot distinguish between direct and indirect (chemical) desorption. This is not an important issue in our work, because we do not consider high UV fields and we focus on regions with $A_V>1$\,mag, where chemical desorption always dominates over photodesorption.

\subsection{Thermodynamics}
\label{sec:thermodynamics}
In order to properly address the thermodynamics of the gas cloud, we compute the heating and cooling rates time dependently that go hand-in-hand with the chemistry calculations. In this way, by solving the thermal balance, the gas and the dust temperature is obtained. Since the dust temperature is directly linked to the chemistry, we compute this at each iteration of the convergence sub-cycle of the chemistry time-step.

For our models we treat the most important heating and cooling rates that pertain to molecular clouds within a large temperature range ($T_{\rm g} = 5-5000$\,K). The used heating and cooling processes are summarized here with the details provided in appendix\,\ref{app:BB}. The references in the parentheses point to the appropriate appendix sections where they are further discussed. The gas-phase heating processes we consider are:

\begin{description}[style=multiline,leftmargin=1.05cm]%,font=\normalfont]
 \item[(\ref{H1})] Photoelectric heating,
 \item[(\ref{H2})] H$_2$ photodissociation heating,
 \item[(\ref{H3})] H$_2$ collisional de-excitation heating,
 \item[(\ref{H4})] CR heating,
 \item[(\ref{HC})] Gas-grain collisional heating,
\end{description}
while the cooling processes in the gas phase are:

\begin{description}[style=multiline,leftmargin=1.05cm]%,font=\normalfont]
 \item[(\ref{C1})] Electron recombination cooling,
 \item[(\ref{C2})] Lyman-$\alpha$ cooling,
 \item[(\ref{C3})] Metastable [OI]-630 nm cooling,
 \item[(\ref{C4})] [OI]-63 $\mu$m fine-structure line cooling,
 \item[(\ref{C4})] [CI]-609 $\mu$m fine-structure line cooling,
 \item[(\ref{C4})] [CII]-158 $\mu$m fine-structure line cooling,
 \item[(\ref{C5})] H$_2$ ro-vibrational line cooling,
 \item[(\ref{C5})] H$_2$O ro-vibrational line cooling,
 \item[(\ref{C5})] CO ro-vibrational line cooling,
 \item[(\ref{C5})] OH ro-vibrational line cooling,
 \item[(\ref{HC})] Gas-grain collisional cooling.
\end{description}

The most important heating processes for the clouds in this work come from photoelectric heating at cloud edge and from compressional heating together with the CR heating in deeper regions. Main cooling is coming from the atomic finestructure lines at low densities, while CO ro-vibrational line cooling takes over once molecules start to form. Eventually, at densities above $10^{4.5}$\,\cmcube, gas-grain collisional coupling will dominate the heat transfer between the two phases. Several of the heating and cooling processes are treated in an approximate fashion, which is explained in the relevant sections of the appendix. Molecular cooling by formaldehyde and methanol, which could potentially be significant, is not treated in this work.

\subsection{Dust temperature}
\label{sec:tdust}
For surface reactions the dust temperature is essential. Since many reactions rates depend exponentially on the dust temperature, a precise knowledge of this variable is highly preferred. This is especially important in the colder parts of the cloud with a temperature of around 10\,K, where a single Kelvin temperature fluctuation can imply an order of magnitude difference in the rates. It is advisable to note that there are also great uncertainties in the surface binding and diffusion energies of many species. Fortunately, such uncertainties do not necessarily translate into as big fluctuations in the final abundances, since calculated abundances are more sensitive to the relative reaction rates between those who compete with each other rather than the absolute reaction rates.

The dust temperature also (indirectly) affects the gas temperature. This is because many of the important cooling rates depend on the abundances of key species that are influenced by dust, and also because of the strong heat transfer due to the gas-grain collisional coupling at densities above a few 10$^4$ \cmcube. In the end, the reaction rates will drive the formation of species, build-up of ices, and the depletion of gas-phase molecules.

To calculate the dust temperature we adopt the formulation presented by \cite{2001A&A...376..650Z}. This was extended to encompass also the low $A_V$ by \cite{2011ApJ...735...15G}. The adopted equations for $A_V < 10$ \citep{2011ApJ...735...15G} and for $A_V \geq 10$ \citep{2001A&A...376..650Z} are, in this order, formulated as
\begin{eqnarray}
T_{\rm dust} (A_V) = 18.67 - 1.637\,A_V + 0.07518\,A_V^2 - 0.001492\,A_V^3,
\end{eqnarray}
\begin{eqnarray}
T_{\rm dust}^{5.6} (A_V) = \left( 6.2 - 0.0031\,A_V \right)^{5.6} + \left( 43\,A_V^{-0.56} - 77\,A_V^{-1.28} \right)^{5.6} + 
\\ \nonumber
\left( 1.8 - 0.098\,A_V^{0.5} + 7.9\times10^{-5}\,A_V^{1.5} \right) \times \left( 7.9\,A_V^{-0.089} \right)^{5.6}.
\label{eq:zucc2}
\end{eqnarray}

By adopting the above mentioned way of calculating the dust temperature, we stepped down from using the analytical expression of \cite{1991ApJ...377..192H} as we did in Paper\,I, which is a major difference with the this work.

The \cite{2001A&A...376..650Z} dust temperature is obtained from a fit to the observed dust temperature of the prestellar core L1544 at various $A_V$. We believe that this reproduces the dust temperature much better at low and at high $A_V$ for $G_0=1$. In an upcoming work (Hocuk et al., in preparation), we will perform a much more detailed analysis of the dust temperature in which we explore the impact of the refractory material, ice growth, coagulation, and heating by CRs and provide our own formulation.

\section{Results}
\label{sec:results}
We present our results on the impact of dust grains on cloud dynamics and star formation. For the three models, \textit{pure gas}, \textit{gas+freeze-out}, and \textit{full chemistry}, as described in Sect.\,\ref{sec:models}, we report in the coming sections the chemical evolution, the thermal profile, the gas compressibility, fragmentation, star formation, and the mass distribution. In all our models, we allow the clouds to evolve until the cloud free-fall time of $t_{\rm ff} \equiv \sqrt{3\pi/32G\rho_0}$ = $1.63$ Myr is reached. The presented results are obtained from these final stages.

\subsection{General behavior}
\begin{figure}
\centering
\includegraphics[scale=0.4]{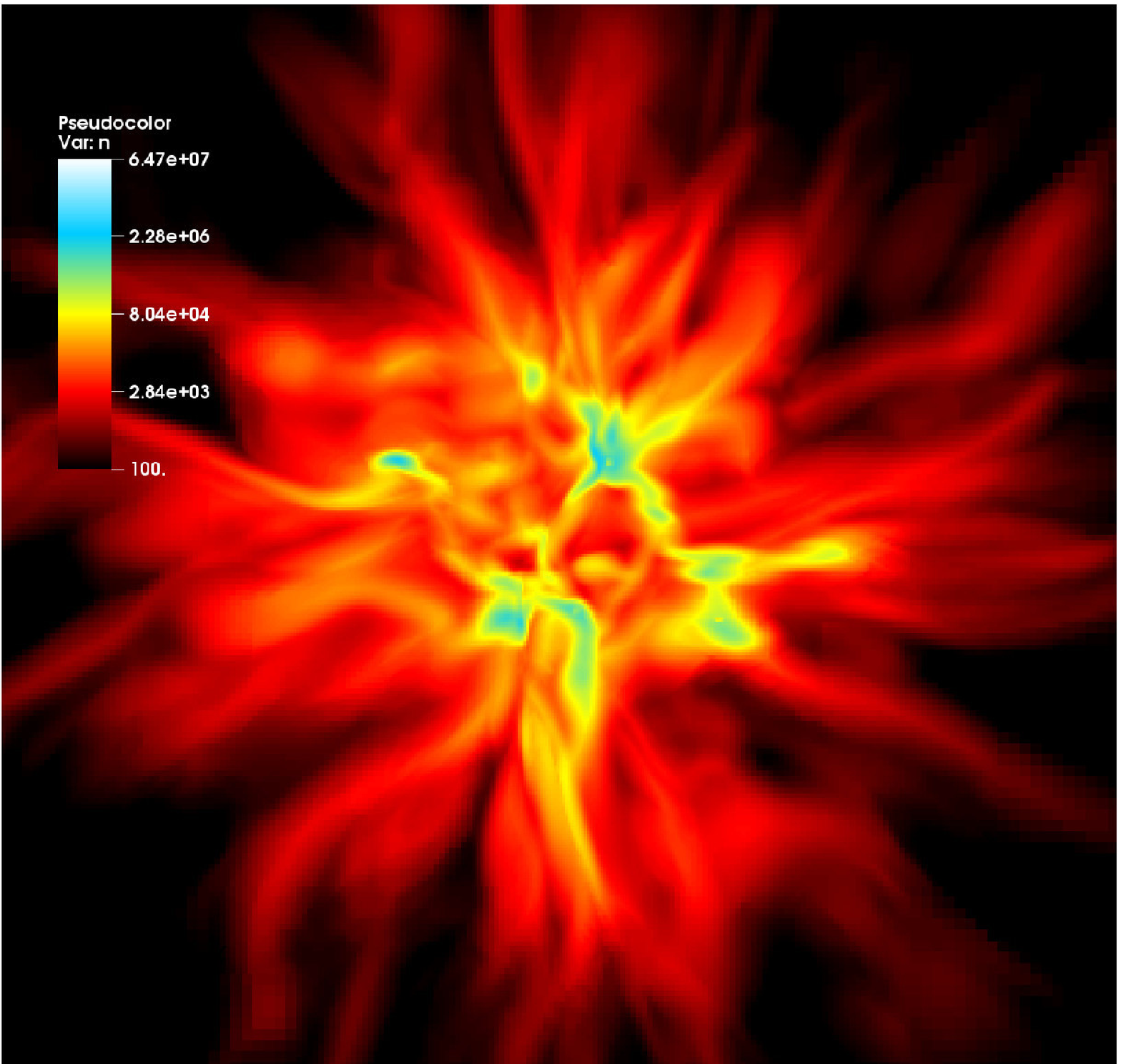}
\includegraphics[scale=0.4]{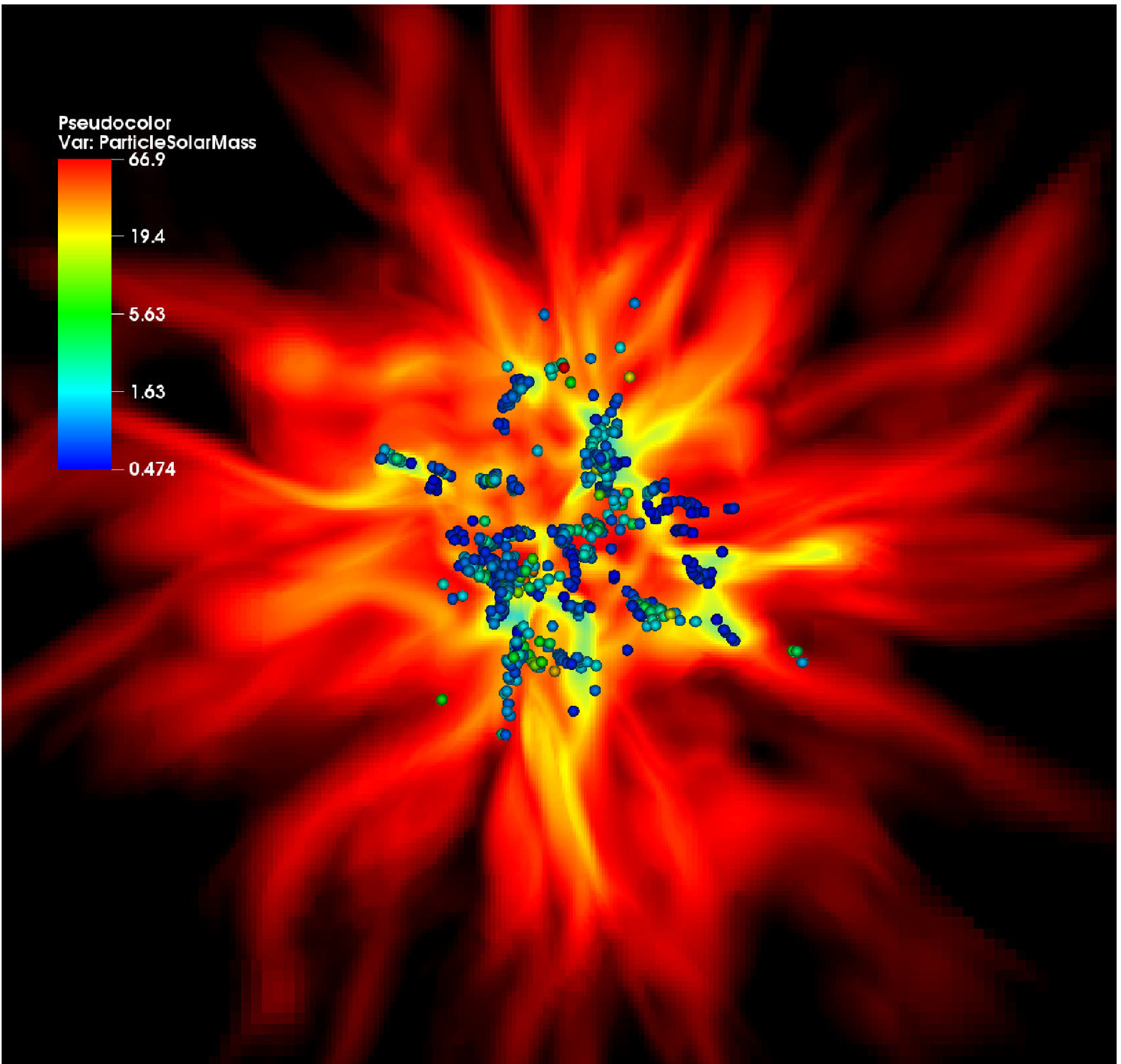}
\caption{A molecular cloud just before collapse. The images show the cloud in the \textit{pure gas} model at a time of 0.99 \tff. On the top, a density slice through the center along the Z-axis is shown. On the bottom, sink particles are over plotted on top of the slice with a slight position shift to make most of the sink particles visible. The legends show the number density in \cmcube (top) and the particle mass of the sink particles in \msol (bottom).}
\label{fig:endgame}
\end{figure}

All clouds reach the density of \nhtot $\gtrsim 10^7$\cmcube, while sink particle formation already occurs in clumps with gas densities between \nhtot $= 3\times 10^5 - 10^7$\cmcube. The gas temperature lingers around in the range 7-10 K in the denser regions at which point the dust temperature is strongly coupled to the gas. Sink particles form in all three models. In every case, the interstellar cloud collapses without delay. With this comes a caveat, which is that simulated non-magnetized clouds collapse too fast once gravitational contraction kicks in. As a consequence, the star formation efficiencies are generally too high \citep{2007ApJ...654..304K, 2015ApJ...800L..11V}. Our velocities and efficiencies will therefore be higher compared to observations. We show in Fig.\,\ref{fig:endgame} density slices of the \textit{pure gas} model, representative for all three models, in the final moments before cloud collapse.

\subsection{Chemical evolution}
\begin{figure*}
\includegraphics[scale=0.329]{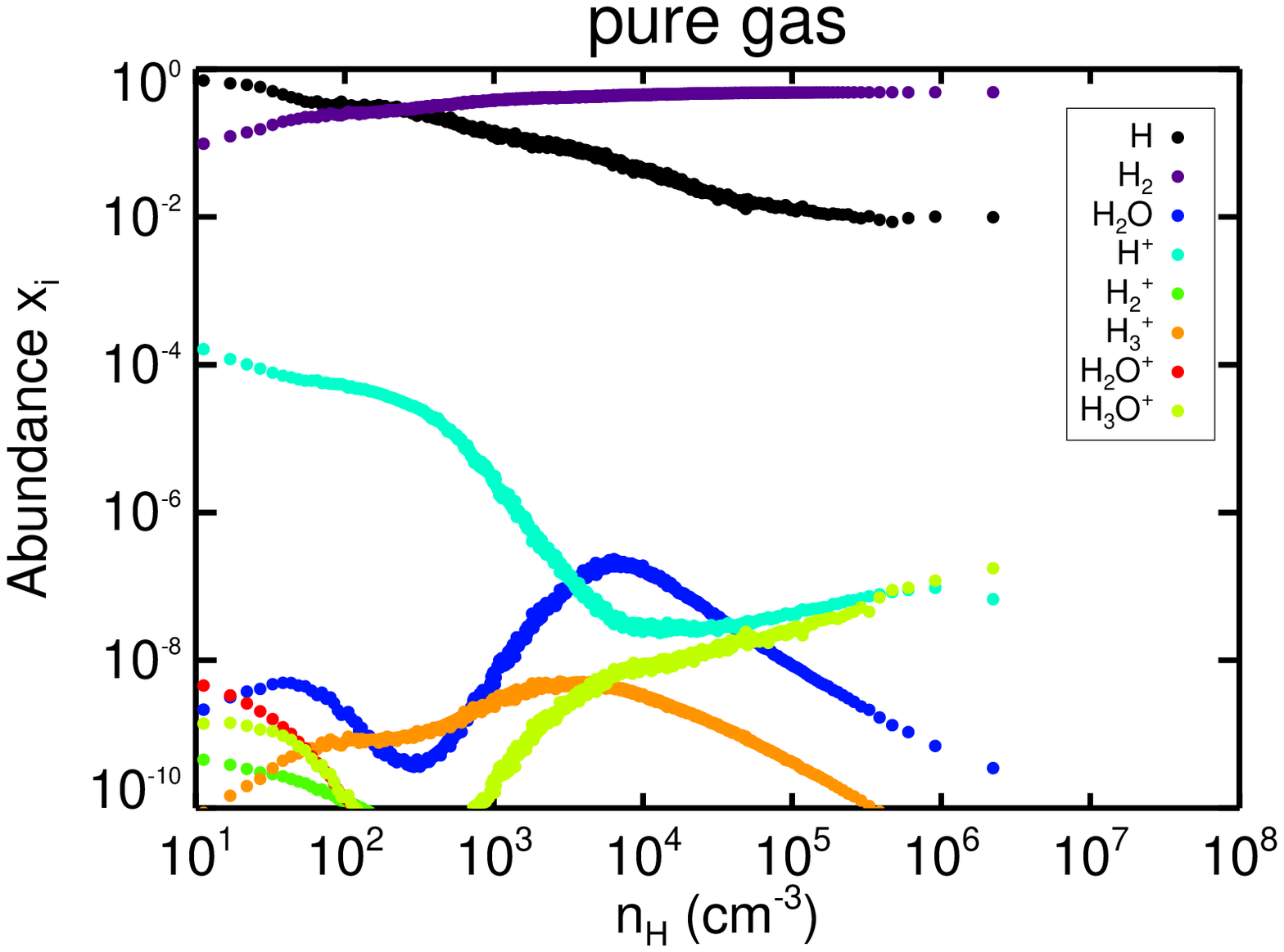}
\includegraphics[scale=0.329]{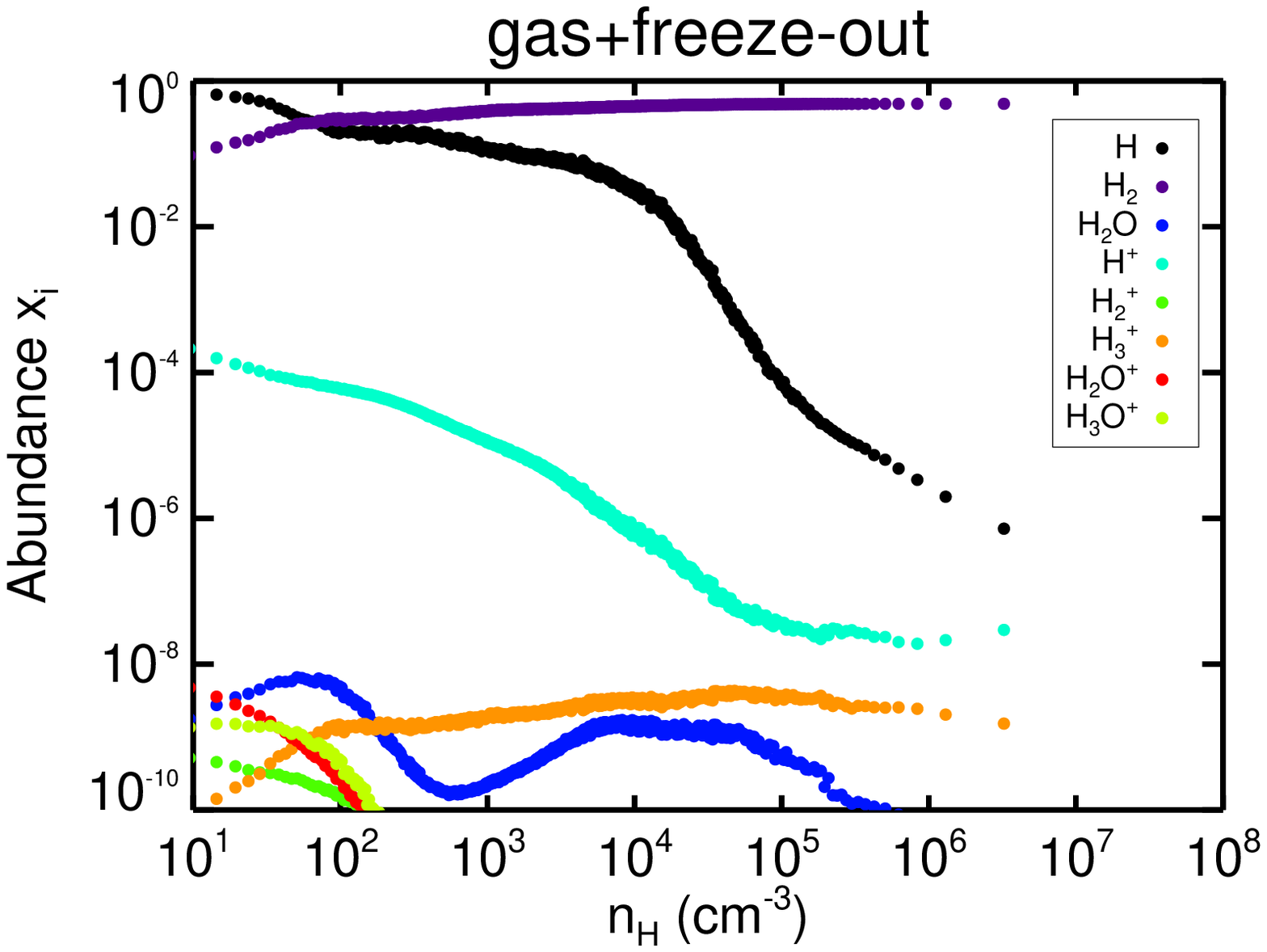}
\includegraphics[scale=0.329]{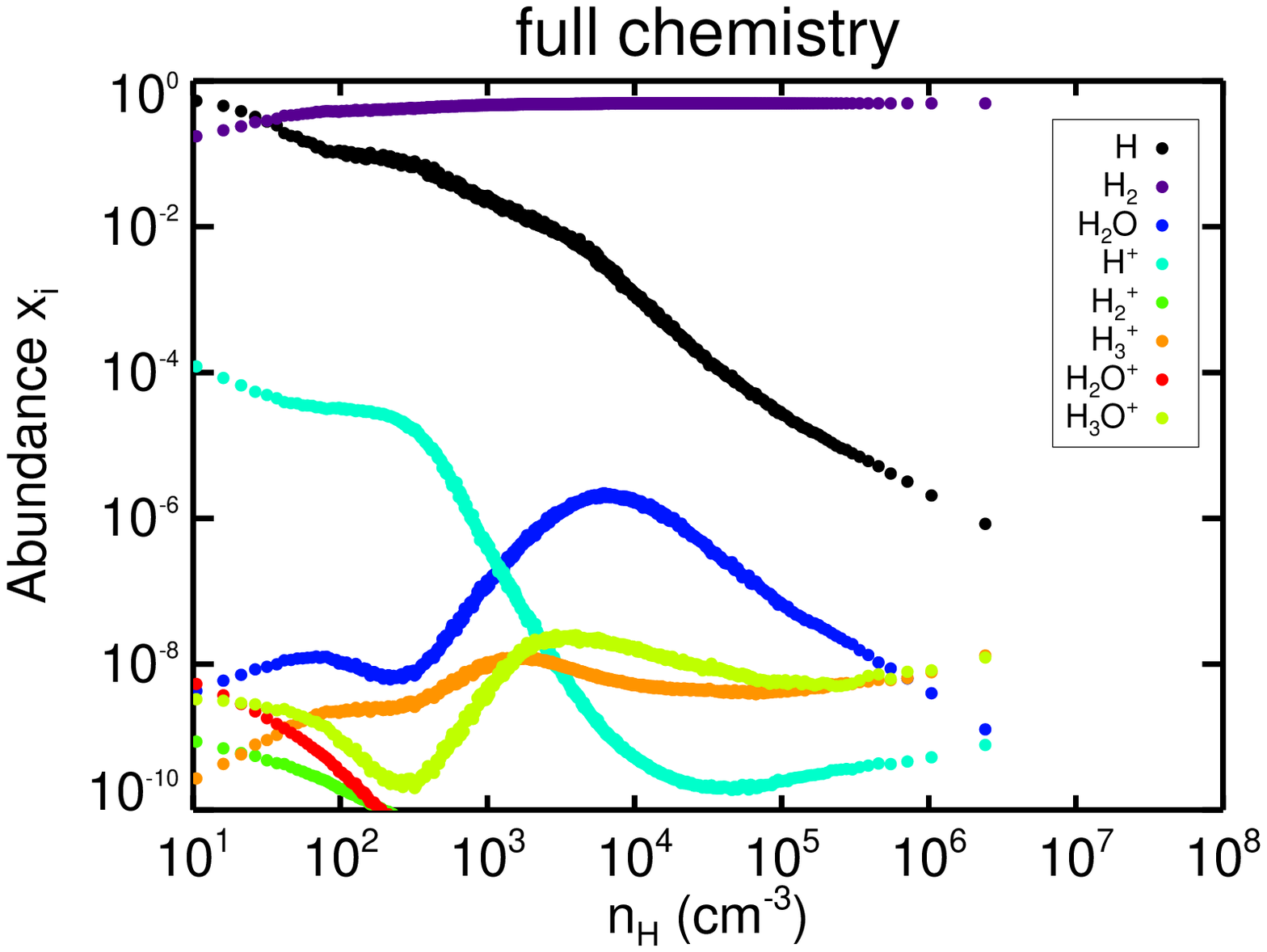}
\includegraphics[scale=0.329]{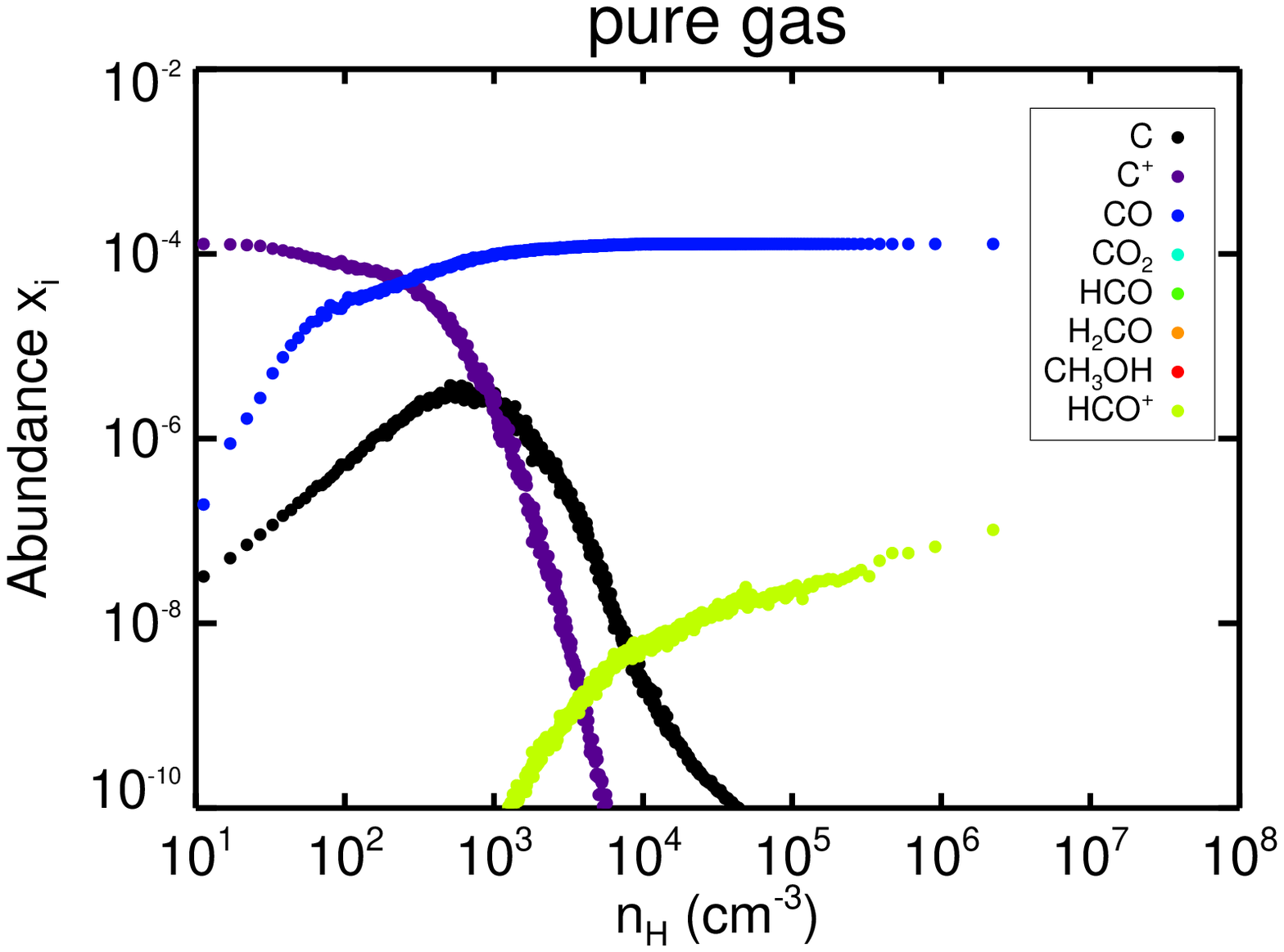}
\includegraphics[scale=0.329]{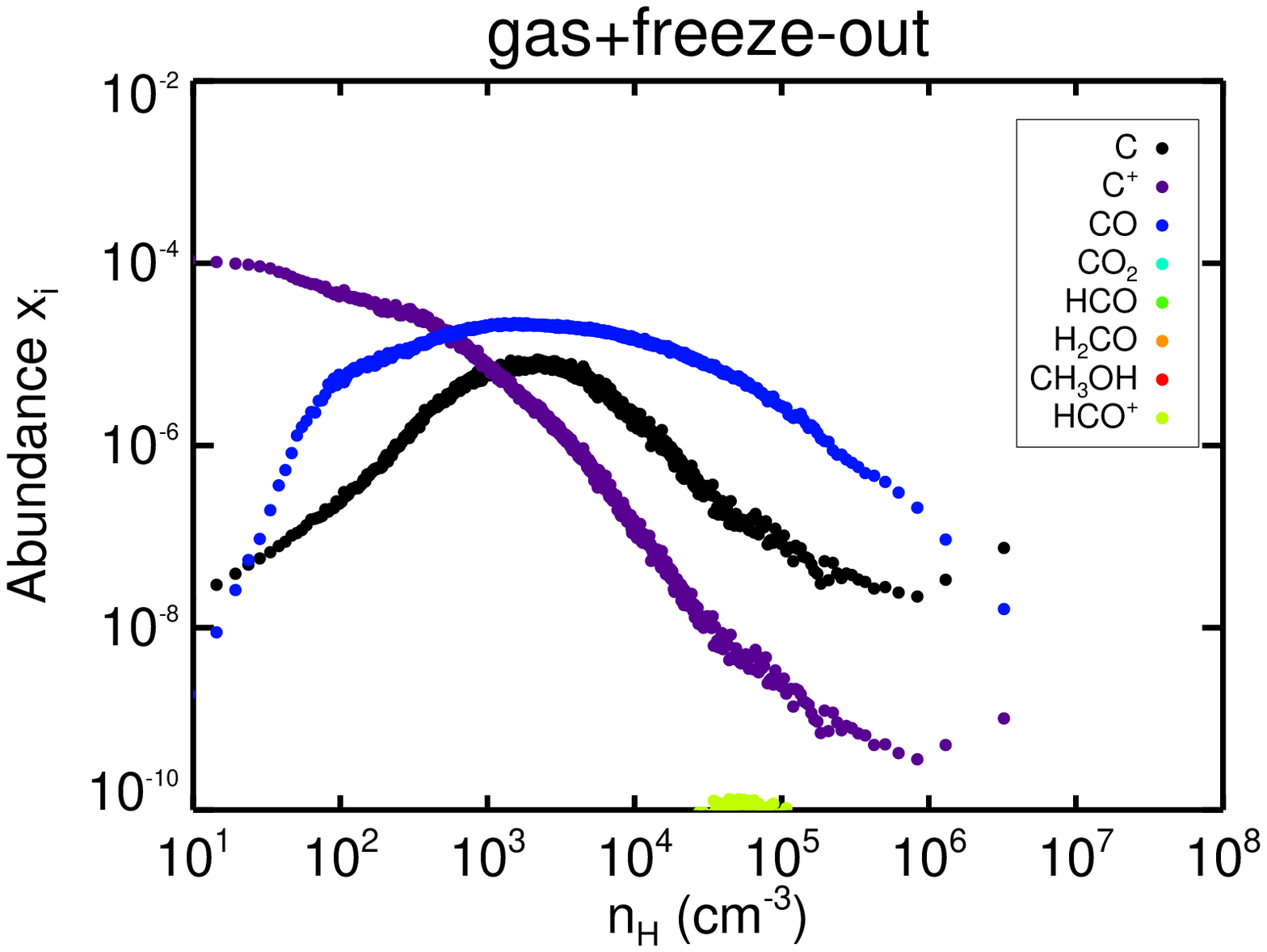}
\includegraphics[scale=0.329]{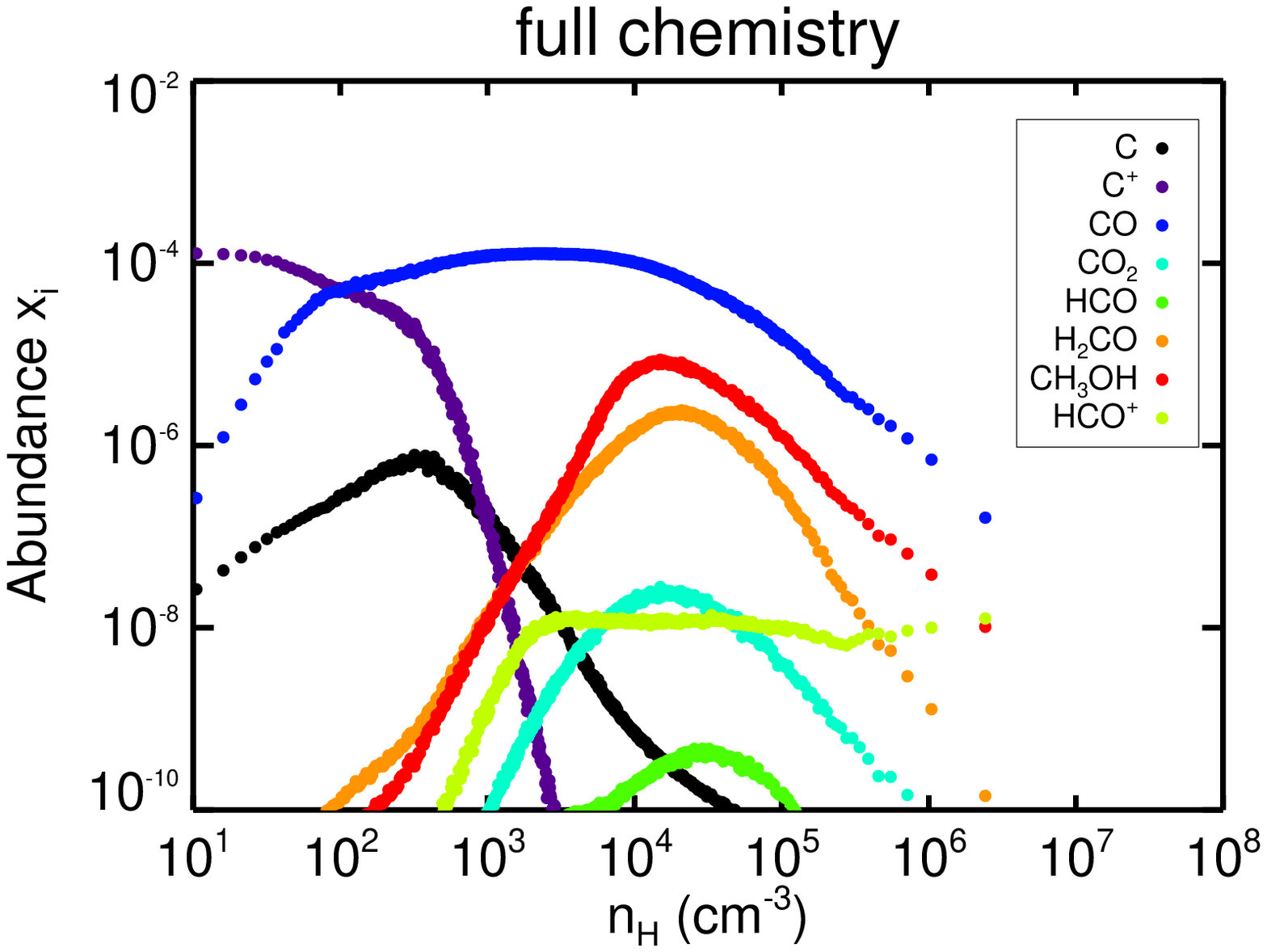}
\includegraphics[scale=0.329]{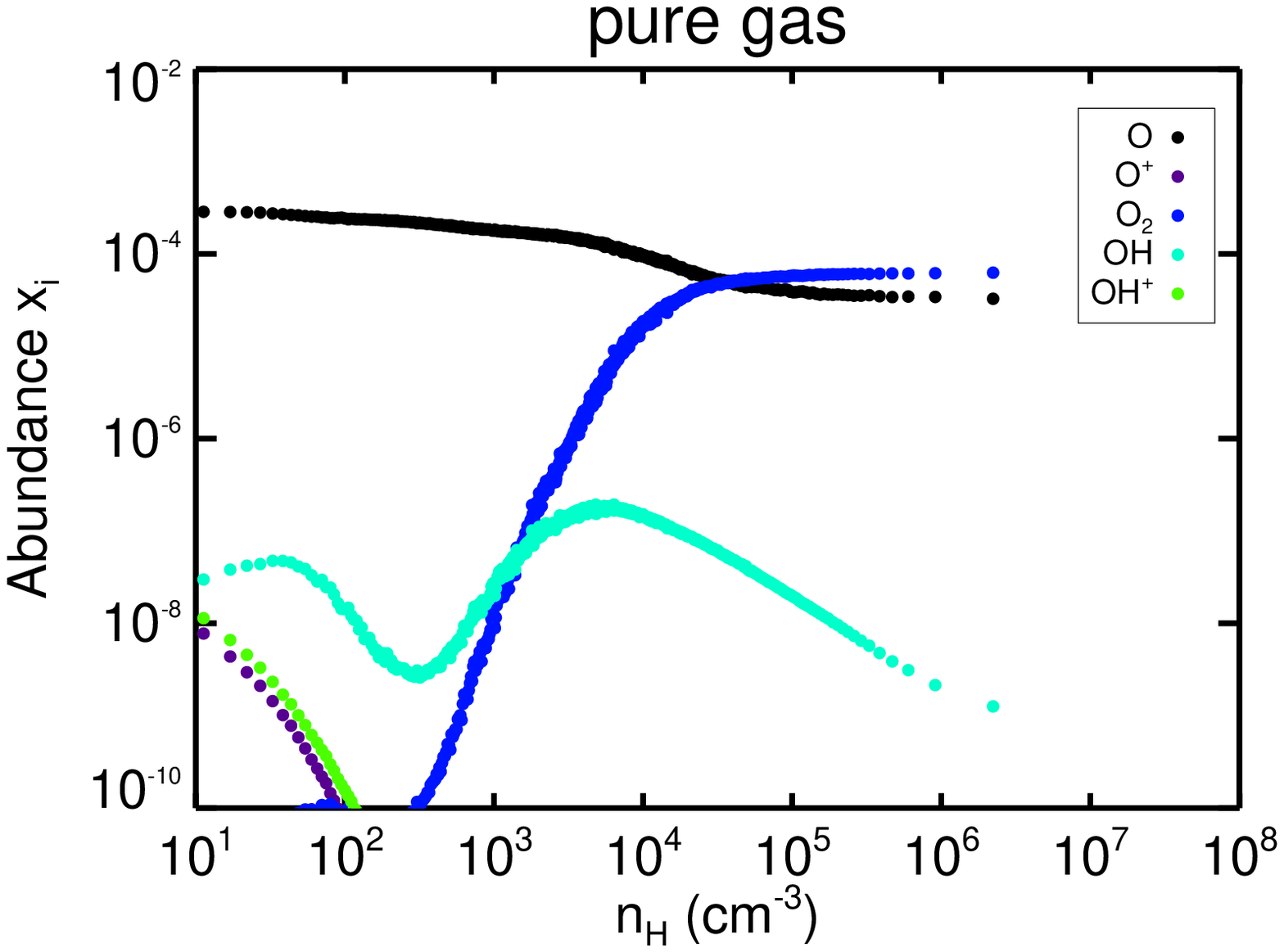}
\includegraphics[scale=0.329]{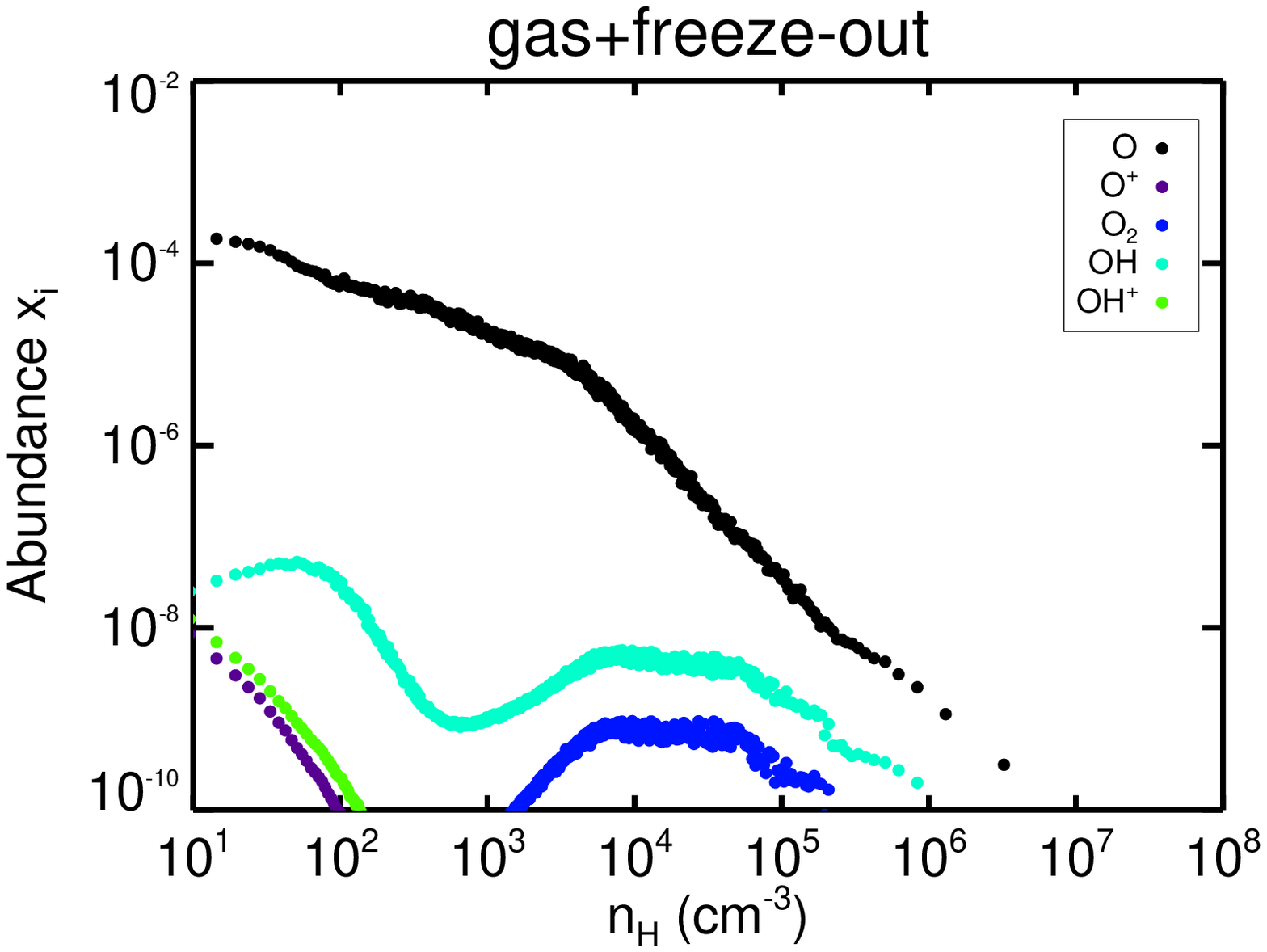}
\includegraphics[scale=0.329]{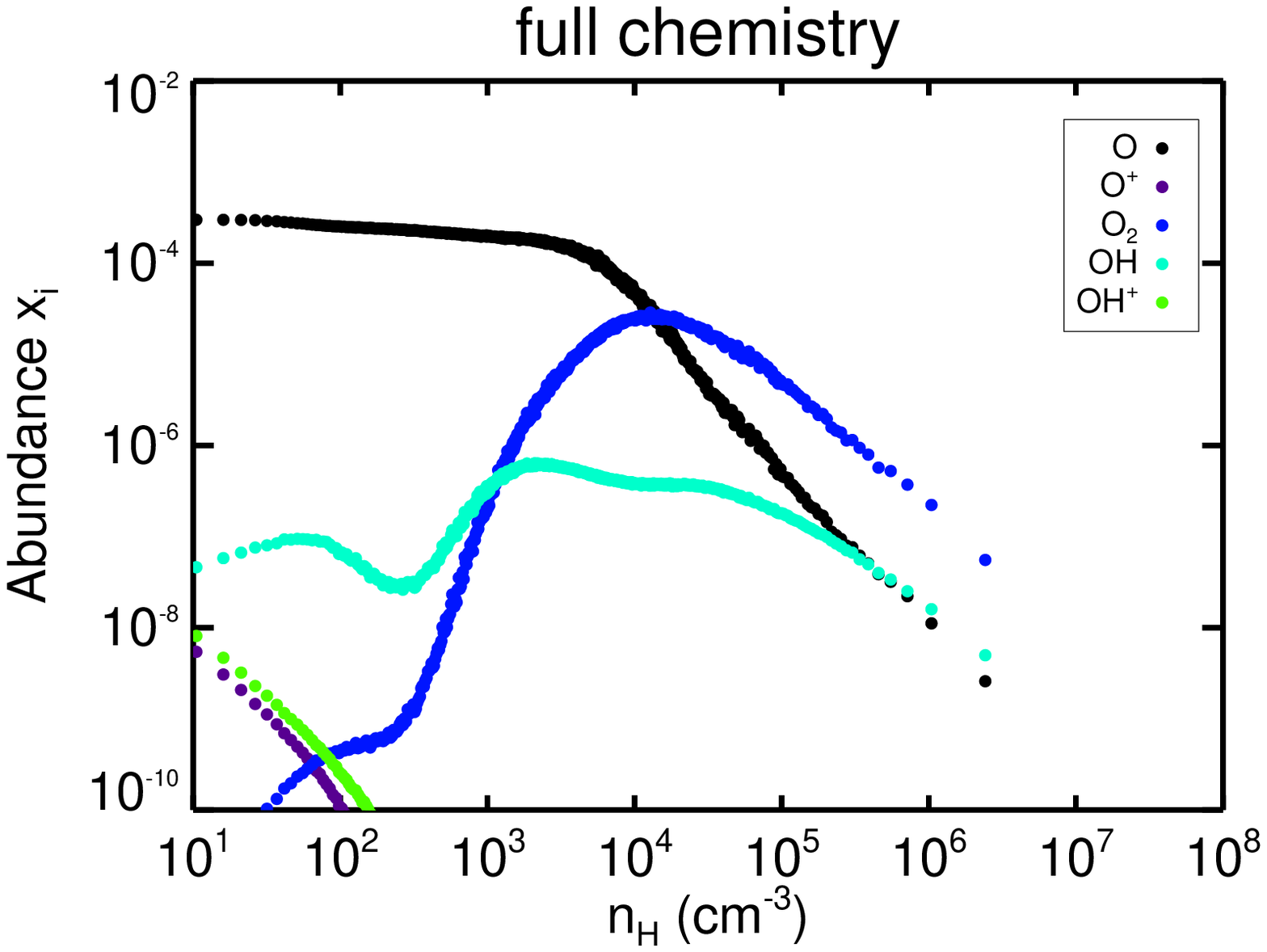}
\includegraphics[scale=0.329]{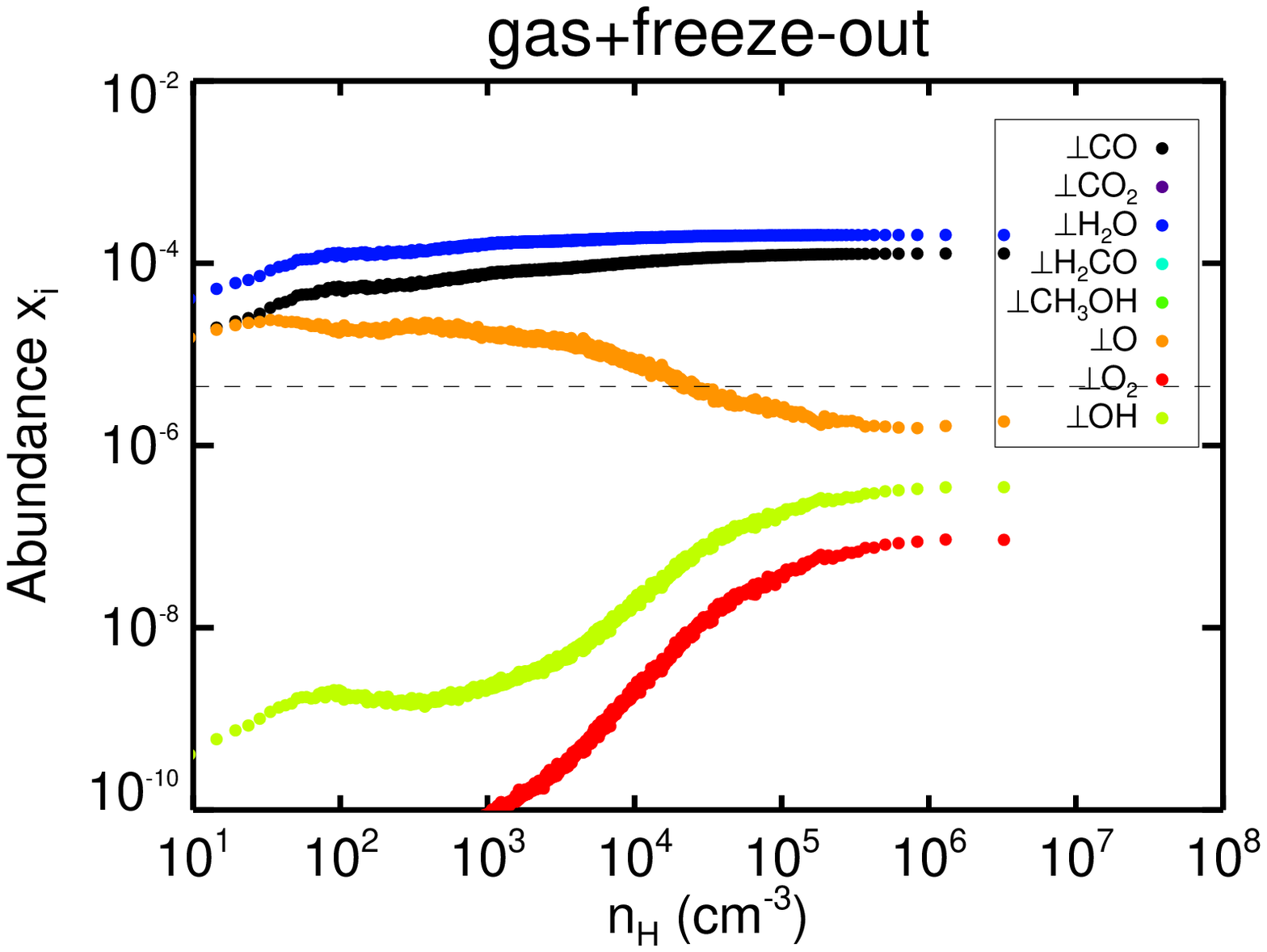}
\includegraphics[scale=0.329]{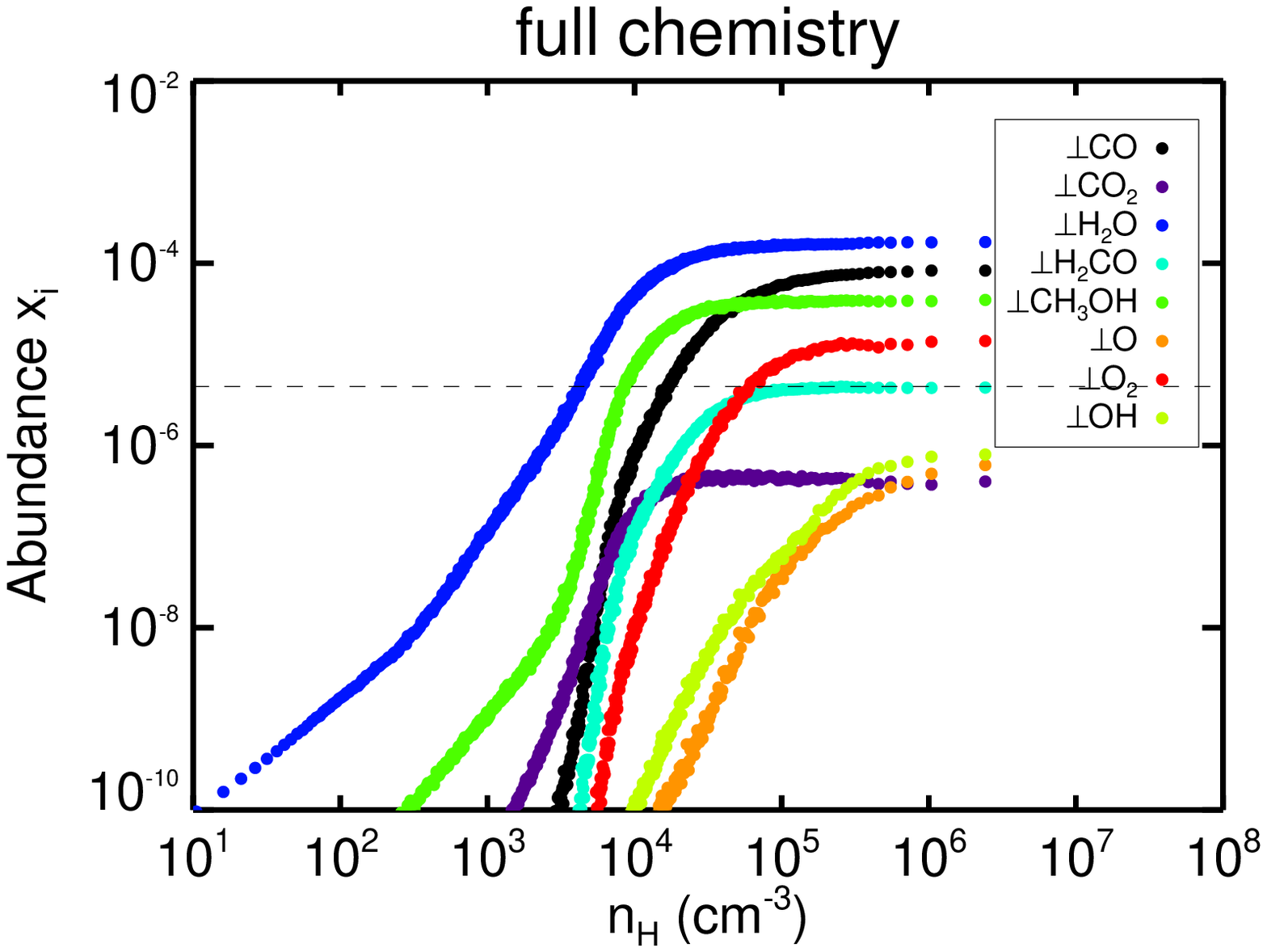}
\caption{Abundances of gas and ice species. From left to right, the models \textit{pure gas}, \textit{gas+freeze-out}, and \textit{full chemistry} are shown. From top to bottom, species related to H/H$_2$O, C/CO, and O/OH are shown. The undermost panels show the ice species for the models \textit{gas+freeze-out}, and \textit{full chemistry}.}
\label{fig:abun}
\end{figure*}

The biggest impact of dust on an interstellar cloud is on the abundances of species. The model clouds strongly differ in their chemical evolution and there are several fundamental differences that we highlight here. In Fig.\,\ref{fig:abun} we show the abundances of a select number of species as a function of number density.

\vspace{0.15cm}
\textbf{The \textit{pure gas} model} evolves by converting H into H$_2$ and C into CO. These transitions occur in the density regime around 250\,\cmcube, with the H abundance stabilizing at \nhtot$= 10^5$\,\cmcube. The H to H$_2$ transition can be a bit deceptive here, since we already start the cloud molecular towards the center, but an equilibrium in creation and destruction of H$_2$ is reached irrespective of starting conditions. We can see that O drops about an order of magnitude to reach a stable abundance of $n(\rm O)/n_{\rm H} = 10^{-4.5}$, while O$_2$ becomes more abundant than O in the gas phase. A lot of atomic hydrogen and oxygen is still present at high densities (\nhtot$\geq10^5$\,\cmcube). This means that there are no efficient (gas-phase) channels to deplete the atomic species in cold, $T_{\rm g} \approx 10$\,K, environments. O$_2$, which is usually overpredicted in numerical models \citep[see, e.g.,][]{2013A&A...558A..58Y} is also clearly too high when surface chemistry is not considered. We obtain a value of $n(\rm O_2)/n_{\rm H} = 7\times10^{-5}$ at the end of the run. Water abundance peaks at a density around \nhtot=10$^{3.6}$\,\cmcube, with a value of a few 10$^{-7}$. This then drops below 10$^{-9}$ at higher densities. In this region, the destruction of water is mostly due to the positively charged ions. The cations H$^+$, H$_2^+$, and H$_3^+$ are efficiently reacting with water molecules to form H$_2$O$^+$ and H$_3$O$^+$. The last mentioned can convert back to water when encountering anions or electrons. There is no molecule formation more complex than water or HCO$^+$. The abundances of species such as formaldehyde or methanol are negligible in this run. This suggests that surface chemistry is required to form complex (organic) molecules, at least, in a bottom-up chemical model. A top-down model stems from breaking dust grains into smaller pieces \citep{2015ApJ...800L..33G}.

\vspace{0.15cm}
\textbf{In the \textit{gas+freeze-out} model} all gas-phase species, aside from H and H$_2$, either remain below the abundance of 10$^{-7.5}$ or dive under it at densities above \nhtot$\geq10^6$\,\cmcube. Only few cations, like H$^+$ and H$_3^+$, have a notable presence, which follows from the high abundances of H and H$_2$ in the gas phase. In contrast to the \textit{pure gas} simulation, the abundance of atomic hydrogen drops nearly linearly with increasing density, corresponding to a constant number density of roughly $n(H)\approx1$\,\cmcube in every region where \nhtot$\geq 10^5$\,\cmcube. We can see that this is also true for the \textit{full chemistry} model, even at lower densities, i.e., \nhtot$\geq 10$\,\cmcube where 1\,\cmcube$\leq n(H)\leq10$\,\cmcube. While the abundance of OH in the \textit{gas+freeze-out} model is only slightly lower in comparison to the \textit{pure gas} model, O and O$_2$ are greatly reduced. The abundance of CO, after reaching a value of a few 10$^{-5}$, drops rapidly, because it freezes out. CO is, however, able to survive in the gas phase at low densities, because of the low adsorption rates. The timescale to CO freeze-out, which can be estimated by $t_D \approx 10^9/n_{\rm H}$ yr, is longer at \nhtot\,$< 600$\,\cmcube than the cloud dynamical time. The density at which freeze-out starts to occur noticeably is around \nhtot$\geq 10^3$\,\cmcube. This is rather significant, because it is the regime where CO begins to dominate the thermal balance. Several observations are in agreement with this result by seeing significant ($\geq 50\%$) CO depletion around densities of $n_{\rm H}=10^4$\,\cmcube \citep{2012MNRAS.423.2342F, 2013ApJ...775L...2L}. So, the premise is valid.

\begin{figure}
\includegraphics[scale=0.48]{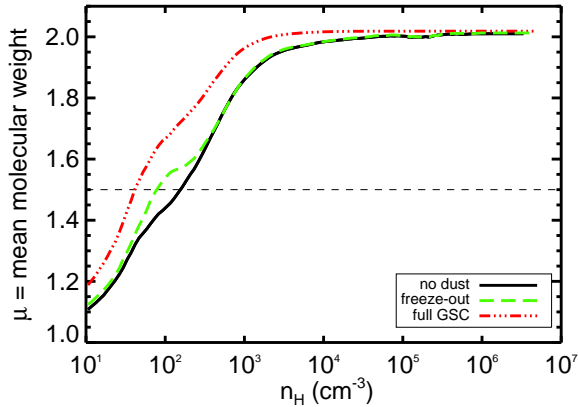}
\caption{Mean molecular weights. The horizontal dashed line indicates the transition from an atomic cloud to a molecular one.}
\label{fig:molmass}
\end{figure}

\vspace{0.15cm}
\textbf{The \textit{full chemistry} model} exhibits a much richer chemistry. The abundances of most species are much higher, and, as a result, the mean molecular weight $\mu$ increases. This is demonstrated in Fig.\,\ref{fig:molmass}. 
The biggest difference in $\mu$ comes from the dust catalyzed H$_2$ formation early on. We can see in Fig.\,\ref{fig:abun} that the H to H$_2$ transition occurs earlier when GSC is included. The transition happens around a density of \nhtot = 30\,\cmcube, slightly higher but consistent with observations of the Perseus molecular cloud \citep{2015ApJ...809..122B}. H$_2$ formation is either enhanced by surface reactions or by the ER process. That the ER mechanism plays a role in the H$_2$ formation at high temperatures is proven by the fact that the H to H$_2$ transition is also shifted to lower densities in the \textit{gas+freeze-out} model, where the ER process was the only extra formation channel in comparison to the \textit{pure gas} model. This shows that the high temperature formation route is a viable pathway to form H$_2$, in accordance with \cite{2015A&A...579A..72B}. We can also infer this from the bump between the green and the black lines around $\mu=1.5$ in Fig.\,\ref{fig:molmass}. Above \nhtot$=10^5$\,\cmcube, when nearly all hydrogen is in molecular form, the difference in $\mu$ is mostly generated due to the increased presence of the more massive molecules, which results in a much smaller difference in $\mu$ due to their relative low abundance. We reach a plateau at a $\mu$ of slightly above 2. This is lower than 2.3, because Helium is not included in our network. 

Water is forming abundantly in this model. The gas-phase abundance reaches a value of 10$^{-6}$ after which it seems to settle down at 10$^{-9}$. Carbon dioxide, formaldehyde, and methanol, are now also present in the gas phase. The oxygen abundance in this run maintains its descend after the onset of freeze-out until it reaches 10$^{-9}$ at \nhtot = 10$^7$\cmcube. Similar to atomic hydrogen, the oxygen number density stays constant at $n(\rm O)\approx 10^{-2}$\cmcube in every region where \nhtot$\geq 10^4$\,\cmcube. O$_2$ formation is slightly catalyzed by dust at low densities, giving higher abundances than those obtained through gas-phase reactions, and peaks around \nhtot$=10^4$\,\cmcube, but decreases afterwards due to freeze-out. This shows that O$_2$, like water and CO, is susceptible to freeze-out even if this does not make abundant O$_2$ ice. The total amount of O$_2$, in any phase, decreases after \nhtot = 10$^5$\cmcube. OH is more abundant compared to the other models and is no longer depleted below \nhtot$=10^5$\,\cmcube. New channels to form OH are provided by grain surfaces, with a high probability to chemically desorb. The drop in OH between \nhtot\,$\sim10^5-10^7$\,\cmcube is only caused by the depletion of O from the gas phase. The abundance of OH settles around a few $10^{-9}$. This is in contrast with the recent work by \cite{2014MNRAS.445L..56T} on the fact that OH is not tracing the central regions of dense cores because of freeze-out.

One major difference with respect to the other models is the presence of CO$_2$. From this we can conclude that GSC is a prerequisite for the formation of CO$_2$ as it is for formaldehyde, methanol, and other complex species.

\subsection{Temperature profile}
\begin{figure*}
\includegraphics[scale=0.40,angle=0]{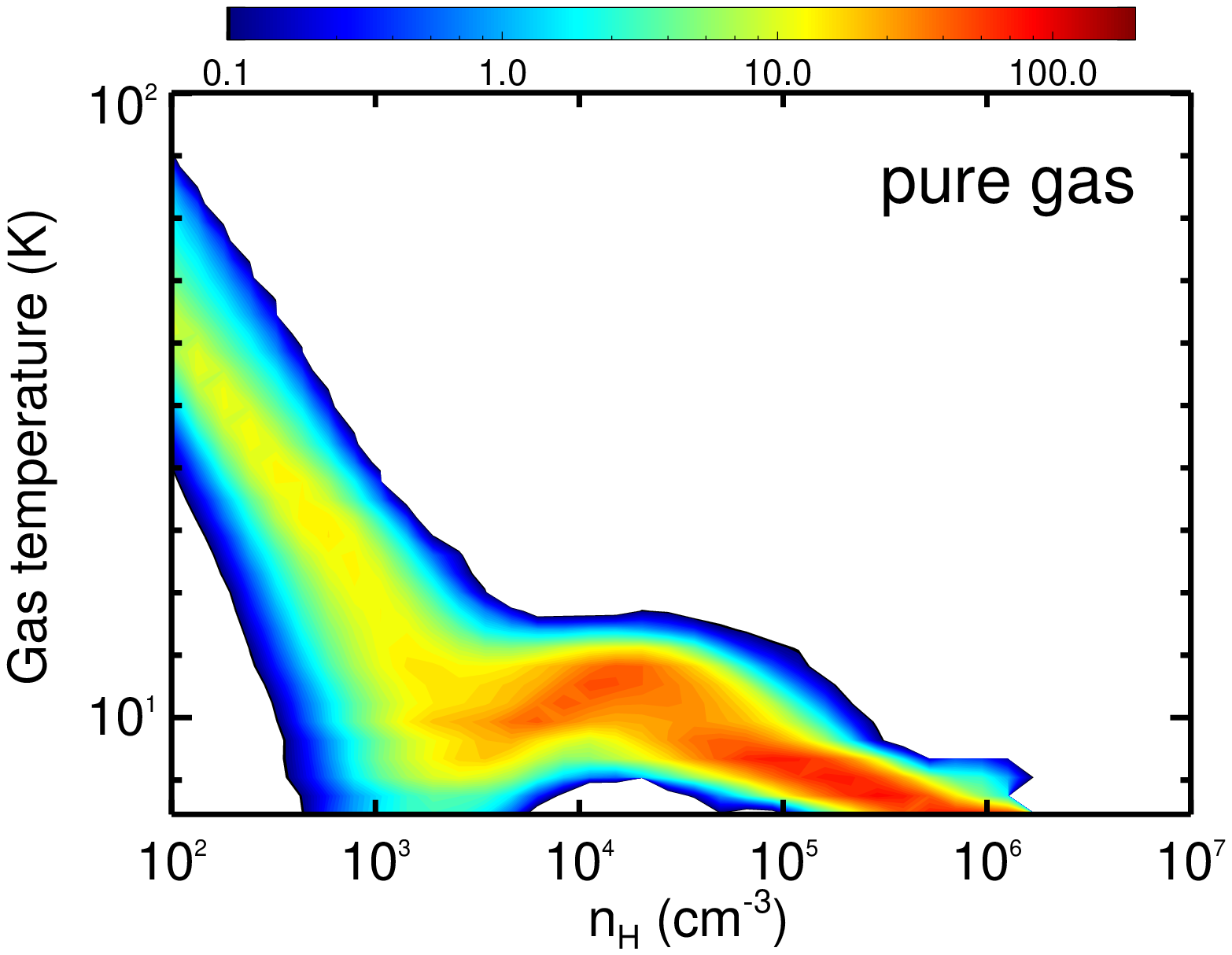} \\
\includegraphics[scale=0.40,angle=0]{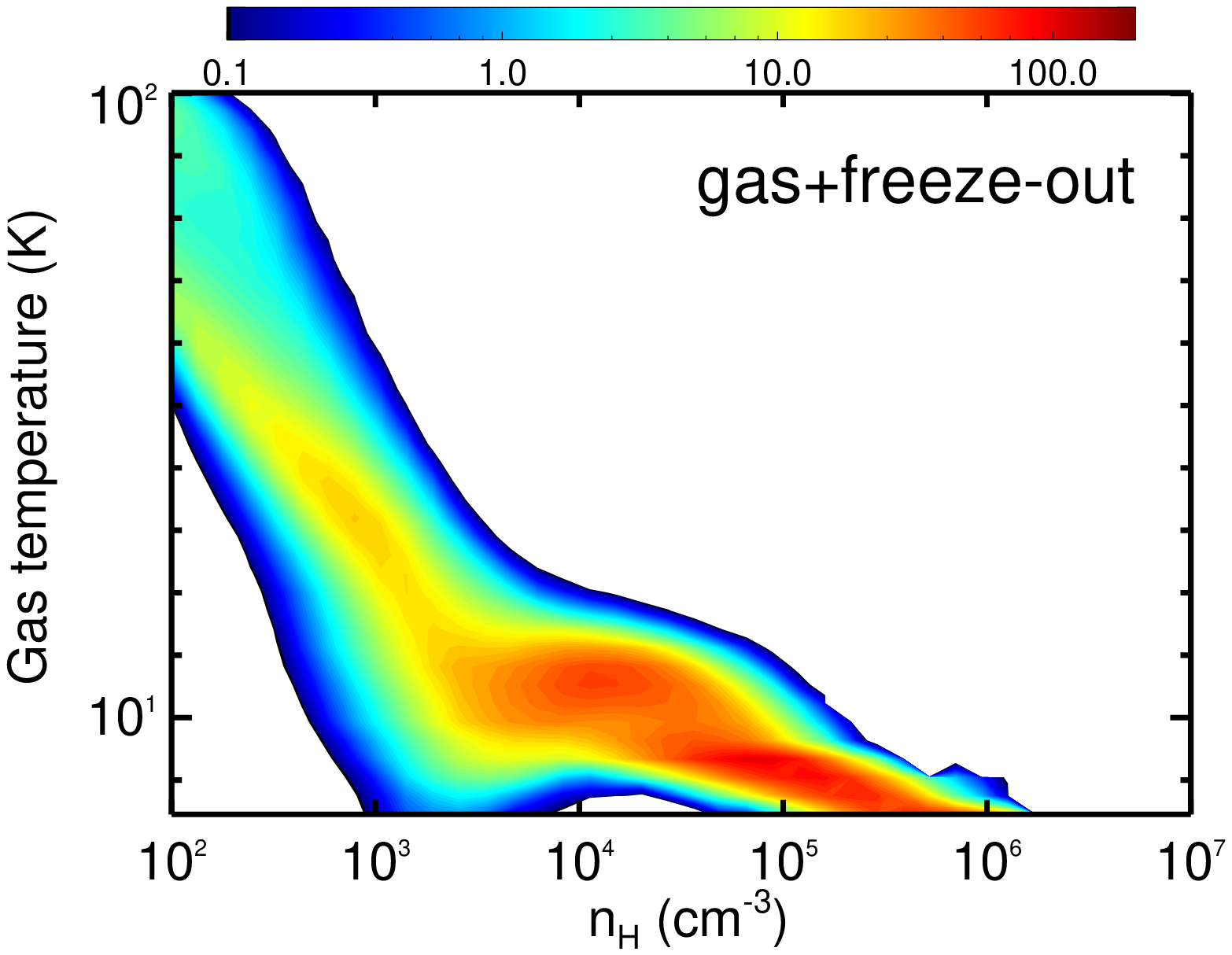}
\includegraphics[scale=0.40,angle=0]{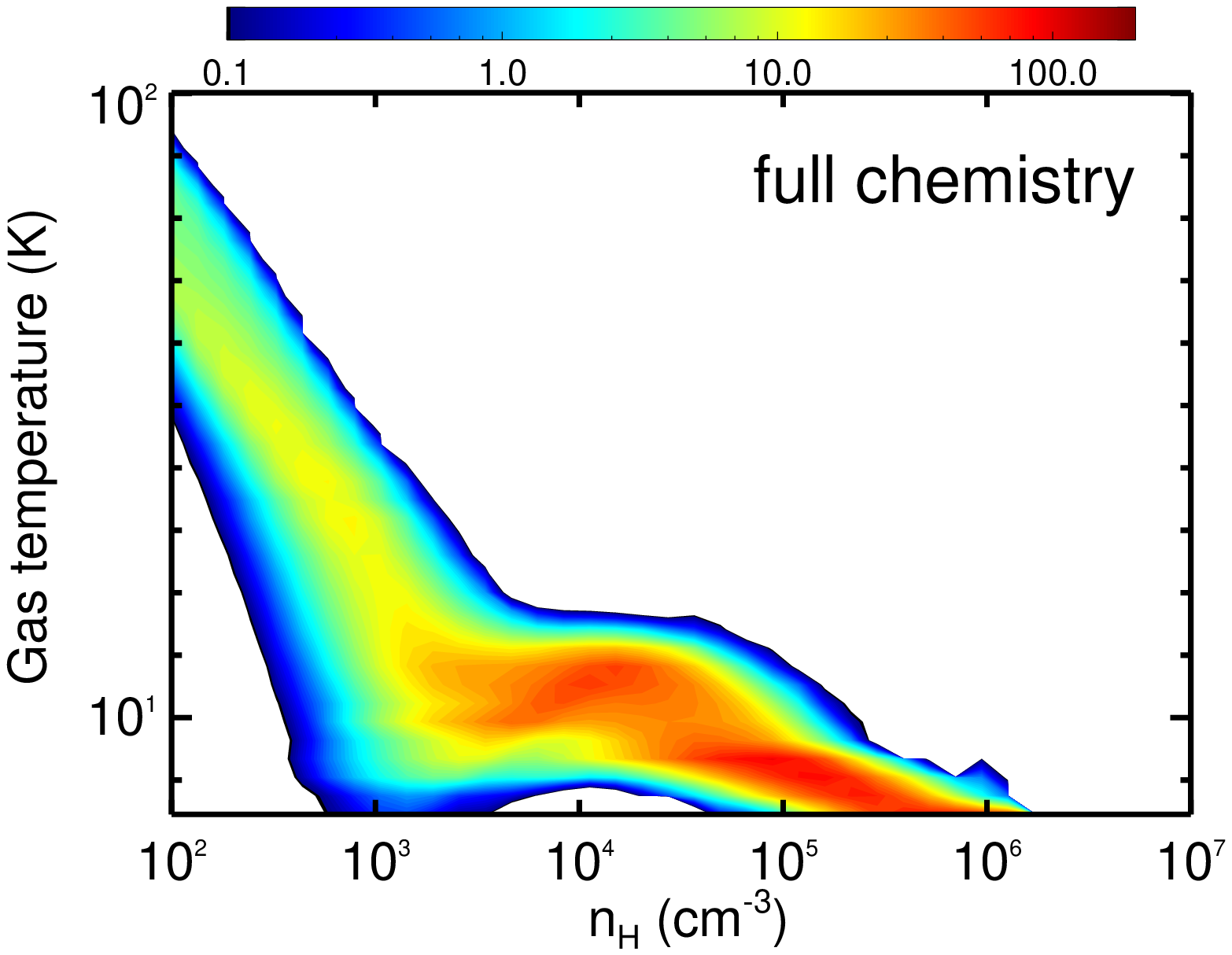}
\caption{Temperature-density phase diagrams. For the collapsing cloud models (\textit{pure gas}, \textit{gas+freeze-out}, \textit{full chemistry}), the gas temperature is plotted against the number density in colored contours. All images are displayed at $t = t_{\rm ff} = 1.63$ Myr. The colors indicate the amount of mass lying in a contoured region, ranging from $0.1$ \msol (dark blue) to $200$ \msol (dark red).}
\label{fig:phases}
\end{figure*}

By affecting the chemistry and changing the molecular abundances in the gas phase, dust grains have an indirect impact on the thermal balance of the cloud. The gas temperature varies between the models since the cooling rates are especially a function of the chemical abundances. The cloud temperature typically drops from 100\,K at $n_{\rm H}\simeq 10^2 \,\rm cm^{-3}$ to below 10\,K at densities above $n_{\rm H}=10^{4}$\,\cmcube. In between, the gas temperature is mostly regulated by the gaseous CO abundance, which can fluctuate. This is where freeze-out plays its most significant role. We show the temperature evolution for our three model clouds in Fig.\,\ref{fig:phases}.

In agreement with the findings of Paper\,I, the \textit{gas+freeze-out} model displays a different evolution in temperature than the other cloud models. The key difference is that the temperature in the \textit{gas+freeze-out} model is higher in the $10^3 - 10^4$\,\cmcube range. The eye can be guided by the red colored regions in Fig.\,\ref{fig:phases} where most of the matter lies. We find that the temperature is a few Kelvin higher in the \textit{gas+freeze-out} model around the density of \nhtot = 10$^{4}$\cmcube compared to the other models. Although seemingly a small difference, this is a critical regime where the cloud is prone to fragmentation and it may still have a sizeable impact. We discuss this further in Sect.\,\ref{sec:eos}. At densities above \nhtot = 10$^{4.5}$\cmcube cooling by gas-grain collisions dominates in all cases, rendering the thermal differences between the models negligible. 

It is advisable to note that the process of freeze-out has a less profound impact on the thermal evolution as compared to Paper\,I. There are several reasons for this, but the main one being the difference in the adopted dust temperature, which is much higher now (e.g., $T_{\rm d}$ at cloud edge is 18\,K now as opposed to 12\,K). To confirm this, we simulated a cloud with the previously used \cite{1991ApJ...377..192H} dust temperature calculation method and kept everything else the same, except the resolution, which was lower. We get much more CO depletion, as indicated, and the gas temperature does experience a larger fluctuation.

The phase diagram of the \textit{full chemistry} model (Fig.\,\ref{fig:phases} bottom right) remarkably resembles the \textit{pure gas} model. We verify that this is due to non-thermal desorption that is reducing the effectivity of freeze-out. This already suggests that only small differences will emerge between these two models during the formation of stars.

\subsection{Equation of state}
\label{sec:eos}
The EOS tells us how compressible the gas is and how susceptible the cloud is to fragmentation. The stiffness of the EOS is crucial in this matter, which depends on how the change in pressure (or temperature) scales with the change in density. Instead of looking at absolute quantities like the Jeans mass, following the derivative of the pressure as a function of density is more meaningful.

\begin{figure*}
\includegraphics[scale=0.41]{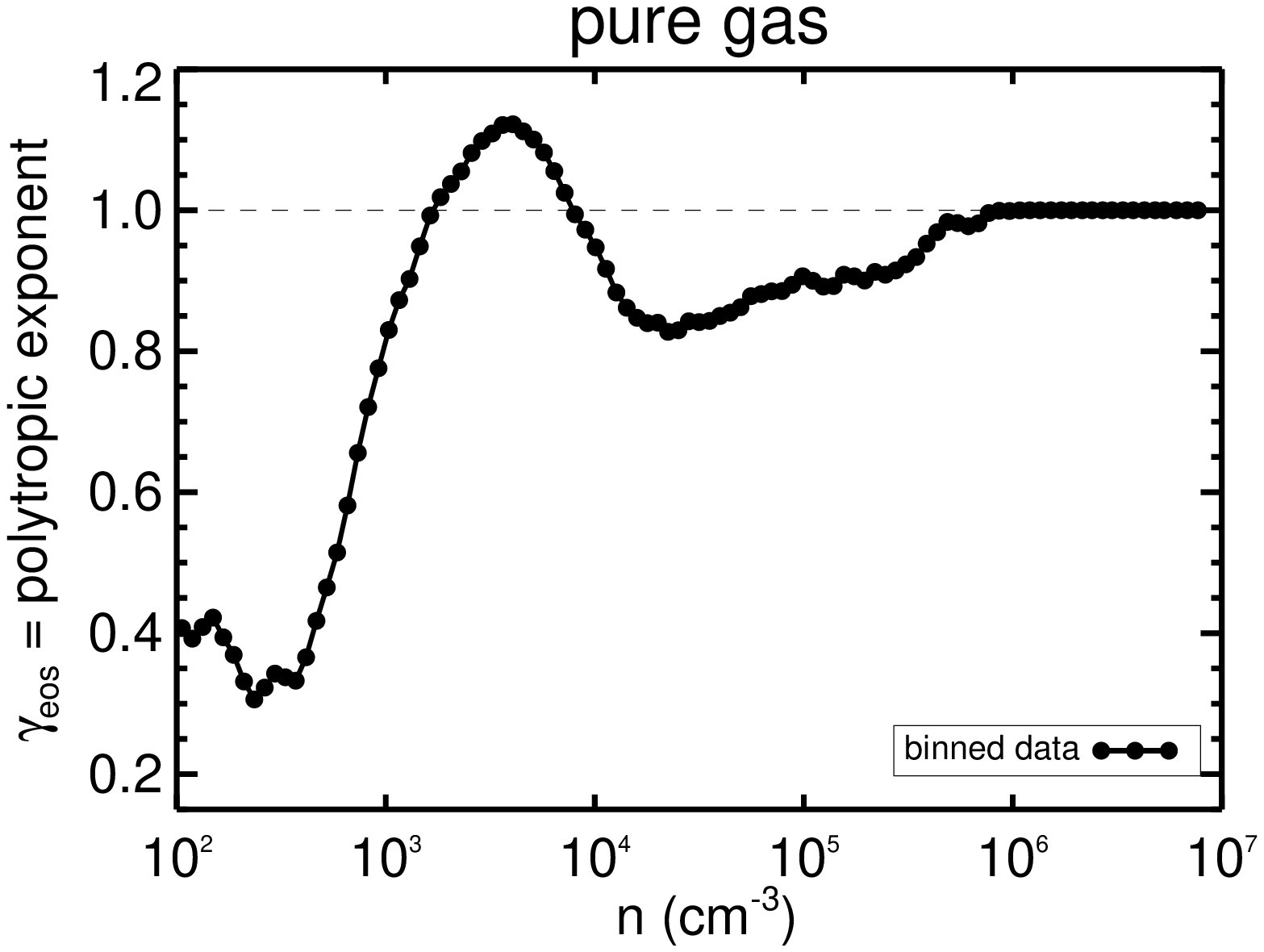} \\
\includegraphics[scale=0.41]{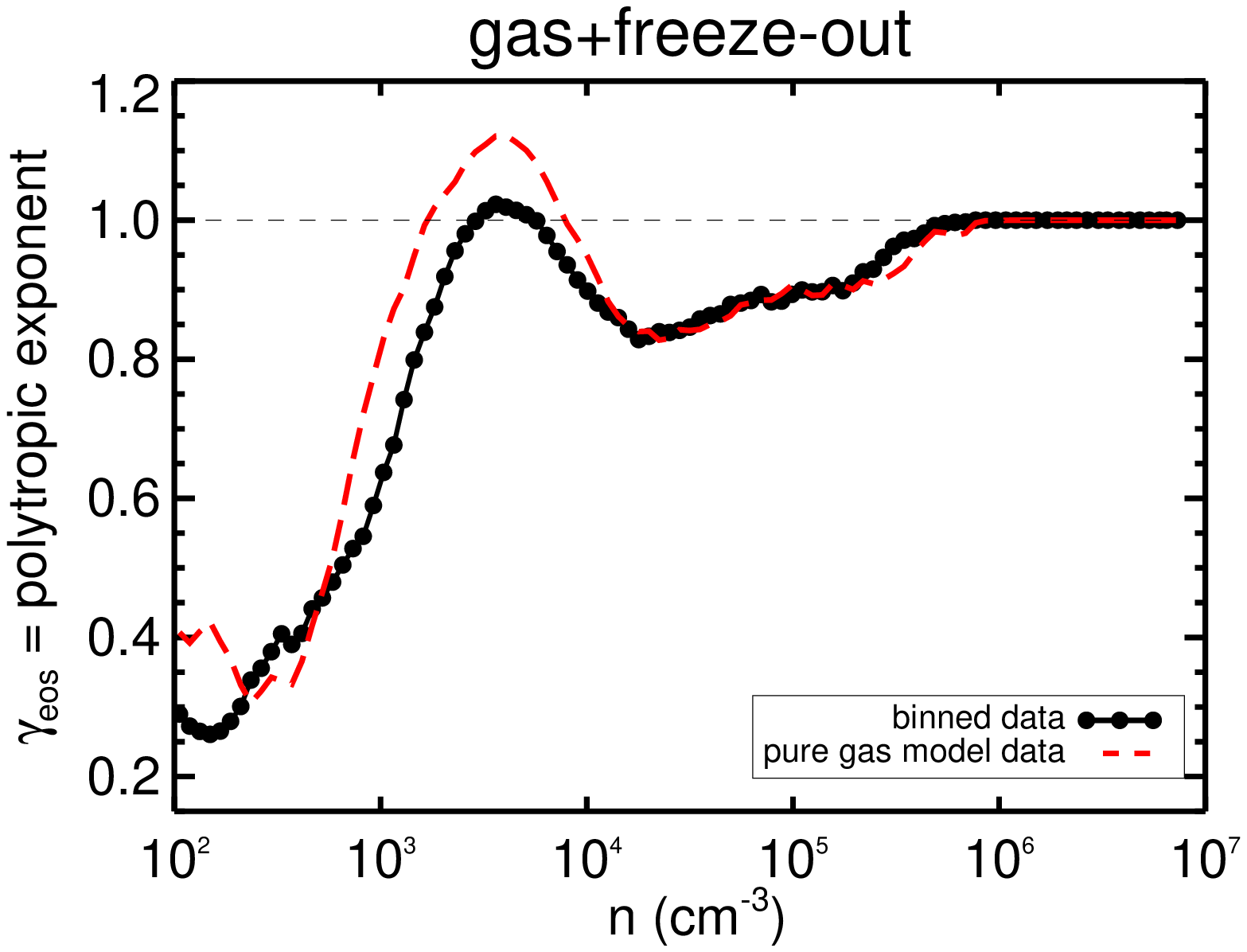}
\includegraphics[scale=0.41]{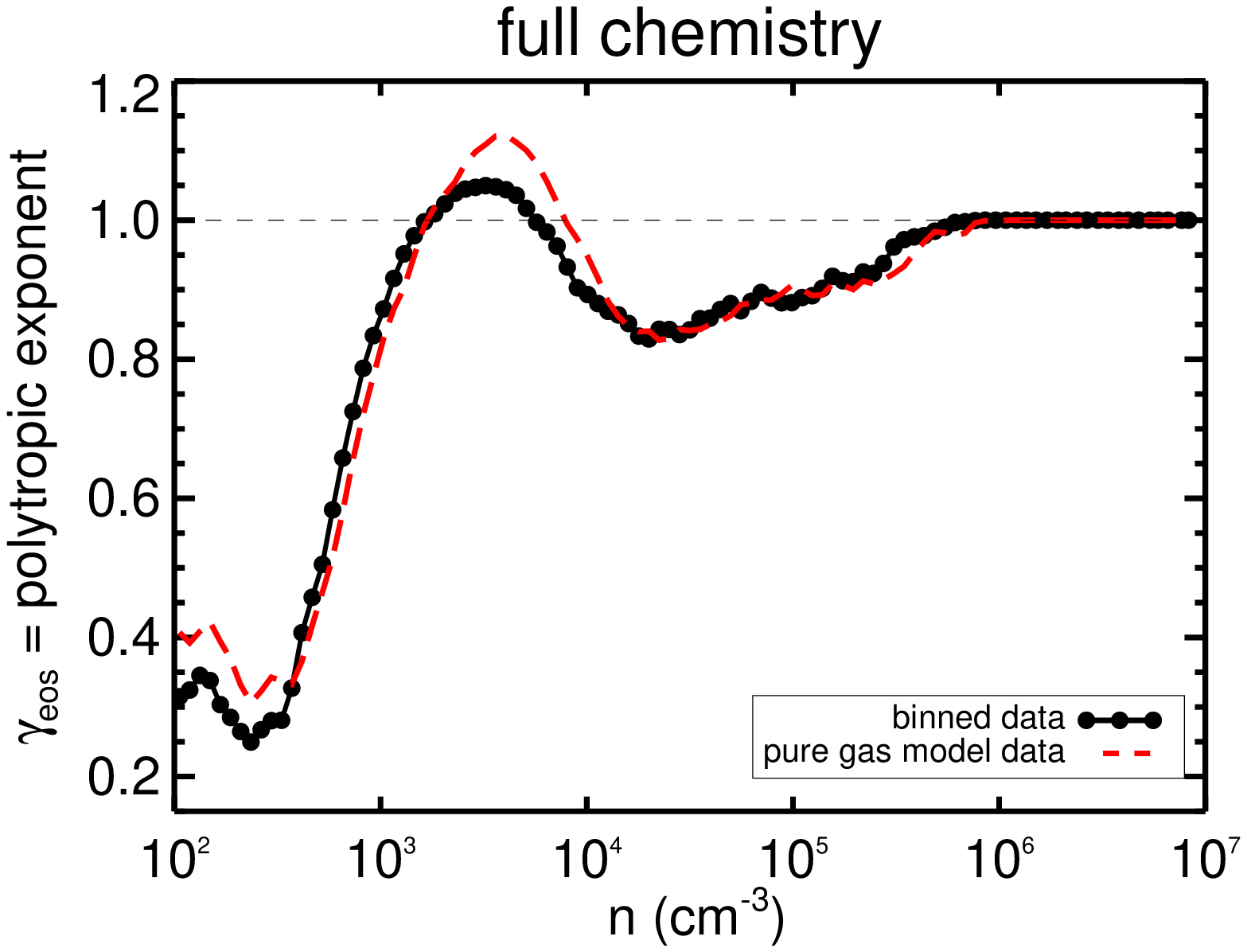}
\caption{The scaling of the EOS. The polytropic exponent $\gamma_{\rm eos}$ is plotted against number density ($n \equiv n_{\rm H}/\mu$) in black circles, where the \textit{pure gas} model is overlaid in red for the GSC models to highlight the difference. The dashed line refers to the isothermal case.}
\label{fig:eos}
\end{figure*}

In our models we employed a polytropic EOS \citep{2000ApJ...538..115S}. This EOS has a proportionality of the form 
\begin{eqnarray}
P \propto \rho^{\gamma_{\rm eos}},
\end{eqnarray}
with $\gamma_{\rm eos}$ as the polytropic exponent. Since the EOS of an ideal gas is defined as $P \propto \rho T_{\rm g}$, the polytropic exponent can be written as
\begin{eqnarray}
\gamma_{\rm eos} = \frac{d\,{log}\,P}{d\,{log}\,\rho} = 1 + \frac{d\,{log}\,T_{\rm g}}{d\,{log}\,n},
\end{eqnarray}
which is related to the logarithmic derivatives of the heating and cooling functions of the gas. Here $n \equiv n_{\rm H}/\mu$. We obtain the EOS by observing the slope of the temperature-density phase diagrams (Fig.\,\ref{fig:phases}), that is, by taking the derivative of the gas temperature as a function of $n$. We bin the obtained data using 20 bins per dex. In Fig \ref{fig:eos} we display $\gamma_{\rm eos}$ as a function of $n$ for the three models.

One can see that the polytropic exponent of the EOS peaks at a density around $n = 10^{3.4}$ \cmcube. The \textit{gas+freeze-out} model peaks at a slightly higher density and has a smaller peak width than the other two models. This is because of the slower cooling, implying longer times to reach minimum temperatures. The \textit{gas+freeze-out} model also has a lower $\gamma_{\rm eos}$ and stays below the isothermal $\gamma_{\rm eos}=1$ line. This makes the gas more prone to fragmentation and compact object formation, since the gas temperature, which we can formulate as $T_{\rm g} \propto \rho^{\gamma_{\rm eos}-1}$, will continually decrease with increasing density during contraction.

In all three models $\gamma_{\rm eos}$ reaches a local minimum at a density of $n = 2.0-2.5\times10^4$ \cmcube. Multiplying by $\mu$ this corresponds to \nhtot = $4.0-5.0\times10^4$ \cmcube, matching the regime where gas and dust become collisionally coupled. Once again this shows that fragmentation is set at this important density regime, since $d\gamma/d\rho$ is consistently larger than 0 at higher densities. At the density of $n = 5.9-6.0\times10^5$ \cmcube the gas becomes isothermal, meaning that the gas temperature is strongly coupled to the dust temperature. The latter stays nearly constant beyond these densities, as we can see from the formulation (Eq. \ref{eq:zucc2}). A slight increase in temperature, i.e., $T_{\rm g} \propto \rho^{0.09}$ with $\gamma_{\rm eos}\simeq1.09$, in the density regime \nhtot$ \approx 10^6 - 10^{11}$\cmcube is normally predicted for solar metallicity \citep{2000ApJ...534..809O, 2005ApJ...626..627O}.

\subsection{Filamentary structure}
\label{sec:fragment}
Due to turbulent motions, the spherical clouds become filamentary after initiating the runs. We measure the filament sizes to be on average 0.12 pc in diameter in our simulations, but there are hints that this size may be resolution limited \citep[see e.g.,][]{2013A&A...556A.153H}. The filaments are resolved by 6-7 cells given the spatial resolution of 0.015\,pc. We measure the filaments by taking the FWHM of random local maxima at different time-steps. We do notice that the filaments are slightly more compact at the beginning and grow in size as time progresses and density accumulates. The total increase is about a factor of two.

\begin{figure}
\includegraphics[scale=0.48]{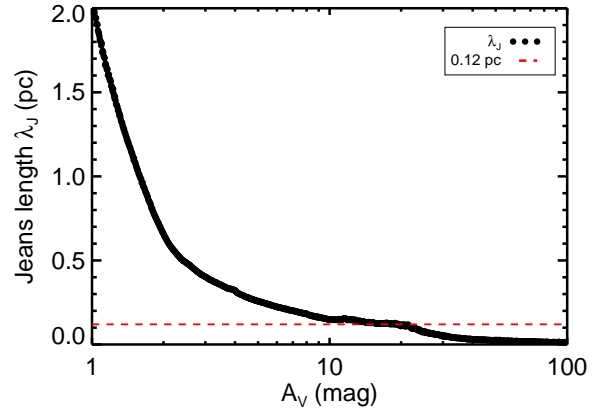}
\caption{Tracing the Jeans length. $\lambda_{\rm J}$ is plotted against $A_V$ for the model \textit{full chemistry}. Other models show the same behaviour. This figure shows that $\lambda_{\rm J}$ stalls between $A_V = 10-20$ at the critical length scale of $\sim$0.1\,pc.}
\label{fig:jeanslen}
\end{figure}

It is often observed that filaments are 0.1\,pc in size \citep{2011A&A...529L...6A, 2014prpl.conf...27A, 2015MNRAS.452.3435K}, though, tentative detections of compact filaments have been made with higher resolution observations \citep{2015Natur.518..213P}. We point out that the size of 0.1\,pc is the typical Jeans length, $\lambda_{\rm J}$, when the density is around $n = n_{\rm H}/\mu = 3\times10^4$ \cmcube. This is the density where the gas and dust start to be well coupled and the cloud enters a collisionally dominated pressure regime. We can see this when we formulate the Jeans length for gas-grain collisionally coupled conditions, i.e.,
\begin{eqnarray}
\lambda_{\rm J} &=& \left( \frac{\pi c_s^2}{G \rho} \right)^{\frac{1}{2}} = 0.10 \rm \,pc \left( \frac{c_s}{0.16 \rm \,km\,s^{-1}} \right)  \left( \frac{n}{3\times10^4\,\rm cm^{-3}} \right)^{-\frac{1}{2}},
\\ \nonumber
&=& \left( \frac{\pi k_{\rm B} T_{\rm g} \gamma_{\rm eos}}{G \rho \mu m_{\rm H}} \right)^{\frac{1}{2}} = 0.10 \rm \,pc \left( \frac{T_{\rm g}}{8 \rm \,K} \right)^{\frac{1}{2}}  \left( \frac{n_{\rm H}/\mu}{3\times10^4\,\rm cm^{-3}} \right)^{-\frac{1}{2}},
\label{eq:jeanslen}
\end{eqnarray}
where $c_s$ is the sound speed, $\rho$ is the mass density and $k_{\rm B}$ is the Boltzmann constant. At these densities one can safely assume that everything is in molecular form and that $\mu\simeq2.3$ (note that we have $\mu=2$ from our models). For a near isothermal case, i.e. $\gamma_{\rm eos}=0.9$, the sound speed becomes 0.16 $\rm km ~s^{-1}$ at a gas temperature of 8\,K, assuming the gas temperature is close to the dust temperature. For densities of $n \gtrsim 3\times10^4$\,\cmcube this is likely to be the case. This size becomes 0.12 pc if the gas temperature does turn out to be a bit higher, say 10\,K, or when $\mu=2.0$. In Fig.\,\ref{fig:jeanslen} we plot $\lambda_{\rm J}$ as a function of $A_V$, which is representative for all models. %sqrt(8*pi*1.38e-16*0.9/(3e4*6.67e-8*1.67e-24^2*2.33^2))/3.086e18 = 0.104

Length scales below this size will not gravitationally contract until higher densities or lower temperatures are reached to overcome the internal gas pressure. However, the temperature cannot decrease much further as it is tightly coupled to the dust. It will, in fact, be efficiently heated by the dust when the gas cools below the dust temperature as long as the two phases stay coupled. The only way to compress further is to increase its density. This will not decrease the Jeans length as strongly since the dependence is only by the square root. A possible reason why these length scales are more easily observed is because it just takes more time to decrease the Jeans length any further. In Fig.\,\ref{fig:phases}, we indeed see mass piling up around these critical densities. Moreover, filaments accreting surrounding material will see an increase in their widths. If this scales about equally as the $\rho^{1/2}$ decrease in Jeans length, the filament sizes may remain almost constant until accretion stops.

Our concept is a different one than the reasoning given by \cite{2005MNRAS.359..211L}, where the idea was that the transition through an isothermal EOS, $\gamma_{\rm eos}=1$, provides enough thermal support to stop a cylinder from gravitational contraction along the radial direction. We reach such a condition at the density of $n_{\rm H} \simeq 1.2\times10^6$\,\cmcube, where the Jeans length would be on the order of 0.022\,pc. Here, we argue that dust cooling causes the EOS to reach a minimum at lower densities ($n_{\rm H}/\mu = 2\textnormal{-}3\times10^4$\,\cmcube) and that this is the moment where $\lambda_{\rm J} \simeq 0.1$\,pc. From this point on, \textbf{the change in} \boldmath $\gamma_{\rm eos}$ \unboldmath is positive towards higher densities, thereby suppressing further fragmentation and making this size scale a preferred one for fragmentation. Thus, no thermal support is required. It is worth noting that also other explanations have been suggested recently, i.e., ambipolar diffusion \citep{2015IAUGA..2252264N} and turbulence \citep{2015arXiv151005654F}.

We find that the gas already reaches cold $\sim$10\,K temperatures at a density of \nhtot\,$=3000$\,\cmcube (Fig.\,\ref{fig:phases}). The cooling slows down after this point and the heating even prevails for a short period, causing the gas temperature to rise. Henceforth, the Jeans mass $M_{\rm J}$, with its strong temperature scaling, also stalls. At this phase $M_{\rm J} \approx 13$\,\msol. Sink particles only start to form when the density increases by at least two orders of magnitude, i.e., where the Jeans mass drops below 1\,\msol. This means that the density still has to increase and it does so either through gas accretion on to filaments, or, and mainly, through filaments crossing one another. This is in agreement with the suggested scenario by observers working both in massive star-forming regions \citep[e.g.,][]{2012A&A...543L...3H, 2014MNRAS.440.2860H} and in low-mass star-forming regions \citep[e.g.,][]{2013A&A...554A..55H, 2015A&A...574A.104T} on the fact that star formation in filamentary structures is mainly found at the intersection or within the merging of filaments (or bundles within filaments).

\subsection{Star formation}
\begin{figure}
\includegraphics[scale=0.42]{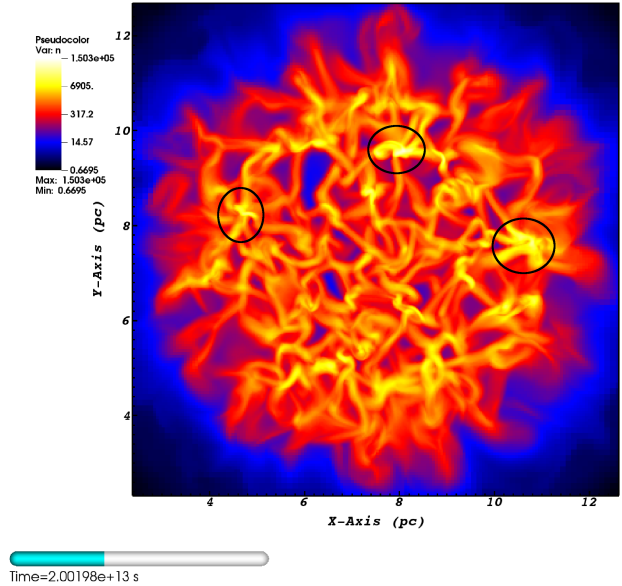}
\caption{Fragmenting cloud. The image shows a density slice though the center of the filamentary cloud. Star formation is occurring in the intersections between the filaments (knots), where the encircled regions highlight the locations of the first sink particles.}
\label{fig:fragment1}
\end{figure}

We see the first sink particles forming in the intersections between the filaments. This happens after a cloud evolution of $8.9\times10^5$ yr, that is 0.55 of the cloud free-fall time. The primary sink particle locations at the onset of star formation is highlighted in Fig.\,\ref{fig:fragment1}.
The densities are higher in these regions and the first sink particle forms within the northern clump at an average clump density around the center of \nhtot $= 3\times10^5$ \cmcube, see the enclosed region by the middle black circle in Fig.\,\ref{fig:fragment1}. Generally, however, sink particles form at clump densities of around \nhtot $= 10^6$\,\cmcube. Multiple sink particles are forming out of these knots, and the more massive a clump the more sink particles it produces, while only occasional sink particle forms along the filaments. The most massive sink particle sits in the largest clump at the center of the deepest potential well. This is not the center of the cloud, the density is not particularly high there, but the northern clump. The mass of the most massive particle grows fast, from 1 \msol to 25 \msol in 0.5 Myr, but most of this comes from capturing the protostars in the process of forming around it, all conceived within the same clump. We generally only see merging events for stars that are forming in the same parental clump, which concerns only about 9-10\,\% of the stars during the entire run for each model.

Lower mass stars have a higher chance of surviving when the most massive star is less massive, thereby avoiding merging because of the lower escape velocity, or if they are kicked out due to gravitational interactions. This means that more massive stars are more isolated than others. To coin a phrase ``it is lonely at the top'' holds up here. Indeed, we see that as more stars are forming, and the interactions between the stars increase, less mergers occur and more sub-solar mass stars survive. It is at this moment that the IMF is being shaped. The low-mass star population can continue to grow in this way, enhanced by the fact that as gaseous matter gets depleted, low-mass star formation is preferred over high-mass star formation, but stellar feedback will slow down and regulate the low-mass star population \citep[e.g.,][]{2010ApJ...713.1120K, 2010MNRAS.405..401D}. On the other hand, stellar feedback also reduces accretion, which mostly affects the high-mass stars. Stellar feedback is not included in our models, but we do not continue our simulations beyond a cloud free-fall time where it would have the biggest impact.

The first stars have a mass of around 0.5 \msol. Numerical resolution limits us from forming lower mass stars. After their formation they start to accrete surrounding material. In the end, on average 47\% of the total mass comes from accretion, as measured at $t=t_{\rm ff}$. The rest is acquired at formation. Thus, $\int_{0}^{t_{\rm ff}} \dot{M}_{\rm acc} dt \simeq 0.9M_{\rm form}$. By counting the number of formed sink particles at any given time, we have an idea on the star formation rate. Dividing the total mass of the sink particles by the total mass inside the cloud, we obtain the star formation efficiency, i.e., SFE = $M_{\rm sinks}/\left(M_{\rm sinks} + M_{\rm gas}\right)$. Measured at $t=t_{\rm ff}$, this becomes the normalized quantity SFE/\tff. We show in Table \ref{tab:sfes} the SFEs and other quantities for the three models.
\begin{table}
\caption{Star formation parameters}
\begin{tabular}{l|ccc}
\hline
\hline
Parameter             &  \textit{pure gas} & \textit{gas+freeze-out} & \textit{full chemistry} \\
\hline
SFE at $t=t_{\rm ff}$ & 18.0\%      & 18.0\%      & 18.7\%      \\
Accreted fraction     & 48\%        & 47\%        & 47\%        \\
Number of stars       & 721         & 769         & 768         \\
Mean mass             & 1.98\,\msol & 1.87\,\msol & 1.90\,\msol \\
Minimum mass          & 0.47\,\msol & 0.45\,\msol & 0.46\,\msol \\
\hline
\end{tabular}
\label{tab:sfes}
\end{table}
Here we can see that the number of formed sink particles is greater in the GSC included runs, \textit{gas+freeze-out} and \textit{full chemistry}, but that in general the differences are too little to make hard statements.

\subsection{The IMFs}
\label{sec:imf}
\begin{figure*}
\includegraphics[scale=0.44]{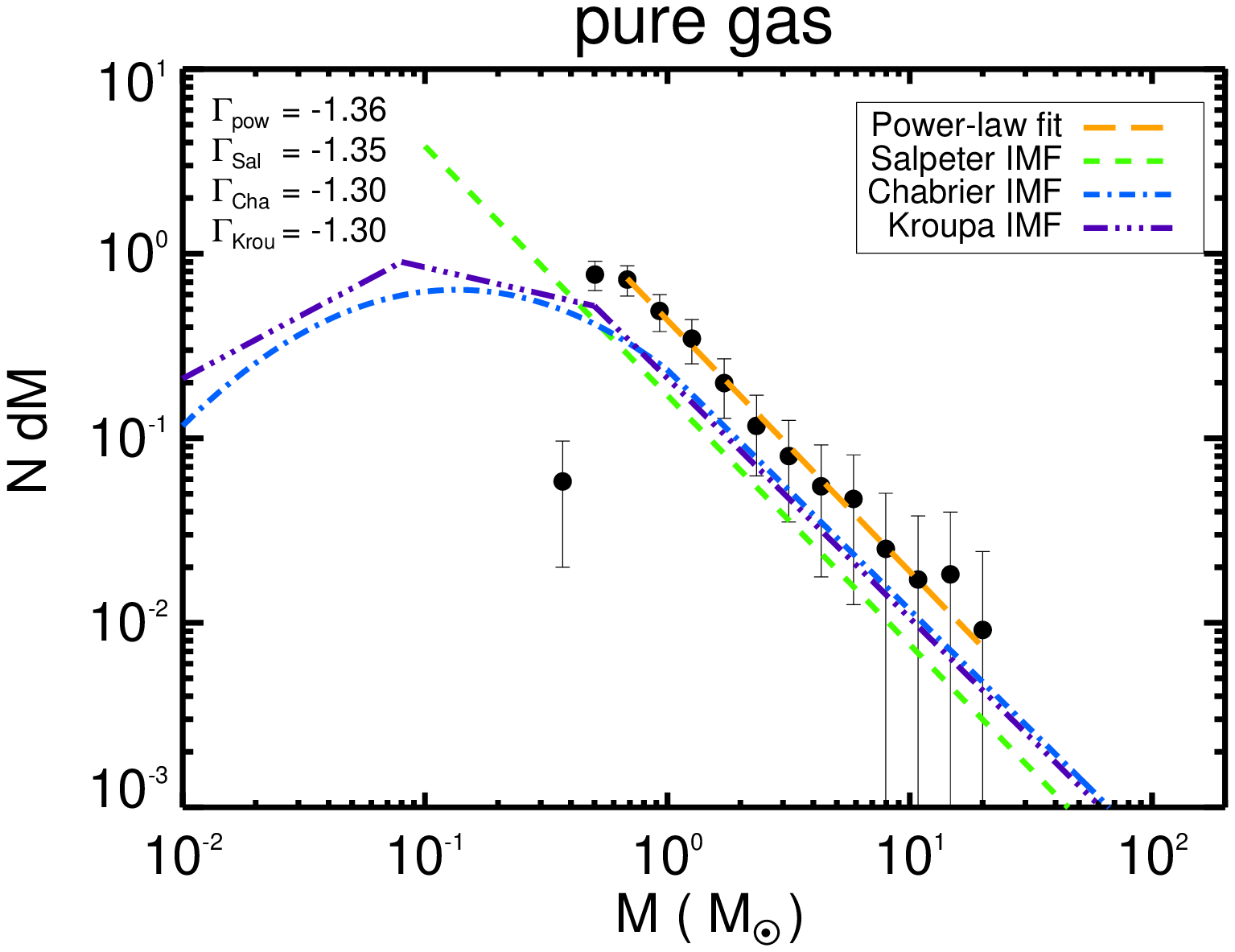} \\
\includegraphics[scale=0.44]{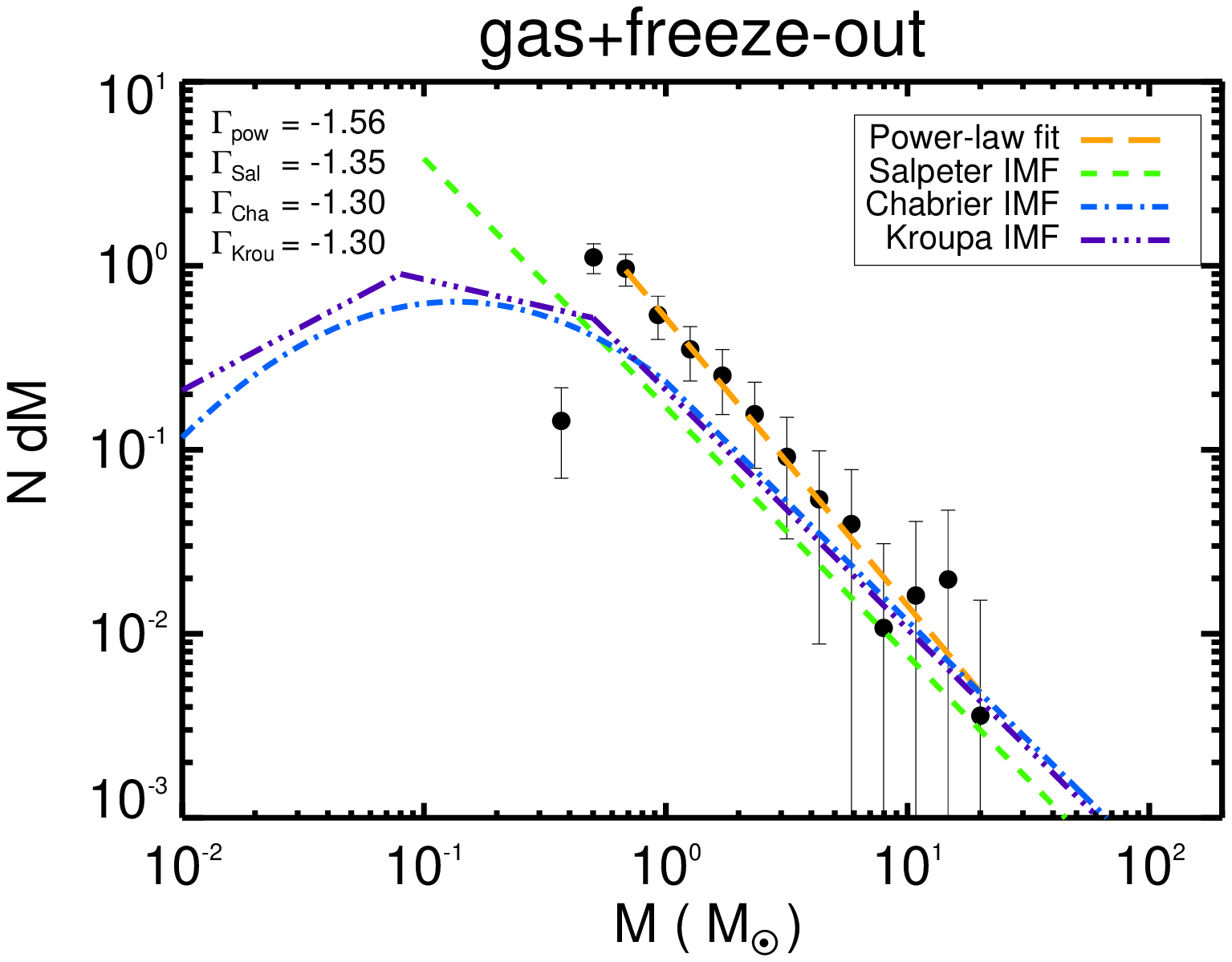}
\includegraphics[scale=0.44]{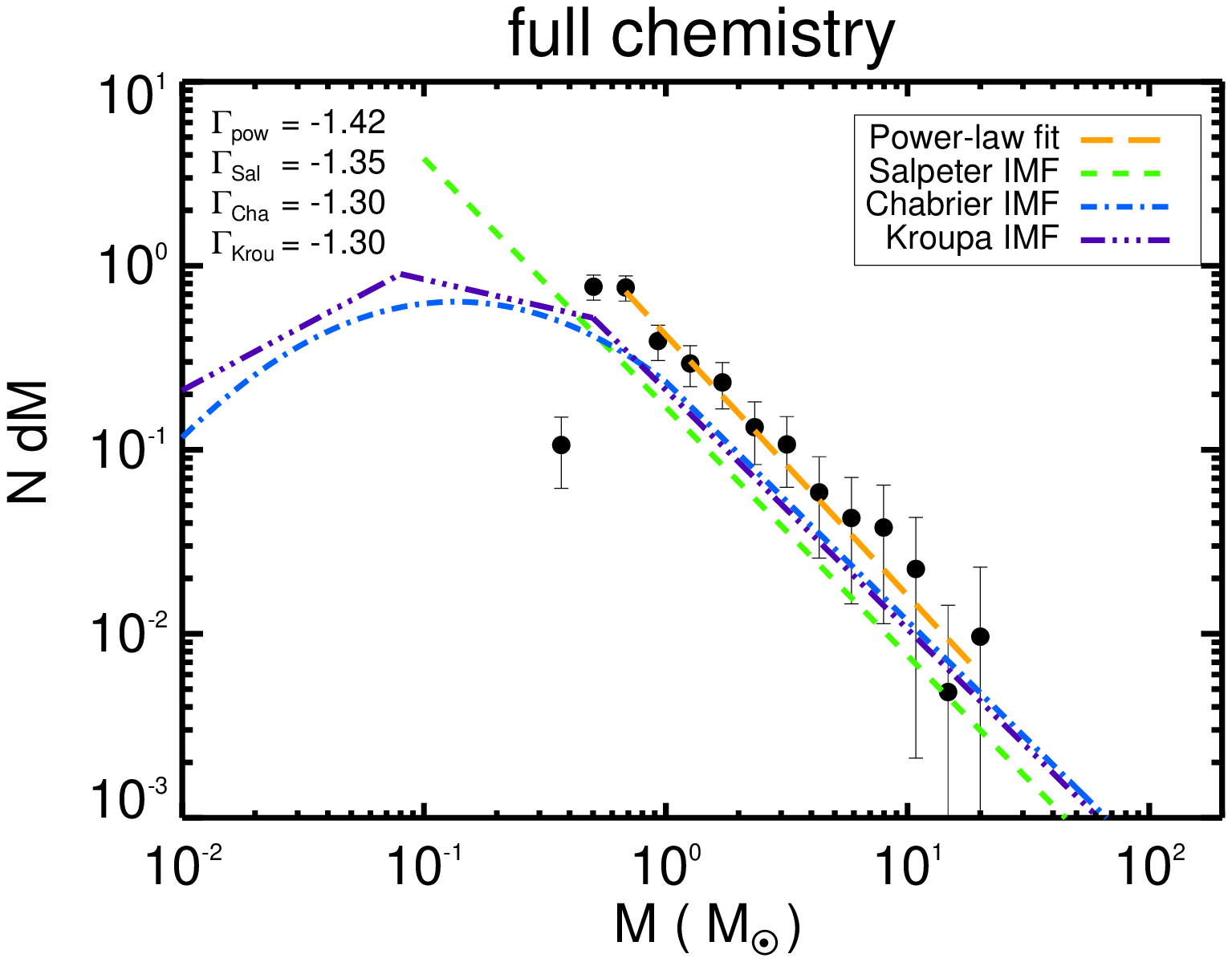}
\caption{Stellar IMFs. The number of sink particles is plotted against binned masses with a bin size of 0.133 \msol. A power-law is fitted to the numerical data above the turn-over mass (orange long dahsed). In these figures, three mainstream IMFs are plotted for reference, the Salpeter IMF (green short dashed), the Chabrier IMF (blue dot-dashed), and the Kroupa IMF (purple triple dot-dashed). All functions are normalized by the integral under the curves. The power-law slopes of the respective functions are given in the upper left corner.}
\label{fig:imfs}
\end{figure*}

With the emergence of a sizable number of stars, we can construct the IMFs. The IMFs are logarithmically binned histograms with a fixed number of bins between 0.2 and 20 \msol, where the number of stars within a bin is plotted against its mass. In Fig.\,\ref{fig:imfs}, we show the mass distributions for the three cloud models. 
To act as a reference, we overlay three well-known IMFs in our figures. These are the \cite{1955ApJ...121..161S}, \cite{2001MNRAS.322..231K}, and \cite{2003PASP..115..763C} IMFs.

In order to compile the IMFs, and to be epoch independent, we took the sink-particles from all time frames of the simulation and normalized them. To verify, we also made the comparison at a single epoch, i.e., $t=t_{\rm ff}$, which happens to be the same time when the clouds converted equal amount of gas into sink particles. In either case, the results are similar. The error bars for the IMFs are obtained from the Poisson distribution, i.e., error =$\sqrt{N_{\rm bin}}$. A power-law is fitted to the data points, given by the orange dashed line, above the most likely turn-over mass. The characteristic mass was not fixed and was allowed to change for each model. 

There are no great differences between the models \textit{pure gas} and \textit{full chemistry}. The only thing that stands out is that the \textit{gas+freeze-out} model is somewhat different from the others. The main difference is that the slope is steeper and diverges from the Salpeter slope. There are more solar and sub-solar mass stars compared to the other models. This also shows from the increased number of sink particles in this model. Despite the fact that the temperatures were higher, the resulting softer EOS makes star formation easier with more low-mass stars forming. The more massive stars do not grow as much because of the increased fragmentation. This is in agreement with the theory of fragmentation induced starvation \citep{2010ApJ...725..134P}. Knowing that the \textit{gas+freeze-out} model is an extreme case, the main result is that for the simulated conditions in this work no significant deviation is found in the IMF by changing the chemistry details, as long as the dust temperature is properly calculated.

\section{Conclusions}
\label{sec:conclusion}
With the implementation of detailed surface chemistry calculations in 3D hydrodynamical simulations, we have explored the impact of dust chemistry on the EOS, molecular cloud structure, star formation, and on the IMF. Our three models comprised three levels of complexity in the chemical network by increasing the gas-grain interaction. We simulated collapsing clouds with two extreme depletion scenarios, one where everything remains in the gas phase in a pure gas-phase model and the other where everything that is adsorbed will remain frozen as long as the temperature of the dust remains cold in a gas plus freeze-out model. The third model had the full gas and grain surface chemistry implemented, with non-thermal channels like chemical and photodesorption.

\textsc{Chemistry:} The impact of surface reactions on cloud chemistry is unquestionable. Without GSC, we do not see any significant molecule formation heavier than HCO at temperatures below 100\,K in a cloud lifetime. When we do include GSC, we see a lot of methanol and formaldehyde forming and the abundances of water and OH are enhanced. The impact of non-thermal desorption is also fundamental. Without it, species quickly freeze out on dust surfaces at already diffuse cloud conditions, while molecules are efficiently released into the gas phase through chemical and photodesorption. We find significant increase in the water, formaldehyde, and methanol abundances in the gas phase when non-thermal desorption is included. The H$_2$ abundance is enhanced from both the H$_2$ formation on grain surfaces as well as the ER mechanism at high temperatures. We see the H to H$_2$ transition occurring at lower densities, i.e., \nhtot = 30\,\cmcube, in simulations with the full chemistry implemented. We conclude that chemistry, with a detailed consideration of the gas-grain interplay, is needed to explain the observed gas-phase molecular abundances, the ice abundances, the presence of complex (organic) molecules, the H to H$_2$ transition, and the CO depletion factors in the ISM.

\begin{figure}
\includegraphics[scale=0.24]{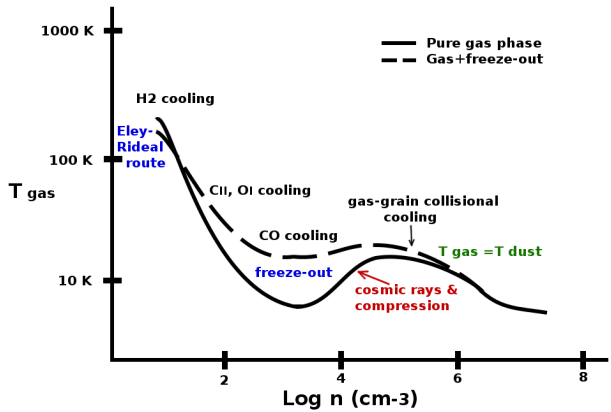}
\caption{Illustrating the impact of freeze-out on gas temperature in a collapsing cloud. The dashed line portrays how the gas temperature changes when freeze-out is considered during cloud evolution. The main cooling channels (black) and the heating by CRs and compression (red) together with the main effects of freeze-out (blue) are given in the figure.}
\label{fig:mycool}
\end{figure}

\textsc{Thermal balance:} While grain surface chemistry has a profound impact on the abundances of gas-phase species, its impact on the thermal balance is more conservative. This is partly because self-shielding works in the opposite direction of cooling when the abundances of species are changed. We also find that when one cooling channel is deprived, other cooling pathways take over to cover the loss to some degree. Still, a respectable difference in the thermal profiles can be seen. The changes in the thermal balance also affect the EOS. When considering the freeze-out process, regardless of the $2-3$\,K higher gas temperatures, the EOS becomes softer, i.e. lower values of $\gamma_{\rm eos}$ due to cooling deprivation, which we show in Fig.\,\ref{fig:eos}. The reason being that because gas cools less rapidly as a result of CO freeze-out, the temperature does not increase as much (or at all) in the regime $n_{\rm H}=10^{3.5}-10^{4.5}$\,\cmcube where heating gradually dominates over cooling due to CRs and adiabatic compression. This process also does not allow for the gas temperature to drop (much) below the dust temperature. The second drop comes from the gas being efficiently cooled by dust grains at $n_{\rm H}>10^{4.5}$\,\cmcube, strongly linking the temperature in the two phases together. In Fig.\,\ref{fig:mycool}, we sketch the general effects of the freeze-out process on the gas temperature by giving two extreme examples, pure gas phase reactions and freeze-out without the ability of non-thermal desorption.

\textsc{Cloud structure:} The arising fluctuations in the EOS have ramifications on cloud fragmentation. At densities above \nhtot\,$>10^{4.5}$\,\cmcube, the Jeans length is on the order of 0.1\,pc and $\gamma_{\rm eos}$ growing close to the order of unity. The gas temperature will not cool much further at this stage, or heat up, since it is anchored to the dust temperature, suspending the Jeans length at this size scale. Only through the increase in density, the Jeans length ($\propto\rho^{-1/2}$) can further decrease. As a result, it will not change as greatly by further compression from this point onward. This delay allows the gas to fragment at this scale more preferably. We find typical filament sizes of 0.12\,pc in our simulations, albeit, that this may be resolution limited. The filaments are narrower early on in the simulation, but get thicker towards the formation of stars, possibly because of accretion and as a result of the collapsing cloud. Therefore, the filament thickness may be an indication of cloud evolution time.

\textsc{Star formation:} Star formation occurs mostly in the intersections between filaments, where the densities are higher and the potential wells are deeper. There is also significant inflow of gas to these knots. The first star formation events occur around half a cloud free-fall time, with sink particle masses on the order of 0.5\,\msol. The efficiency of star formation is somewhat on the high side, SFE = 18\%. This is not entirely unexpected as it is known that simulated non-magnetized clouds generally collapse too fast once gravitational contraction kicks in \citep{2007ApJ...654..304K, 2015ApJ...800L..11V}. We find that the number of formed stars is higher in the models that include freeze-out, owing to the softer EOS. This causes the IMFs to slightly differ between the three models. The increased star formation in the dust included runs result in a slightly steeper high-mass slopes, which deviate from Salpeter, but only marginally. We conclude that surface chemistry enhances star formation, if only a little, by increasing the number of low-mass stars, but that there is negligible impact as far as the IMF is concerned. The distribution of stellar masses are robust to the chemical changes in the very early phases of cloud evolution.

\subsection{Caveats}
\label{sec:discussion}
Despite the decrease in mergers with increasing interaction between stars and the depletion of gas, favoring low-mass star formation, the slope of the IMF flattens with time (beyond a free-fall time) as a result of continual and strong accretion by massive stars. This can be prevented if sellar feedback is introduced for future generation of stars. We note that when plotting the IMFs just after $t=t_{\rm ff}$, the GSC models now reproduce the Salpeter slope rather well, while the \textit{pure gas} model becomes top-heavy. From this point of view, one could conclude that surface chemistry is actually needed to reproduce the universal IMF. Thus, instead of concluding based on absolute values, such as the slope, which can depend on the moment picture, we conclude on the relative differences between the models as these are unaffected. The qualitative results are robust against time and manner of plotting.

Since solid-state chemistry is still a developing field, many of the species binding energies, diffusion energies, and activation barriers are yet uncertain. These values can change in the future. We realize that this may affect our chemistry and thus the cloud evolution. However, since our main focus is on the impact of freeze-out, especially of CO, this process is better known than others. While the fine details may affect the species abundances, it is not expected to see any significant changes in star formation and the IMF.

The processes of photodesorption and chemical desorption are also not yet fully understood, especially for molecules residing on icy surfaces. The substrate of the grain mantle, the size of the grain, or the neighboring species may significantly affect the desorption rates. These will change the results of the \textit{full chemistry} model to be either more comparable to the \textit{gas+freeze-out} model or to the \textit{pure gas} model. However, observed depletion factors do imply that desorption, by any means, is not dominant enough to completely negate the process of freeze-out \citep[e.g.,][]{2012MNRAS.423.2342F, 2012ApJ...759L..37C, 2013ApJ...775L...2L, 2014A&A...570A..65G}.

\section*{Acknowledgments}
SH thanks G. M. Mu{\~n}oz Caro for his insights on photodesorption, and A. Ivlev on discussions about CRs. PC, MS, and SC acknowledge the financial support of the European Research Council (ERC; project PALs 320620). The software used in this work was developed in part by the DOE NNSA ASC- and DOE Office of Science ASCR-supported Flash Center for Computational Science at the University of Chicago. The simulations have been run on the super computer `Hydra' at the Rechenzentrum Garching of the Max-Planck-Gesellschaft \& IPP. Some of the kinetic data used in this work has been downloaded from the online database \textit{KiDA} \citep[][http://kida.obs.u-bordeaux1.fr]{2015ApJS..217...20W}.

\bibliography{Hocuk-IMF-2015.bib}

\appendix
%========

\section{Chemistry}
\label{app:AA}
Time-dependent rate equations are solved at each grid cell that include both gas phase and grain surface reactions. Here are the grain surface reactions described in more detail, which are borrowed from \cite{2010A&A...522A..74C} and \cite{2015A&A...576A..49H}. The reactions involving dust grains are subdivided into five types.

\subsection{Adsorption on to dust grains}
\label{sec:accretion}
Species in the gas phase can be accreted on to grain surfaces. This depends on the motion of the gas species relative to the dust particle. Since the motions are dominated by thermal velocity, the adsorption rate depends on the square root of the gas temperature as $v_{\rm th} = \sqrt{8 k_{\rm B} T_{\rm g}/ \pi m}$. Once the gas species is in contact with the dust, there is a probability for it to stick on the surface of the grain. The sticking coefficient is calculated as
\begin{equation}
S(T_{\rm g},T_{\rm d}) = \left( 1 + 0.4\left( \frac{T_{\rm g} + T_{\rm d}}{100} \right)^{0.5} + 0.2\frac{T_{\rm g}}{100} + 0.08\left( \frac{T_{\rm g}}{100} \right)^2 \right)^{-1},
\label{eq:sticking}
\end{equation}
where $T_{\rm g}$ is the gas temperature and $T_{\rm d}$ is the dust temperature \citep{1979ApJS...41..555H}. It should be noted that this coefficient is inherently contrived for H atoms. Using this, the adsorption rate becomes
\begin{eqnarray}
 \nonumber
R_{ads} &=& n_{x_i} k_{ads} S(T_{\rm g},T_{\rm d}) ~~~\rm cm^{-3} s^{-1},
\\
k_{ads} &=& n_d \sigma_d v_{x_i} ~~~\rm s^{-1},
\label{eq:eqchem1}
\end{eqnarray}
where $R_{ads}$ is the adsorption rate, $k_{ads}$ is the adsorption rate coefficient, $n_d$ is the number density of dust grains, $\sigma_d$ is the cross section of the grain, $v_{x_i}$ is the thermal velocity of species $x_i$, that is, $v_{x_i} = \sqrt{8 k_{\rm B} T_{\rm g}/\pi m_{x_i}}$, with $m_{x_i}$ the mass in grams, and $n_{x_i}$ is the number density of the engaging species. In this equation, $n_d \sigma_d$ represents the total cross section of dust in cm$^{-1}$, which is obtained by integrating over the grain-size distribution. 

The MRN grain-size distribution \citep{1977ApJ...217..425M} is adopted here, with a value of $\alpha_{\rm MRN} = \langle n_d \sigma_d / n_{\rm H} \rangle_{\rm MRN} = 1\times10^{-21}$ cm$^{2}$. This distribution is chosen rather than the one of \cite{2001ApJ...548..296W}, which has a three times larger total cross section, $\alpha_{\rm WD} = 3\times10^{-21}$ cm$^{2}$, because MRN does not include poly-cyclic aromatic hydrocarbons (PAHs). The freeze-out of species on PAHs to form ices is not known.

\subsection{Thermal desorption, evaporation}
\label{sec:evaporation}
After species are bound on grain surfaces, they can evaporate back into the gas. The evaporation rate depends exponentially on the dust temperature and on the binding energies of the species with the substrate. The binding energy of each species differs according to the type of substrate. Two possible substrates are considered here, bare surfaces (assuming carbon substrate) and water ice substrate, since ices are mostly made of water. Species adsorbed on water ice substrate have in most cases binding energies higher than on bare dust or other ices \citep[e.g., see][]{2007ApJ...668..294C}. Species on top of CO, which can attain a significant coverage on dust, have binding energies that more closely resemble the binding energies of bare dust \citep[e.g.,][]{1988Icar...76..201S, 2014ApJ...781...16K}. Therefore, the binding energies on CO are chosen to be the same as on bare dust.

The fraction of the dust covered by (water) ice $\mathcal{F}_{ice}$ and bare $\mathcal{F}_{bare}$ is computed to distinguish between the two. Together with the deposited amount of water ice, this fraction depends on the total number of possible attachable sites on grain surfaces per cubic cm of space, designated as $n_{d} n_{sites}$, which is defined as 
\begin{equation}
n_{d} n_{sites} = n_{d} \frac{4\pi r_d^2}{a_{\rm pp}^2} = n_{d}\sigma_d \frac{4}{a_{\rm pp}^2} \simeq 4.44\times10^{-6} n_{\rm H} ~~~{\rm cm^{-3} mly^{-1}},
\label{eq:ndns}
\end{equation}
where the radius of dust is given by $r_d$ and the typical separation between two adsorption sites on a grain surface is given by $a_{\rm pp}$, which is assumed to be 3\,\AA. A full monolayer (mly) is reached when all the possible sites on a grain surface are occupied by an atom or a molecule. To convert this from monolayers to number densities, one needs to multiply by $n_{d} n_{sites}$. For the dust-to-gas mass ratio $\epsilon_d$ intrinsic to the dust number density $n_{d}$, the typical value of 0.01 is adopted.

The fractions, $\mathcal{F}_{ice}$ and $\mathcal{F}_{bare}$, are obtained in the following manner; if the grain is covered by less than 1 mly of water ice,
\begin{equation}
\mathcal{F}_{ice} = \frac{n_{\bot H_2O}}{n_{d} n_{sites}}.
\label{eq:icefrac}
\end{equation}
When the grain is covered by more than one layer of water ice, $\mathcal{F}_{ice} \equiv 1$. The bare fraction of the dust is obtained by
\begin{equation}
\mathcal{F}_{bare} = 1 - \mathcal{F}_{ice}.
\label{eq:barefrac}
\end{equation}

At this stage it is assumed that the bound species are homogeneously distributed and the cases where stratified layers of species can form are neglected. This assumption can lead to an over- or underestimation of reaction rates because the species abundances on different substrates can differ. It is expected, however, that since the situation $\mathcal{F}_{bare} \approx \mathcal{F}_{ice}$ is not very common, the over- or underestimation will only be marginal. With these definitions, it is possible to formulate the evaporation rate as follows:
\begin{eqnarray}
\nonumber
R_{evap} &=& n_{x_i} k_{evap} ~~~\rm cm^{-3} s^{-1},
\\
k_{evap} &=& \nu_0 \left( \mathcal{F}_{bare} \exp\left(-\frac{E_{bare,i}}{T_{d}}\right) + \mathcal{F}_{ice} \exp\left(-\frac{E_{ice,i}}{T_{d}}\right)\right) ~~\rm s^{-1},
\label{eq:eqchem2}
\end{eqnarray}
where $R_{evap}$ is the evaporation rate, $k_{evap}$ is the evaporation rate coefficient, $\nu_0$ is the oscillation frequency, which is typically $10^{12}$ s$^{-1}$ for physisorbed species, $E_{bare,i}$ and $E_{ice,i}$ are the binding energies of species $x_i$ on bare grains and ices. The species specific binding energies can be found in Table \ref{tab:bindingenergies}.

\begin{table}
\caption{Binding energies for the substrates bare grain and water ice.}
\begin{tabular}{lll|lll}
\hline
\hline
& \multicolumn{2}{c}{Substrate} & & \multicolumn{2}{c}{Substrate} \\
Species                 & Bare (K)      & Ice (K)       & Species           & Bare (K)        & Ice (K) \\
\hline
$\bot$ H                & 500$^{~eg}$   & 650$^{~nd}$   & $\bot$ CO$_2$     & 3000$^{~hr}$  & 2300$^{~hqr}$ \\
$\bot$ H$_{\rm c}$      & 10000$^{~c}$  & 10000$^{~c}$  & $\bot$ H$_2$O     & 4800$^{~aj}$  & 5700$^{~fa}$ \\
$\bot$ H$_2$            & 300$^{~js}$   & 300$^{~j}$    & $\bot$ HO$_2$     & 4000$^{~j}$   & 4300$^{~d}$ \\
$\bot$ O                & 1580$^{~po}$  & 1420$^{~po}$  & $\bot$ H$_2$O$_2$ & 6000$^{~j}$   & 6000$^{~j}$ \\%H2O2_ice=5640, O_ice=1410
$\bot$ O$_2$            & 1255$^{~jh}$  & 1160$^{~hd}$  & $\bot$ HCO        & 1600$^{~m}$   & 1600$^{~m}$ \\%O2_bare=1260
$\bot$ O$_3$            & 2230$^{~pj}$  & 2200$^{~pd}$  & $\bot$ H$_2$CO    & 3700$^{~ip}$  & 3200$^{~ip}$ \\%O3_bare=2500
$\bot$ OH               & 4600$^{~j}$   & 2850$^{~gd}$  & $\bot$ CH$_3$O    & 3700$^{~l}$   & 3700$^{~l}$ \\
$\bot$ CO               & 1200$^{~bhg}$ & 1300$^{~hk}$  & $\bot$ CH$_3$OH   & 3700$^{~l}$   & 3700$^{~l}$ \\%CO_bare=1250
\hline
\end{tabular}
\label{tab:bindingenergies}
\raggedright
$^{\textbf{a}}$\,\cite{1988Icar...76..201S}, 
$^{\textbf{b}}$\,\cite{2003Ap&SS.285..633C}, 
$^{\textbf{c}}$\,\cite{2004ApJ...604..222C}, 
$^{\textbf{d}}$\,\cite{2007ApJ...668..294C}; O$_2$(ice) = 1000\,K, O$_3$(ice) = 1800\,K, OH(ice) = 3500\,K, 
$^{\textbf{e}}$\,\cite{2010A&A...522A..74C}; H(bare) = 550\,K, 
$^{\textbf{f}}$\,\cite{2001MNRAS.327.1165F}, 
$^{\textbf{g}}$\,\cite{2011ApJ...735...15G}; H(bare) = 450\,K, CO(bare) = 1150\,K, 
$^{\textbf{h}}$\,\cite{2012MNRAS.421..768N}, 
$^{\textbf{i}}$\,\cite{2012A&A...543A...5N}, 
$^{\textbf{j}}$\,\cite{2013NatSR...3E1338D}; O$_3$(bare) = 2100\,K, 
$^{\textbf{k}}$\,\cite{2014ApJ...781...16K,2014A&A...569A.107K}; CO(ice) = 1000-1600\,K, 
$^{\textbf{l}}$\,associated with H$_2$CO binding energies, 
$^{\textbf{m}}$\,\cite{2006A&A...457..927G}, 
$^{\textbf{n}}$\,\cite{2007MNRAS.382.1648A}, 
$^{\textbf{o}}$\,\cite{2015ApJ...801..120H}; O(bare,ice) = (1850\,K,1660\,K), 
$^{\textbf{p}}$\,\cite{minissaleTez2014}; H$_2$CO(bare) = 3400\,K, 
$^{\textbf{q}}$\,\cite{2014PCCP...1615630K}; CO$_2$(ice) = 2670\,K, 
$^{\textbf{r}}$\,\cite{1990Icar...87..188S}; CO$_2$(bare,ice) = (2690\,K,2860\,K), 
$^{\textbf{s}}$\,\cite{1997ApJ...483L.131P}. \\
Some references report decreasing binding energies with coverage. Low coverage values are adopted for these cases.
\end{table}

\subsection{Two-body reactions on dust grains}
\label{sec:twobody}
While species are attached to grain surfaces, they can move around by thermal diffusion and meet other species with which they can react. The mobility of the species depends on the oscillation frequency $\nu_0$. Only physisorbed species are considered here, except for neutral hydrogen where chemisorption is also taken into account. In addition to the mobility, the reaction rate depends on the specific binding energy with the substrate and on the dust temperature. When two species meet, they can immediately react if there is little to no reaction barrier. If the reaction barrier is high, however, it might be crossed by tunneling. The probability of overcoming the reaction barrier by tunneling is given by
\begin{equation}
P_{\rm reac} = \exp\left( -(2a/\hslash)\sqrt{2 \, m_{\rm red} \, k_{\rm B} \, E_a} \, \right),
\label{eq:tunneling}
\end{equation}
where $a=1$\,\AA, $m_{\rm red}$ is the reduced mass of the two engaging species, that is, $m_{\rm red}=(m_i \times m_j) / (m_i + m_j)$, $\hslash=h/2\pi$ with $h$ the Planck constant, and $E_a$ is the activation energy of the barrier. The probability is defined as $P_{\rm reac} \equiv 1$ if there is no barrier for the reaction to take place. The reaction diffusion competition \citep{1982A&A...114..245T, 2007A&A...469..973C, 2011ApJ...735...15G}, which would make the tunneling probabilities substantially higher, is not considered here.

When a reaction occurs, the products either remain on the surface or immediately desorb into the gas phase depending on the exothermicity of the reaction. The probabilities of desorption are given by $\delta_{\rm bare}$ and $\delta_{\rm ice}$ for the two substrates. The fraction that remains on the surface will be the complement, i.e., $1 - \delta_{\rm ice, bare}$.
Non-exothermic reactions that do not desorb are by definition multiplied by 1.
With all this information, the two-body reaction rate on grain surfaces can be formulated as
\begin{eqnarray}
R_{2body} &=& \frac{n_{x_i} n_{x_j}}{n_{d} n_{sites}} P_{\rm reac} k_{2body} ~~~\rm cm^{-3} s^{-1},
\\ \nonumber
k_{2body} &=& \nu_0 \mathcal{F}_{bare} \left( \exp\left(-\frac{2}{3}\frac{E_{bare,i}}{T_{\rm d}}\right) + \exp\left(-\frac{2}{3}\frac{E_{bare,j}}{T_{\rm d}}\right)\right)(1\textrm{-}\delta_{\rm bare})
\\ \nonumber
& + & \nu_0 \mathcal{F}_{ice~~} \left( \exp\left(-\frac{2}{3}\frac{E_{ice,i~~}}{T_{d}}\right) + \exp\left(-\frac{2}{3}\frac{E_{ice,j~~}}{T_{d}}\right)\right)(1\textrm{-}\delta_{\rm ice}), %~\rm s^{-1}
\label{eq:eqchem3}
\end{eqnarray}
where $R_{2body}$ is the two-body reaction rate and $k_{2body}$ is the two-body rate coefficient. The exponent in this equation represents the diffusion of species on the surface. The diffusion is assumed to occur with a barrier of two-thirds of the binding energy (in comparison, 67\% \citealt{2013NatSR...3E1338D}, 40\% \citealt{2003Ap&SS.285..633C}, 90\% \citealt{2007ApJ...658L..37B}). The desorption rate is obtained by multiplying the rate coefficient with $\delta_{\rm ice, bare}$ instead of $1-\delta_{\rm ice, bare}$.

\subsection{CR processes on grain surfaces}
\label{sec:crs}
CR reaction rates on grain surfaces are assumed to be the same as the rates found in the gas phase. CR processes are usually inefficient destruction mechanisms, but can dominate the destruction rates deep inside the cloud. These reaction rates depend on the CR ionization rate per H$_2$ molecule, $\zeta_{\rm H_2} = 5\times10^{-17}$\,s$^{-1}$ \citep{2007ApJ...671.1736I, 2011A&A...536A..41H, 2012A&A...537A.138C}. The CR reaction rate is formulated as
\begin{eqnarray}
\nonumber
R_{CR} &=& n_{x_i} k_{CR} ~~~\rm cm^{-3} s^{-1},
\\
k_{CR} &=& z_{x_i} \zeta_{\rm H_2} ~~~\rm s^{-1},
\label{eq:eqchem4}
\end{eqnarray}
where $R_{CR}$ is the CR reaction rate, $k_{CR}$ is the CR rate coefficient, and $z_{x_i}$ the CR ionization rate factor that is subject to the ionizing element (see \textit{KiDA} database).

CR-induced UV (CRUV) photons are also considered using the same equation (Eq. \ref{eq:eqchem4}). In this case, $z_{x_i}$ is replaced by $z_{\rm CRUV}$, the UV photon generation rate per CR ionization. UV photons from CRs do not suffer from radiation attenuation as normal UV photons do \citep{2011MNRAS.414.1705P}. Hence, the lack of dependence on optical depth in this formula.

\subsection{Photoprocesses on grain surfaces}
\label{sec:uvreactions}
When UV photons arrive on a dust particle, they can interact with the adsorbed species and either photodissociate or photoevaporate them. The same formula is used for both types of photoprocesses. Photoreactions scale linearly with the local radiation flux (erg\,cm$^{-2}$\,s$^{-1}$). The radiation field strength is necessarily a function of extinction, which is given by $\xi_{x_i} A_V$, where $\xi_{x_i}$ is the extinction factor that is contingent on the relevant species. $A_V$ is obtained by dividing the column density $N_{\rm H}$ over the scaling factor, that is, $A_V = N_{\rm H}$/2.21$\times$10$^{21}$ mag \citep{2009MNRAS.400.2050G}. The general photo-process rate equation is defined as
\begin{eqnarray}
\nonumber
R_{phot} &=& n_{x_i} f_{\rm ss} k_{phot} F_{\rm UV} ~~~\rm cm^{-3} s^{-1},
\\
k_{phot} &=& \alpha_{x_i} e^{-\xi_{x_i} A_V} ~~~\rm s^{-1}.
\label{eq:eqchem5}
\end{eqnarray}
where $R_{phot}$ is the photo-process reaction rate, $k_{phot}$ is the photo-process rate coefficient, $\alpha_{x_i}$ is the unattenuated rate coefficient, $f_{\rm ss}$ is the self-shielding factor, and $F_{\rm UV}$ is the UV flux in units of $1.71\,G_0$. The factor 1.71 arises from the conversion from the Draine field \citep{1978ApJS...36..595D} to the Habing field for the far ultraviolet (FUV) intensity. The same $\alpha_{x_i}$, $\xi_{x_i}$, and $f_{\rm ss}$ is applied for the gas phase and the surface reactions.

When UV photons arrive on an icy surface with multiple layers only the first two layers are allowed to be penetrated by UV photons. As shown by \cite{2006JChPh.124f4715A}, \cite{2010JChPh.132r4510A}, and \cite{2010A&A...522A.108M}, the uppermost few layers alone contribute to photodesorption. Photodissociation does seem to occur deeper into the ice, but trapping and recombination of species tend to dominate \citep{2008A&A...491..907A}. This means that the highest number density that the photons can see is $n_{x_i} = {\rm min}(n_{x_i},2n_{d}n_{sites})$. This restriction is also enforced for reactions with CRUV photons, as given in Sect.\,\ref{sec:crs}.

Self-shielding, denoted as $f_{\rm ss}$, is taken into account for H$_2$ and CO molecules. These molecules can shield the medium against photoprocesses on grain surfaces as well as for species in the gas phase. The self-shielding factor for H$_2$ is obtained from \cite{1996ApJ...468..269D} equation 37, which is formulated as
\begin{equation}
f_{\rm ss} = \frac{0.965} {(1 + x/b_5)^2} + \frac{0.035} {(1 + x)^{0.5}}  \times {\rm exp}\left[-8.5\times 10^{-4} (1 + x)^{0.5}\right],
\label{eq:selfshield}
\end{equation}
where $x \equiv N_{\rm H_2} / 5\times 10^{14}$ cm$^{-2}$, $b_5 \equiv b/10^5$ cm\,s$^{-1}$, and $b$ is the line broadening of H$_2$ lines, which is taken as 3 km\,s$^{-1}$. This factor is solely a function of H$_2$ column density $N_{\rm H_2}$. For CO molecules, self-shielding is established by incorporating the self-shielding tables from \cite{2009A&A...503..323V} into the code. Given an H$_2$ column and a CO column, the factor $f_{\rm ss}$ is obtained from the tables. For all other species $f_{\rm ss} = 1$.

\section{Thermodynamics}
\label{app:BB}
Here the heating and cooling rates are described and their relevance to the models explained. To be used in many of the upcoming equations, $G_x$, the dust attenuated UV flux is defined as
\begin{equation}
G_x = G_0 \, {\rm exp}\left( -~\gamma_{\lambda} A_V \right),
\label{eq:g1}
\end{equation}
where $G_0$ is the radiation flux in Habing field units \citep{1968BAN....19..421H}, $A_V \equiv 1.086\,\tau_V$ is the extinction in visible wavelengths with $\tau_V$ the opacity in visible wavelengths, and $\gamma_{\lambda} ~(=A_{\lambda}/A_V)$ is the wavelength specific extinction normalized to visible. The value of $\gamma_{\lambda}$ can be obtained from, e.g., \cite{1989ApJ...345..245C}. Here the Planck averaged value of $\gamma_{\lambda}=1.8$ is adopted.

\subsection{Heating rates}
The most prominent non-conserving, non-equilibrium heating processes that are relevant to this work are introduced here. The heating processes include photoelectric heating, H$_2$ photodissociation heating, H$_2$ collisional de-excitation heating, CR heating, and, in situations when the gas temperature is lower than the dust temperature, gas-grain collisional heating. Compressional heating and shock heating are taken into account through the hydrodynamics, the EOS, and the shock routines of {\sc flash}. The heating functions and their rates are mainly obtained from \cite{1979ApJS...41..555H, 1985ApJ...291..722T, 1994ApJ...427..822B, 2005A&A...436..397M, 2009A&A...501..383W}, with some additions and adjustments of our own.

\subsubsection{Photoelectric heating}
\label{H1}
UV photons can eject electrons from atomic shells with excess kinetic energy upon photoabsorption. Electrons ejected in this way can collide with other atoms to transfer heat into the system. Neutral and positively charged \citep{2015ApJ...812..135I} dust particles are efficient electron suppliers. Photoelectric emission by small dust grains and PAHs is a dominant heating source in the diffuse ISM, which is an important part of the parameter space. This rate depends on the total density \nhtot, the electron number density $n_e$, and the radiation field strength $G_x$. The expressions provided by \cite{1994ApJ...427..822B, 2005A&A...436..397M} and \cite{2008ApJ...683..238K} are used here to calculate this type of heating. The photoelectric heating rate is defined as
\begin{equation}
\Gamma_{\rm phe} = 10^{-24} \, \epsilon \,n_{\rm H} \times G_x ~~~\rm erg ~cm^{-3} ~s^{-1},
\label{eq:heat1}
\end{equation}
where the heating efficiency $\epsilon$, which inherently estimates the grain charge, is given by
\begin{equation}
\epsilon = \frac{0.0487}{\left(1+4\times10^{-3} \psi^{0.73}\right)} + \frac{0.0365\,\left(T_{\rm g}/10^4\right)^{0.7}}{(1+2\times10^{-4} \psi)}.
\label{eq:heat1b}
\end{equation}
In here, $\psi$ is the grain charge parameter that estimates the ratio of ionization over recombination. $\psi$ is defined as
\begin{equation}
\psi \equiv G_x \sqrt{T_{\rm g}}/n_e.
\label{eq:heat1c}
\end{equation}

The first term of the efficiency equation (Eq. \ref{eq:heat1b}) describes the efficiency of PAHs, while the second term approximates the larger grains. This expression can be extended to account for very large grains (VLGs $> 1 \mu$m) as provided by \cite{2000A&A...353..276K} and \cite{2009A&A...501..383W} using equations 93 and 94 in the latter paper. Since it is not expected to have much VLGs during the evolution of translucent clouds, this has been excluded here.

It can be noted that by taking He into account in the number density, as \cite{2015MNRAS.449.2643B} advocate, the photoelectric heating rate needs to be multiplied by a factor 1.33. Whereas \cite{2003ApJ...587..278W} add a factor 1.3 due to the increased PAH abundances, following the \cite{1999ESASP.427..579T} PAH analysis. However, \cite{2003ApJ...587..278W} also modify the electron number density in $\psi$ with a factor $\phi_{\rm PAH}=0.5$ to scale the electron-PAH collision rates, which reduces the heating efficiency.

\subsubsection{H$_2$ photodissociation heating}
\label{H2}
Energetic UV photons can also dissociate H$_2$ molecules into neutral hydrogen atoms. The process involves exciting the H$_2$ molecule by Lyman-Werner band UV photons which can spontaneously decay into the electronic ground state. About 10\% of the excited H$_2$ molecules decay by dissociating the molecule, thereby releasing 0.4 eV in the form of kinetic energy \citep{1973ApJ...186..165S, 1996A&A...307..271S}. To obtain the photodissociation rate, one has to take into account the H$_2$ self-shielding $f_{\rm ss}$ that is described in Eq. \ref{eq:selfshield}.
Photodissociation heating by H$_2$ is one of the main photodissociation heating rates in the ISM due to the relative high H$_2$ abundance. This heating rate is given by
\begin{equation}
\Gamma_{\rm h2p} = 2.179\times10^{-23} \, f_{\rm ss} \, n_{\rm H_2} \, G_x ~~~\rm erg ~cm^{-3} ~s^{-1}.
\label{eq:heat2}
\end{equation}

\subsubsection{H$_2$ collisional de-excitation heating}
\label{H3}
UV pumped H$_2$ can also cascade down to an excited, vibrational level ($v > 0$), only to collisionally decay back to the lowest ground state. Radiative energy is converted into thermal energy in this way. A simplified, two-level approximation version of this cascade process is used here, similar to the rate calculated by \cite{1979ApJS...41..555H} and other more recent publications \citep{2005A&A...436..397M, 2009A&A...501..383W}. A two-level system is an accurate approximation below $T_{\rm g} = 4000$\,K since the excitation energy of the first vibrational state is under 6000\,K, while collisional dissociation of H$_2$ dominates the H$_2$ cooling above 4000\,K. Aside from the gas temperature and the number density of vibrationally excited molecular hydrogen $n_{\rm H^{\star}_2}$, the heating rate depends on the atomic hydrogen number density $n_{\rm HI}$ and molecular hydrogen number density $n_{\rm H_2}$, the two main collisional partners that de-excite H$_2$. The condensed version of this rate is expressed as
\begin{eqnarray}
\Gamma_{\rm h2c} &=& n_{\rm H^{\star}_2} \Delta E_{ul} \left(n_{\rm HI}C_{ul}^{\rm HI} + n_{\rm H_2}C_{ul}^{\rm H_2}\right) ~~~ \rm erg ~cm^{-3} ~s^{-1},
\\ \nonumber
&=& 4.166\times10^{-24} \, n_{\rm H^{\star}_2} \sqrt{T_{\rm g}}
\\ \nonumber
& & \times\left( n_{\rm HI} \, {\rm exp}\left(-\frac{1000}{T_{\rm g}}\right) + 1.4\,n_{\rm H_2} \, {\rm exp}\left(-\frac{18,100}{T_{\rm g}+1200}\right) \right),
\label{eq:heat3}
\end{eqnarray}
where $C_{ul}^{\rm HI}$ and $C_{ul}^{\rm H_2}$ [$\rm cm^3 \, s^{-1}$] are collisional de-excitation coefficients \citep{1985ApJ...291..722T} and $\Delta E_{ul}=2.6$ eV is the effective energy of the pseudo vibrational level \citep{1978ApJ...225..405L}. 

In this work $n_{\rm H^{\star}_2}$ is not directly calculated, but approximated by $n_{\rm H^{\star}_2} = 10^{-7} n_{\rm H_2}$. It seems that $n_{\rm H^{\star}_2}/n_{\rm H_2} \leq 10^{-6}$ is a commonly found value for various ISM with $A_V > 1$, see \cite{2010ApJ...713..662A}. However, the choice of $n_{\rm H^{\star}_2}$ is not vital as this heating rate is typically orders of magnitude lower than other heating rates at cold ($T_{\rm g} < 100 \rm \, K$) ISM conditions. The gas temperature in the current simulations is always below 100 K at densities above 100 \cmcube. On a separate note, a correction may be applied to $n_{\rm H^{\star}_2}$ by subtracting the collisional excitation of H$_2$, i.e., $n_{\rm H^{\star}_2} - n_{\rm H_2} \exp(-{\Delta E_{ul}}/{k_{\rm B}T_{\rm g}})$, following \cite{2009A&A...501..383W}.

\subsubsection{CR heating}
\label{H4}
Gas heating by CRs is considerably lower compared to heating by UV photons at low opacities. They can, however, penetrate high column densities while experiencing very little extinction, i.e., $N_H > 10^{25}$\,cm$^{-2}$ or 96\,g\,cm$^{-2}$ \citep{2009A&A...501..383W}, which is about 5 orders of magnitude larger than UV photons can penetrate \citep{2007ARA&A..45..339B, 2011MNRAS.414.1705P}. Because of this, CR heating can become a dominant heating source deep inside a molecular cloud. Atomic and molecular hydrogen releases about 3.5 and 8 eV, respectively, into the gas upon primary ionization \citep{2004A&A...428..511J}. Combining both atomic and molecular heating, the CR heating rate due to H$_2$ ionization including the small He ionization contribution \citep{1985ApJ...291..722T, 2005A&A...436..397M} together with the HI ionization \citep{2004A&A...428..511J, 2009A&A...501..383W}, is given by
\begin{equation}
\Gamma_{\rm CR} = \zeta_{\rm CR} \left( 1.5\times10^{-11}n_{\rm H_2} +  0.55\times10^{-11}n_{\rm HI}\right) ~~~\rm erg ~cm^{-3} ~s^{-1}, 
\label{eq:heat4}
\end{equation}
where $\zeta_{\rm CR}$ is the primary CR ionization rate for which the value of $\zeta_{\rm CR} = 5\times10^{-17}$ s$^{-1}$ is adopted \citep{2000ApJ...538..115S}. Higher ionization rates of up to $10^{-15}$\,s$^{-1}$ have also been found
\citep{1998ApJ...499..234C, 2008ApJ...688..306G, 2012ApJ...745...91I}, however, in general there seems to be a sharp drop at about $N_{\rm H_2} = 10^{22}$\,cm$^{-2}$, i.e., $A_V \approx 10$ to an ionization rate that lies in between $10^{-17}-10^{-16}$\,s$^{-1}$ \citep{2015ApJ...812..135I}.

\cite{2004A&A...428..511J} used a prefactor of $2.5\times10^{-11}$\,erg for the first term of Eq. \ref{eq:heat4}, which means that they either assume more energy is converted to heat by the free electrons or that secondary ionizations with significant energy are also considered. In this case the CR heating will be about 67\% higher in molecular clouds.

\subsection{Cooling rates}
The cooling of the gas is insured by several different types of non-conserving, non-equilibrium processes. These processes are: electron recombination with PAHs cooling, electron impact with H (i.e., Ly $\alpha$) cooling, metastable transition of [OI]-630 nm cooling, fine-structure line cooling of [OI]-63 $\mu$m, [CI]-609 $\mu$m and [CII]-158 $\mu$m, molecular cooling by H$_2$, CO, OH, and H$_2$O, and gas-grain collisional cooling. The last mentioned is the same function as given in heating except that the gas in this case is warmer than the dust. The cooling functions and their rates are gathered from \cite{1989ApJ...338..197S}, \cite{1993ApJ...418..263N}, \cite{1994ApJ...427..822B}, \cite{1995ApJS..100..132N}, \cite{2005A&A...436..397M}, \cite{2009A&A...501..383W}, \cite{2010ApJ...722.1793O}, and Paper\,I. Cooling by adiabatic expansion is handled by the EOS routines of {\sc flash}.

\subsubsection{Electron recombination with PAHs cooling}
\label{C1}
Electrons can recombine with atoms and molecules, and in particular with PAHs, which cool the gas as a result. The electron's kinetic energy is radiated away in this case. Each recombination results in a loss of $E_{\rm rec} \lesssim \frac{3}{2} k_{\rm B} T_e$, on average, which is the mean kinetic energy of the recombining electrons. In reality this is typically lower and about $E_{\rm rec} \sim 0.7 k_{\rm B} T_e$, because the slower electrons are more easily captured by protons. This form of cooling is important at high temperatures, i.e., $T_{\rm g} \gtrsim 5000$ K and therefore not critical to the results in this work. 

The cooling scales with the gas temperature and with the parameter $\psi$ (Eq. \ref{eq:heat1c}) due to its effect on Coulomb interactions. Following the analytical expression of \cite{1994ApJ...427..822B}, the electron recombination cooling rate is approximated by
\begin{equation}
\Lambda_{\rm rec} = 3.49\times10^{-30} T_{\rm g}^{0.944} \psi^{\beta_{\rm rec}} \, n_{\rm e} n_{\rm H} ~~~\rm erg ~cm^{-3} ~s^{-1}, 
\label{eq:cool1}
\end{equation}
where $\beta_{\rm rec} \equiv 0.735/T_{\rm g}^{0.068}$.

\subsubsection{Electron impact with H: Ly $\alpha$ cooling}
\label{C2}
Electrons impacting hydrogen atoms can result in the excitation of the atom to the second electronic state. The decay to the ground state releases a Ly $\alpha$ photon which can escape the local medium. This form of cooling is highly efficient at gas temperatures above 8000\,K, which increases with temperature. The rate depends on the total hydrogen and electron number densities. Neglecting Ly $\alpha$ line trapping, which is a good approximation for cold or optically thin regions, the cooling rate is given by \citep{1978ppim.book.....S}
\begin{equation}
\Lambda_{\rm Ly \alpha} = 7.3\times10^{-19} n_{\rm e} n_{\rm H} \, {\rm exp} \left( -\frac{118,400\,\rm K}{T_{\rm g}} \right) ~ \rm erg ~cm^{-3} ~s^{-1},
\label{eq:cool2}
\end{equation}
where 118,400\,K = 10.2\,eV is the excitation energy to the second electronic state and the prefactor contains the transition energy and the collisional de-excitation coefficient.

\subsubsection{Metastable [OI]-630 nm cooling}
\label{C3}
At temperatures in excess of a few 1000 K, the $^1$D--$^3$P metastable line emission due to neutral oxygen is also efficient. This form of cooling arises as a result of hydrogen atoms exciting neutral oxygen atoms through collisions.
Similar to expression Eq. \ref{eq:cool2}, the [OI]-630 nm metastable line cooling is defined as 
\begin{equation}
\Lambda_{\rm OI-630} = 1.8\times10^{-24} n_{\rm OI} n_{\rm H} \, {\rm exp} \left( -\frac{22,800\, \rm K}{T_{\rm g}} \right) ~\rm erg ~cm^{-3} ~s^{-1},
\label{eq:cool3}
\end{equation}
where $n_{\rm OI}$ is the neutral oxygen number density \citep{1989ApJ...338..197S}. Oxygen can also be excited by impinging electrons resulting in an enhanced cooling rate \citep[Eq. D12,][]{1989ApJ...338..197S}
\begin{equation}
\Lambda^e_{\rm OI-630} = 8.2\times10^{-23} n_{\rm OI} n_{\rm e} \, \frac{ {\rm exp} \left( -{22,800\, \rm K}/{T_{\rm g}} \right) }
{ \sqrt{T_{\rm g}} + 7\times10^{-9} n_{\rm e} T }
~\rm erg ~cm^{-3} ~s^{-1}.
\label{eq:cool3b}
\end{equation}
This rate is generally much smaller compared to cooling through hydrogen collisions, because $n_e/n_{\rm H} \ll 1$.

\subsubsection{Fine-structure line cooling by [OI]-63 $\mu$m, [CII]-158 $\mu$m, and [CI]-609 $\mu$m}
\label{C4}
When most of the gas is in atomic form, the dominant mode of cooling occurs through atomic fine-structure lines for $\rm T_{\rm g} \lesssim 5000 \, K$. This is one of the most important cooling mechanisms in the diffuse ISM. Fine-structure line cooling can have a strong impact at the early stages of cloud evolution. In this, fine-structure transitions such as [OI]-63 $\mu$m and [CII]-158 $\mu$m are the most prominent cooling transitions to consider. 

In the case of optically thin lines, the subthermal, non-LTE two-level line emission cooling is defined as 
\begin{eqnarray}
\Lambda_{\rm ul} (n_{\rm H} \ll n_{cr}) &=& n_{\rm H} n_l \frac{g_u}{g_l} \gamma_{lu} \exp\left( -\frac{h\nu_{ul}}{k_{\rm B}T_{\rm g}} \right) h\nu_{ul}
\\ \nonumber
&=& n_{\rm H} n_l \gamma_{ul} h\nu_{ul} ~~~ \rm erg ~cm^{-3} ~s^{-1},
\label{eq:cool4}
\end{eqnarray}
where the $u$ and $l$ subscripts denote the upper and lower levels, $n_{cr}$ is the temperature dependent critical number density, $\gamma_{ul}$ and $\gamma_{lu}$ are the collisional (de-)excitation coefficients, $g_l$ and $g_u$ are the statistical weights, and $h\nu_{ul}$ is the energy difference between the two levels. For the [OI]-63 $\mu$m and the [CII]-158 $\mu$m fine-structure lines the cooling rates in the optically thin, non-LTE cases are a by \cite{2005fost.book.....S}. Here their expressions are reformulated with more precise values, i.e.,
\begin{eqnarray}
\Lambda_{\rm OI}^{\rm thin}~ &=& 1.2\times10^{-9} \left( \frac{n_{\rm H}}{10^3 \,\rm cm^{-3}} \right)^2 \exp\left( -\frac{227.7 \,\rm K}{T_{\rm g}} \right) ~\frac{\rm eV}{\rm cm^{3} ~s}, %\rm eV ~cm^{-3} ~s^{-1},
\label{eq:cool4a}
\\
\Lambda_{\rm CII}^{\rm thin} &=& 8.6\times10^{-10} \left( \frac{n_{\rm H}}{10^3 \,\rm cm^{-3}} \right)^2 \exp\left( -\frac{91.2 \,\rm K}{T_{\rm g}} \right) ~\frac{\rm eV}{\rm cm^{3} ~s},
\label{eq:cool4b}
\end{eqnarray}
to which the [CI]-609 $\mu$m line cooling is added in the same format
\begin{eqnarray}
\Lambda_{\rm CI}^{\rm thin} &=& 1.3\times10^{-11} \left( \frac{n_{\rm H}}{10^3 \,\rm cm^{-3}} \right)^2 \exp\left( -\frac{23.6 \,\rm K}{T_{\rm g}} \right) ~\frac{\rm eV}{\rm cm^{3} ~s}.
\label{eq:cool4b2}
\end{eqnarray}
In these equations, only the collisional rate coefficients with atomic Hydrogen are used. The main calculations are, however, performed for multiple collisional species. The rates in the above equations are gathered from \cite{1977A&A....56..289L}, \cite{2005ApJ...620..537B}, and \cite{2007ApJ...654.1171A}. Other rates are taken from the Leiden Atomic and Molecular Database \citep[LAMDA, ][]{2005A&A...432..369S} 

In order to account for LTE effects, so when $n_{\rm H} \gg n_{cr}$ and, as a result, the cooling rate conforms to a linear scaling with density as opposed to $n_{\rm H}^2$, the following correction factor is applied
\begin{equation}
f_{lte, x_i} = \frac{n_{cr, x_i}}{n_{cp} + n_{cr, x_i}},
\label{eq:cool4c}
\end{equation}
where $n_{cp}$ is the number density of the collisional partner, which is often either atomic hydrogen or molecular hydrogen. The calculations in this work are performed for both collisional partners as well as collisions with electrons and protons, so $n_{cp} = [n_{\rm HI},n_{\rm H_2},n_{\rm e^-},n_{\rm H^+}]$. The temperature dependence of the rates are also taken care of by scaling the critical densities according to a fit over the LAMDA data or by directly using \cite{1989ApJ...342..306H}.

In order to account for radiation trapping, that is, to also consider optically thick cases, an escape probability is used, which depends on the optical depth parameter $\tau$. The escape probability expression for a sphere is

\begin{equation}
f_{esc, x_i} = \frac{1-e^{-\tau_{x_i}}}{\tau_{x_i}}.
\label{eq:cool4d}
\end{equation}
$\tau_{x_i}$ is obtained by integrating the absorption coefficient $\alpha_{\nu}$ along the path $ds$ and over all frequencies. By making use of the Einstein coefficients to define $\alpha_{\nu}$, thereby adopting a Gaussian line profile function, the following expressions are acquired, i.e.,
\begin{eqnarray}
\tau_{\rm OI}~ &=& 8.4\times10^{-13}  \frac{\Delta s}{\Delta v} \left( \exp{\left( \frac{227.7 \rm \, K}{\rm T_{\rm g}} \right)} - \frac{n_{cp}}{n_{cp} + n_{cr,{\rm OI}}} \right) n_u,
\label{eq:cool4e}
\\
\tau_{\rm CII} &=& 3.4\times10^{-13} \frac{\Delta s}{\Delta v} \left( \exp{\left( \frac{91.2 \rm \, K}{\rm T_{\rm g}} \right)} - \frac{n_{cp}}{n_{cp} + n_{cr,{\rm CII}}} \right) n_u, 
\label{eq:cool4f}
\\
\tau_{\rm CI}~ &=& 6.7\times10^{-13} \frac{\Delta s}{\Delta v} \left( \exp{\left( \frac{23.6 \rm \, K}{\rm T_{\rm g}} \right)} - \frac{n_{cp}}{n_{cp} + n_{cr,{\rm CI}}} \right) n_u, 
\label{eq:cool4g}
\end{eqnarray}
where $\Delta v$ is the quadratic mean of the turbulent and the thermal velocities and $\Delta s$ is the distance at which radiation can escape. This distance is found by making use of the large velocity gradient approximation (LVG).

Hence, by combining Eqs. \ref{eq:cool4a}, \ref{eq:cool4b}, and \ref{eq:cool4b2} with \ref{eq:cool4c} and \ref{eq:cool4d} the atomic fine-structure cooling rates are gained,
\begin{eqnarray}
\Lambda_{\rm OI}~ &=& \Lambda_{\rm OI}^{\rm thin} \times f_{lte,{\rm OI}} \times f_{esc,{\rm OI}},
\label{eq:cool4h}
\\
\Lambda_{\rm CII} &=& \Lambda_{\rm CII}^{\rm thin} \times f_{lte,{\rm CII}} \times f_{esc,{\rm CII}}.
\label{eq:cool4i}
\\
\Lambda_{\rm CI}~ &=& \Lambda_{\rm CI}^{\rm thin} \times f_{lte,{\rm CI}} \times f_{esc,{\rm CI}}.
\label{eq:cool4j}
\end{eqnarray}

Relevant constants, such as the Einstein A, collisional partner densities, as well as the critical species densities are gathered from LAMDA \citep{2005A&A...432..369S}.

\subsubsection{Molecular cooling by H$_2$, CO, H$_2$O, and OH}
\label{C5}
In the presence of molecules at densities of \nhtot $\lesssim$ 3$\times 10^{4}$\cmcube, with some dependence on opacity and temperature, molecular ro-vibrational line cooling will dominate the thermal balance of interstellar gas clouds. Among the typical molecules present in molecular clouds, H$_2$, CO, and H$_2$O are the most prominent coolants for a variety of different physical conditions. In the current calculations, next to the aforementioned molecular transitions, the rotational transitions of OH are also included. Owing to the lower transitional energies of the rotational levels, rotational transition cooling (emitted in $\mu$m wavelengths) will be more effective at lower temperatures, while vibrational transition cooling (emitted in the infrared) is often more efficient at higher temperatures.

A detailed treatment of the radiative cooling processes is implemented as provided by \cite{1993ApJ...418..263N} for the molecules H$_2$, CO (vibrational), and H$_2$O (high T), complemented by \cite{1995ApJS..100..132N} for H$_2$O (low T), and by \cite{2010ApJ...722.1793O} for CO (rotational) and OH. The final cooling functions are calculated and applied using the four-parameter ($L_0$, $\ell_{\rm LTE}$, $ n_{\frac{1}{2}}$, $\alpha_f$) fitting function described and defined by \cite{1993ApJ...418..263N} as
\begin{equation}
\frac{1}{L_{{\rm mol,}x_i}} = \frac{1}{L_0} + \frac{n_{\rm H_2}}{\ell_{LTE}} + \frac{1}{L_0} \left[ \frac{n_{\rm H_2}}{n_{\frac{1}{2}}} \right]^{\alpha_f} \left( 1 - \frac{n_{\frac{1}{2}}L_0}{\ell_{LTE}} \right),
\label{eq:cool5a}
\end{equation}
where $L_{{\rm mol,}x_i}$ is the cooling function with $x_i = [\rm H_2, CO, H_2O, OH]$, $L_0$ is the cooling function coefficient in the low density limit, which depends on $T_{\rm g}$ alone, and $\ell_{\rm LTE}$, $ n_{\frac{1}{2}}$, and $\alpha_f$ are functions of both $T_{\rm g}$ and the column density parameter $\widetilde{N}_{x_i}$ \citep{1993ApJ...418..263N}. The column density parameter is defined as
\begin{equation}
\widetilde{N}_{x_i} = N_{x_i} / \Delta v ~~~\rm cm^{-2} ~km^{-1} ~s.
\label{eq:cool5b}
\end{equation}
The tabulated fit parameters are linearly interpolated to one degree of temperature precision and to one percent of a dex of column precision. To obtain the molecular ro-vibrational cooling rates, one has to multiply the cooling function by the colliding species' number densities. %AM I SURE?? .. and the population density of the rotational or vibrational state of that species
The cooling rate can be written as
\begin{equation}
\Lambda_{{\rm mol,}x_i} = L_{{\rm mol,}x_i} n_{x_i} n_{cp, x_i} ~~~\rm erg ~cm^{-3} ~s^{-1}.
\label{eq:cool5}
\end{equation}

To solve this equation one needs $n_{cp, x_i}$, which is the collisional partner or partners needed to excite species $x_i$. In general, these are H$_{2}$ molecules, but also atomic hydrogen and electrons can be important collisional partners. To also take these into account, $n_{cp, x_i}$ is calculated by combining and simply scaling with each collisional partner, which is different for each type of molecular excitation. This work follows \cite{1995ApJS..100..132N} and \cite{1997PhDT.........4Y} to obtain the scaled collisional partner densities. Since they are hard to come by, the rotational ($n_{cp, rH_2}$, $n_{cp, rCO}$, $n_{cp, rH_2O}$, $n_{cp, rOH}$) and the vibrational ($n_{cp, vH_2}$, $n_{cp, vCO}$, $n_{cp, vH_2O}$) collisional partner densities are provided here, which are
\begin{eqnarray}
n_{cp, rH_2} = n_{\rm H_2} + 7n_{\rm H} + 16n_{\rm e} ~~~\rm cm^{-3},\nonumber
\\
n_{cp, vH_2} = n_{\rm H_2} + 7n_{\rm H} + 16n_{\rm e} ~~~\rm cm^{-3}, \nonumber
\\
n_{cp, rCO} = n_{\rm H_2} + \left( \sqrt{2} \frac{\sigma_{\rm H}}{\sigma_{\rm H_2}} \right) n_{\rm H} + \left( \frac{1.3\times10^{-8}}{v\sigma_{\rm H_2}} \right)  n_{\rm e} ~~~\rm cm^{-3}, \nonumber
\\
n_{cp, vCO} = n_{\rm H_2} + 50n_{\rm H} + \left( \frac{L_{\rm CO_e}}{L_{\rm CO_0}} \right) n_{\rm e} ~~~\rm cm^{-3},
\\
n_{cp, rH_2O} = n_{\rm H_2} + 10n_{\rm H} + \left( \frac{K_{\rm H_2O_e}}{K_{\rm H_2O_{H_2}}} \right) n_{\rm e} ~~~\rm cm^{-3}, \nonumber
\\
n_{cp, vH_2O} = n_{\rm H_2} + 10n_{\rm H} + \left( \frac{L_{\rm H_2O_e}}{L_{\rm H_2O_0}} \right) n_{\rm e} ~~~\rm cm^{-3}, \nonumber
\\
n_{cp, rOH} = n_{\rm H_2} + n_{\rm H} + \left( \frac{K_{\rm OH_e}}{K_{\rm OH_{H_2}}} \right) n_{\rm e} ~~~\rm cm^{-3}, \nonumber
\label{eq:rvall}
\end{eqnarray}
where the parameters $\sigma_{\rm H}$, $\sigma_{\rm H_2}$, $v\sigma_{\rm H_2}$, $L_{\rm CO_e}$, $L_{\rm CO_0}$, $L_{\rm H_2O_e}$, $L_{\rm H_2O_0}$, $K_{\rm H_2O_e}$, $K_{\rm H_2O_{H_2}}$, $K_{\rm OH_e}$, and $K_{\rm OH_{H_2}}$ are defined as
\begin{eqnarray}
\sigma_{\rm H} = 2.30\times10^{-15} ~~~\rm cm^{-2}, \nonumber
\\
\sigma_{\rm H_2} = 3.30\times10^{-16} \left(\frac{T_{\rm g}}{1000}\right)^{-1/4} \rm cm^{-2}, \nonumber
\\
v\sigma_{\rm H_2} = 1.03\times10^4 \sqrt{T_{\rm g}} \, \sigma_{\rm H_2} ~~~\rm cm^{-1} ~s^{-1}, \nonumber
\\
L_{\rm CO_e} = 1.03\times10^{-10} \left( \frac{T_{\rm g}}{300} \right)^{0.938} {\rm exp}\left(-\frac{3080}{T_{\rm g}}\right), \nonumber
\\
L_{\rm CO_0} = 1.14\times10^{-14} \, T_{\rm g} \, {\rm exp} \left(-\frac{68}{T_{\rm g}^{1/3}}\right) {\rm exp}\left(-\frac{3080}{T_{\rm g}}\right), \nonumber
\\
L_{\rm H_2O_e} = 2.60\times10^{-6} \, T_{\rm g}^{-0.5} \, {\rm exp}\left(-\frac{2325}{T_{\rm g}}\right),
\\
L_{\rm H_2O_0} = 0.64\times10^{-14} \, T_{\rm g} \, {\rm exp} \left(-\frac{47.5}{T_{\rm g}^{1/3}}\right) {\rm exp}\left(-\frac{2325}{T_{\rm g}}\right), \nonumber
\\
K_{\rm H_2O_e} = 8.57\times10^{-6} \, T_{\rm g}^{-0.5} \, {\rm exp}\left(-\frac{115.1}{T_{\rm g}}\right)  ~~~\rm cm^{3} ~s^{-1} \nonumber
\\
\times ~{\rm ln} \left[2.12 + 184.32 \, T_{\rm g} \, {\rm exp}\left(-\frac{0.577}{1+230.1/T_{\rm g}}\right) \right], \nonumber
\\
K_{\rm H_2O_{H_2}} = 7.40\times10^{-12} \, \sqrt{T_{\rm g}} ~~~\rm cm^{3} ~s^{-1} \nonumber
\\
K_{\rm OH_e} = 6.43\times10^{-6} \, T_{\rm g}^{-0.5} \, {\rm exp}\left(-\frac{135.9}{T_{\rm g}}\right)  ~~~\rm cm^{3} ~s^{-1} \nonumber
\\
\times ~{\rm ln}\left[ 2.21 + 1.63\times10^{-2} \, T_{\rm g} \, {\rm exp}\left(-\frac{0.577}{1+271.9/T_{\rm g}}\right) \right], \nonumber
\\
K_{\rm OH_{H_2}} = 7.20\times10^{-12} \, \sqrt{T_{\rm g}} ~~~\rm cm^{3} ~s^{-1}. \nonumber
\label{eq:rvall2}
\end{eqnarray}
Several of the equations were reduced to appear in their current trimmed form.

\subsubsection{Gas-grain collisional heat exchange}
\label{HC}
One of the strongest cooling mechanisms for gas in molecular clouds at densities of \nhtot $> 10^4$ \cmcube is the cooling by way of gas-grain collisional heat exchange. Depending on the condition, where the gas temperature falls below the dust temperature, this can also be a source of heating and, therefore, is not directly labeled as a cooling process. The reason to mention this type of heat transfer in this section is because it is mostly a source of cooling in this work. The energy exchange proceeds through inelastic collisions between gas and dust particles. The rate at which this occurs depends on the temperature difference between the two phases, the total cross section of dust in a given volume, and on the density of the gas. The gas-grain collisional heat exchange rate is generally defined as \citep{1979ApJS...41..555H, 1983ApJ...265..223B, 2011ApJ...729L...3D}
\begin{equation}
\Lambda_{\rm gg} = n_{\rm H} n_{d} \sigma_d v_{\rm p} f_v \left( 2k_{\rm B}T_{\rm g} - 2k_{\rm B}T_{\rm d} \right) ~~~\rm erg ~cm^{-3} ~s^{-1},
\label{eq:cool6a}
\end{equation}
where $v_{\rm p} = 1.45\times10^4 \sqrt{T}$\,cm\,s$^{-1}$ is the average speed of hydrogen nuclei (protons), $f_v$ is a measure of the contribution by other species, which is usually taken to be 0.3 or 0.4, and $n_{d} \sigma_d$ is, once again, the total dust cross section. The average time between two successive collisions can also be retrieved from this equation, that is, $t_{\rm coll} = (n_{\rm H} \sigma^*_d v_{\rm p} f_v)^{-1} = (n_{d} \sigma_d v_{\rm p} f_v)^{-1}$, with $\sigma^*_d$ the total cross section per hydrogen nuclei, i.e., $\sigma^*_d = n_{d} \sigma_d / n_{\rm H}$. It is worth noting that this is given wrongly in \cite{2006MNRAS.369.1437S}.

For an MRN distribution, Eq. \ref{eq:cool6a} can be simplified to the one given by \cite{1989ApJ...342..306H}
\begin{eqnarray}
\Lambda_{\rm gg} &=& 1.2\times10^{-32} \, n_{\rm H}^2 \left( \frac{T_{\rm g}}{10\rm \,K} \right)^{1/2} \left( \frac{100\,\rm\mathring{A}}{a_{min}} \right)^{1/2} \times \nonumber
\\
& & \left[ 1 - 0.8 \, {\rm exp}(-75/T_{\rm g}) \right] \left(T_{\rm g} -T_{\rm d}\right) ~~~\rm erg ~cm^{-3} ~s^{-1},
\label{eq:cool6b}
\end{eqnarray}
where $a_{\rm min}$ sets the minimum grain size. The value of 10\,\AA{} for the minimum grain size is maintained throughout this work following \cite{2005A&A...436..397M}. The choice of a low $a_{\rm min}$ follows from the reasoning that one also want to take into account the thermal accommodation by PAHs. In comparison, MRN has a value of $50 \,$\AA. 

Quite large differences exist in the devised rates between various groups, such as the ones cited above Eq. \ref{eq:cool6a}, which can differ by over an order of magnitude. The main discrepancy comes from the choice in the grain size distribution. One has to make this choice with care. The rate given in Eq. \ref{eq:cool6b} is one that is somewhat on the high side. Pre-factors from other studies, minding the factor $\sqrt{10}$, are $2.0\times10^{-33}$ \cite{2001ApJ...557..736G}, $3.16\times10^{-33}$ \cite{2005ApJ...635.1151K}, $2.09\times10^{-33}$ (0.1$\mu$m) and $5.01\times10^{-33}$ (MRN) \cite{2006MNRAS.369.1437S}, $1.58\times10^{-33}$ (0.1$\mu$m) and $3.79\times10^{-33}$ (MRN) \cite{2009A&A...501..383W}, and $0.79\times10^{-33}$ \cite{2015MNRAS.449.2643B}. An MRN distribution ($n_d\sigma_d=1\times10^{-21}n_{\rm H}$) gives a factor 2.4 larger rate than a fixed grain size of $a=0.1 \,\mu$m ($n_d \pi a^2 = 4.18\times10^{-22}$\nhtot cm$^{-1}$). A \cite{2001ApJ...548..296W} distribution will give a factor three larger rate than MRN.

\section{Reaction tables}
\label{app:CC}
Here we tabulate the gas-phase and the surface reactions including the adopted activation barriers in Kelvin, E$_A$, within the relevant tables. The 24 self-evident accretion and evaporation reactions, for example $\rm H \rightarrow \bot H$ and $\rm \bot CO \rightarrow CO$, are omitted.

\begin{table*}
\begin{small}
\caption{Photoprocesses (64).}
\begin{tabular}{@{}c@{}}
\hline
\hline
\begin{tabular}{lll}
\multicolumn{3}{c}{Gas} \\
Reactants & & Products \\
\hline
H         + CR          & $\rightarrow$ & H$^+$     + e$^-$     \\%&
O         + CR          & $\rightarrow$ & O$^+$     + e$^-$     \\%&
CO        + CR          & $\rightarrow$ & C         + O         \\
H$_2$     + CR          & $\rightarrow$ & H         + H             \\%&
H$_2$     + CR          & $\rightarrow$ & H         + H$^+$ + e$^-$        \\%&
H$_2$     + CR          & $\rightarrow$ & H$^+$     + H$^-$        \\
H$_2$     + CR          & $\rightarrow$ & H$_2^+$ + e$^-$        \\%&
O$_2$     + CRUV        & $\rightarrow$ & O         + O         \\%&
OH        + CRUV        & $\rightarrow$ & H         + O         \\
CO$_2$    + CRUV        & $\rightarrow$ & O         + CO        \\%&
H$_2$O    + CRUV        & $\rightarrow$ & H         + OH        \\%&
HCO       + CRUV        & $\rightarrow$ & H         + CO        \\
HCO       + CRUV        & $\rightarrow$ & HCO$^+$   + e$^-$        \\%&
H$_2$CO   + CRUV        & $\rightarrow$ & CO        + H$_2$        \\%&
CH$_3$OH  + CRUV        & $\rightarrow$ & H$_2$     + H$_2$CO      \\
C         + CRUV        & $\rightarrow$ & C$^+$     + e$^-$        \\%&
CH$_3$OH  + Photon      & $\rightarrow$ & H$_2$     + H$_2$CO      \\%&
H$_2^+$   + Photon      & $\rightarrow$ & H         + H$^+$        \\
OH$^+$    + Photon      & $\rightarrow$ & O         + H$^+$        \\%&
H$_3^+$   + Photon      & $\rightarrow$ & H$_2$     + H$^+$        \\%&
H$_3^+$   + Photon      & $\rightarrow$ & H         + H$_2^+$       \\
C$^-$     + Photon      & $\rightarrow$ & C         + e$^-$        \\%&
H$^-$     + Photon      & $\rightarrow$ & H         + e$^-$        \\%&
O$^-$     + Photon      & $\rightarrow$ & O         + e$^-$        \\
C         + Photon      & $\rightarrow$ & C$^+$     + e$^-$        \\%&
CO        + Photon      & $\rightarrow$ & C         + O         \\%&
H$_2$     + Photon      & $\rightarrow$ & H         + H         \\
O$_2$     + Photon      & $\rightarrow$ & O         + O         \\%&
OH        + Photon      & $\rightarrow$ & H         + O         \\%&
OH        + Photon      & $\rightarrow$ & OH$^+$    + e$^-$        \\
CO$_2$    + Photon      & $\rightarrow$ & O         + CO        \\%&
H$_2$O    + Photon      & $\rightarrow$ & H         + OH        \\%&
H$_2$O    + Photon      & $\rightarrow$ & H$_2$O$^+$+ e$^-$        \\
HCO       + Photon      & $\rightarrow$ & H         + CO        \\%&
HCO       + Photon      & $\rightarrow$ & HCO$^+$   + e$^-$        \\%&
H$_2$CO   + Photon      & $\rightarrow$ & H         + H + CO        \\
H$_2$CO   + Photon      & $\rightarrow$ & CO        + H$_2$        \\%&
H$_2$CO   + Photon      & $\rightarrow$ & H         + HCO$^+$ + e$^-$       \\
\end{tabular}
\begin{tabular}{lll}
\multicolumn{3}{c}{Ice} \\
Reactants & & Products \\
\hline
$\bot$H$_2$ + CR        & $\rightarrow$ & $\bot$H + $\bot$H             \\
$\bot$O$_2$ + CRUV      & $\rightarrow$ & $\bot$O + $\bot$O             \\
$\bot$OH + CRUV         & $\rightarrow$ & $\bot$H + $\bot$O             \\
$\bot$CO$_2$ + CRUV     & $\rightarrow$ & $\bot$O + $\bot$CO            \\
$\bot$H$_2$O + CRUV     & $\rightarrow$ & $\bot$H + $\bot$OH            \\
$\bot$HCO + CRUV        & $\rightarrow$ & $\bot$H + $\bot$CO            \\
$\bot$H$_2$CO + CRUV    & $\rightarrow$ & $\bot$H + $\bot$HCO           \\
$\bot$CH$_3$O + CRUV    & $\rightarrow$ & $\bot$H + $\bot$H$_2$CO       \\
$\bot$CH$_3$OH + CRUV   & $\rightarrow$ & $\bot$H + $\bot$CH$_3$O       \\
$\bot$CO + CRUV         & $\rightarrow$ & CO                            \\
$\bot$H$_2$O + CRUV     & $\rightarrow$ & H$_2$O                        \\
$\bot$H$_2$CO + CRUV    & $\rightarrow$ & H$_2$CO                       \\
$\bot$CH$_3$OH + CRUV   & $\rightarrow$ & CH$_3$OH                      \\
$\bot$H$_2$ + Photon    & $\rightarrow$ & $\bot$H + $\bot$H             \\
$\bot$O$_2$ + Photon    & $\rightarrow$ & $\bot$O + $\bot$O             \\
$\bot$OH + Photon       & $\rightarrow$ & $\bot$H + $\bot$O             \\
$\bot$H$_2$O + Photon   & $\rightarrow$ & $\bot$H + $\bot$OH            \\
$\bot$CO$_2$ + Photon   & $\rightarrow$ & $\bot$O + $\bot$CO            \\
$\bot$HCO + Photon      & $\rightarrow$ & $\bot$H + $\bot$CO            \\
$\bot$H$_2$CO + Photon  & $\rightarrow$ & $\bot$H + $\bot$HCO           \\
$\bot$CH$_3$O + Photon  & $\rightarrow$ & $\bot$H + $\bot$H$_2$CO       \\
$\bot$CH$_3$OH + Photon & $\rightarrow$ & $\bot$H + $\bot$CH$_3$O       \\
$\bot$CO + Photon       & $\rightarrow$ & CO                            \\
$\bot$H$_2$O + Photon   & $\rightarrow$ & H$_2$O                        \\
$\bot$H$_2$CO + Photon  & $\rightarrow$ & H$_2$CO                       \\
$\bot$CH$_3$OH + Photon & $\rightarrow$ & CH$_3$OH                      \\
&&\\
&&\\
&&\\
&&\\
&&\\
&&\\
&&\\
&&\\
&&\\
&&\\
&&\\
&&\\
\end{tabular}
\tabularnewline\hline
\end{tabular} \hfill
\label{tab:appendixG1}
\end{small}
\end{table*}

\begin{table*}
\begin{small}
\caption{Gas-phase reactions (146).}
\begin{tabular}{@{}c@{}c@{}}
\hline
\hline
\begin{tabular}{lll}
Reactants & & Products \\
\hline
O$_2$      + C$^+$       & $\rightarrow$ & CO        + O$^+$                 \\
H$_2$O     + C$^+$       & $\rightarrow$ & H         + HCO$^+$               \\
O$_2$      + C$^-$       & $\rightarrow$ & CO        + O$^-$                 \\
CO$_2$     + C$^-$       & $\rightarrow$ & CO        + CO        + e$^-$     \\
O          + HCO         & $\rightarrow$ & H         + CO$_2$                 \\
O          + HCO         & $\rightarrow$ & CO        + OH                  \\
CO         + OH          & $\rightarrow$ & H         + CO$_2$                 \\
OH         + OH          & $\rightarrow$ & O         + H$_2$O                 \\
OH         + HCO         & $\rightarrow$ & CO        + H$_2$O                 \\
OH         + H$_2$CO     & $\rightarrow$ & H$_2$O    + HCO                 \\
O          + H$_2^+$     & $\rightarrow$ & H         + OH$^+$                \\
CO         + H$_2^+$     & $\rightarrow$ & H         + HCO$^+$               \\
H$_2$      + H$_2^+$     & $\rightarrow$ & H         + H$_3^+$                \\
OH         + H$_2^+$     & $\rightarrow$ & H         + H$_2$O$^+$               \\
H$_2$O     + H$_2^+$     & $\rightarrow$ & H         + H$_3$O$^+$               \\
HCO        + H$_2^+$     & $\rightarrow$ & CO        + H$_3^+$                \\
H$_2$CO    + H$_2^+$     & $\rightarrow$ & H         + H$_2$      + HCO$^+$   \\
H$_3^+$    + H$^-$       & $\rightarrow$ & H$_2$     + H$_2$                  \\
HCO$^+$    + H$^-$       & $\rightarrow$ & CO        + H$_2$                  \\
H$_3$O$^+$ + H$^-$       & $\rightarrow$ & H         + H$_2$      + OH      \\
H$_3$O$^+$ + H$^-$       & $\rightarrow$ & H$_2$     + H$_2$O                 \\
OH         + HCO$^+$     & $\rightarrow$ & CO        + H$_2$O$^+$               \\
H$_2$O     + HCO$^+$     & $\rightarrow$ & CO        + H$_3$O$^+$               \\
HCO        + HCO         & $\rightarrow$ & CO        + H$_2$CO                \\
H          + HCO         & $\rightarrow$ & CO        + H$_2$                  \\
H          + OH          & $\rightarrow$ & O         + H$_2$                  \\
H          + H$_2$O      & $\rightarrow$ & H$_2$     + OH                  \\
H          + H$_2$CO     & $\rightarrow$ & H$_2$     + HCO                 \\
H          + O$_2$       & $\rightarrow$ & O         + OH                  \\
H          + CO$_2$      & $\rightarrow$ & CO        + OH                  \\
O          + H$_2$       & $\rightarrow$ & H         + OH                  \\
H$_2$      + OH          & $\rightarrow$ & H         + H$_2$O                 \\
O          + OH          & $\rightarrow$ & H         + O$_2$                  \\
CO         + H$_2$O$^+$  & $\rightarrow$ & OH        + HCO$^+$               \\
H$_2$      + H$_2$O$^+$  & $\rightarrow$ & H         + H$_3$O$^+$               \\
OH         + H$_2$O$^+$  & $\rightarrow$ & O         + H$_3$O$^+$               \\
H$_2$O     + H$_2$O$^+$  & $\rightarrow$ & OH        + H$_3$O$^+$               \\
HCO        + H$_2$O$^+$  & $\rightarrow$ & CO        + H$_3$O$^+$               \\
O          + H$_3^+$     & $\rightarrow$ & H         + H$_2$O$^+$               \\
O          + H$_3^+$     & $\rightarrow$ & H$_2$     + OH$^+$                \\
CO         + H$_3^+$     & $\rightarrow$ & H$_2$     + HCO$^+$               \\
OH         + H$_3^+$     & $\rightarrow$ & H$_2$     + H$_2$O$^+$               \\
H$_2$O     + H$_3^+$     & $\rightarrow$ & H$_2$     + H$_3$O$^+$               \\
CO$_2$     + H$^+$       & $\rightarrow$ & O         + HCO$^+$               \\
HCO        + H$^+$       & $\rightarrow$ & CO        + H$_2^+$                \\
H$_2$      + O$^+$       & $\rightarrow$ & H         + OH$^+$                \\
HCO        + O$^+$       & $\rightarrow$ & CO        + OH$^+$                \\
H$_2$CO    + O$^+$       & $\rightarrow$ & OH        + HCO$^+$               \\
C          + H$_3$O$^+$  & $\rightarrow$ & H$_2$     + HCO$^+$               \\
\end{tabular}
\begin{tabular}{lll}
Reactants & & Products \\
\hline
H$_2$CO    + H$^+$       & $\rightarrow$ & H$_2$     + HCO$^+$               \\
CO         + OH$^+$      & $\rightarrow$ & O         + HCO$^+$               \\
H$_2$      + OH$^+$      & $\rightarrow$ & H         + H$_2$O$^+$               \\
OH         + OH$^+$      & $\rightarrow$ & O         + H$_2$O$^+$               \\
H$_2$O     + OH$^+$      & $\rightarrow$ & O         + H$_3$O$^+$               \\
HCO        + OH$^+$      & $\rightarrow$ & CO        + H$_2$O$^+$               \\
C          + O$_2$       & $\rightarrow$ & O         + CO                  \\
C          + OH          & $\rightarrow$ & H         + CO                  \\
HCO$^+$    + C$^-$       & $\rightarrow$ & C         + H          + CO      \\
H$_3$O$^+$ + C$^-$       & $\rightarrow$ & C         + H          + H$_2$O     \\
H$_3^+$    + C$^-$       & $\rightarrow$ & C         + H          + H$_2$      \\
HCO$^+$    + O$^-$       & $\rightarrow$ & H         + O          + CO      \\
H$_3$O$^+$ + O$^-$       & $\rightarrow$ & H         + O          + H$_2$O     \\
H$_3^+$    + O$^-$       & $\rightarrow$ & H         + O          + H$_2$      \\
O          + H$_2$CO     & $\rightarrow$ & OH        + HCO                  \\
O$_2$      + HCO         & $\rightarrow$ & OH        + CO$_2$                  \\
H          + H$_3^+$     & $\rightarrow$ & H$_2$     + H$_2^+$                 \\
H          + H$_3$O$^+$  & $\rightarrow$ & H$_2$     + H$_2$O$^+$                \\
H$_2$      + H$_3^+$     & $\rightarrow$ & H         + H          + H$_3^+$    \\
H$_2$      + H$_3$O$^+$  & $\rightarrow$ & H         + H          + H$_3$O$^+$   \\
H$_2$      + HCO$^+$     & $\rightarrow$ & H         + H          + HCO$^+$   \\
H$_2$      + H$^+$       & $\rightarrow$ & H         + H          + H$^+$     \\
CO         + O$_2$       & $\rightarrow$ & O         + CO$_2$                  \\
H          + CO          & $\rightarrow$ & C         + OH                   \\
H$_2$      + O$_2$       & $\rightarrow$ & OH        + OH                   \\
O          + CO$_2$      & $\rightarrow$ & CO        + O$_2$                   \\
O          + H$_2$O      & $\rightarrow$ & OH        + OH                   \\
H$_2$      + e$^-$       & $\rightarrow$ & H         + H          + e$^-$     \\
H          + H$_2$       & $\rightarrow$ & H         + H          + H       \\
H$_2$      + H$_2$       & $\rightarrow$ & H         + H          + H$_2$      \\
H          + H$_2^+$     & $\rightarrow$ & H$_2$     + H$^+$                 \\
C$^+$      + C$^-$       & $\rightarrow$ & C         + C                   \\
C$^+$      + H$^-$       & $\rightarrow$ & C         + H                   \\
H$^+$      + C$^-$       & $\rightarrow$ & C         + H                   \\
H$^+$      + H$^-$       & $\rightarrow$ & H         + H                   \\
O$^+$      + C$^-$       & $\rightarrow$ & C         + O                   \\
O$^+$      + H$^-$       & $\rightarrow$ & H         + O                   \\
H$_2^+$    + H$^-$       & $\rightarrow$ & H         + H$_2$                  \\
O          + H$^+$       & $\rightarrow$ & H         + O$^+$                 \\
OH         + H$^+$       & $\rightarrow$ & H         + OH$^+$                \\
H$_2$O     + H$^+$       & $\rightarrow$ & H         + H$_2$O$^+$               \\
H          + O$^+$       & $\rightarrow$ & O         + H$^+$                 \\
H$_2$O     + O$^+$       & $\rightarrow$ & O         + H$_2$O$^+$               \\
HCO        + C$^+$       & $\rightarrow$ & C         + HCO$^+$               \\
HCO        + O$^+$       & $\rightarrow$ & O         + HCO$^+$               \\
OH         + O$^+$       & $\rightarrow$ & O         + OH$^+$                \\
OH         + H$_2^+$     & $\rightarrow$ & H$_2$     + OH$^+$                \\
H$_2$O     + H$_2^+$     & $\rightarrow$ & H$_2$     + H$_2$O$^+$               \\
HCO        + H$_2^+$     & $\rightarrow$ & H$_2$     + HCO$^+$               \\
\end{tabular}
\begin{tabular}{lll}
Reactants & & Products \\
\hline
HCO        + H$_2$O$^+$  & $\rightarrow$ & H$_2$O    + HCO$^+$               \\
HCO        + H$^+$       & $\rightarrow$ & H         + HCO$^+$               \\
H$_2$O     + OH$^+$      & $\rightarrow$ & OH        + H$_2$O$^+$               \\
HCO        + OH$^+$      & $\rightarrow$ & OH        + HCO$^+$               \\
HCO$^+$    + C$^-$       & $\rightarrow$ & C         + HCO                 \\
HCO$^+$    + O$^-$       & $\rightarrow$ & O         + HCO                 \\
H$^+$      + O$^-$       & $\rightarrow$ & H         + O                   \\
H$_2$      + H$^+$       & $\rightarrow$ & H         + H$_2^+$                \\
H          + C$^+$       & $\rightarrow$ & C         + H$^+$                 \\
C          + H$^+$       & $\rightarrow$ & H         + C$^+$                 \\
H          + H$^+$       & $\rightarrow$ & H$_2^+$   + Photon              \\
C          + O           & $\rightarrow$ & CO        + Photon              \\
H          + O           & $\rightarrow$ & OH        + Photon              \\
O          + O           & $\rightarrow$ & O$_2$     + Photon              \\
H          + OH          & $\rightarrow$ & H$_2$O    + Photon              \\
O          + C$^-$       & $\rightarrow$ & CO        + e$^-$                 \\
O$_2$      + C$^-$       & $\rightarrow$ & CO$_2$    + e$^-$                 \\
OH         + C$^-$       & $\rightarrow$ & HCO       + e$^-$                 \\
H$_2$O     + C$^-$       & $\rightarrow$ & H$_2$CO   + e$^-$                 \\
H          + H$^-$       & $\rightarrow$ & H$_2$     + e$^-$                 \\
O          + H$^-$       & $\rightarrow$ & OH        + e$^-$                 \\
CO         + H$^-$       & $\rightarrow$ & HCO       + e$^-$                 \\
OH         + H$^-$       & $\rightarrow$ & H$_2$O    + e$^-$                 \\
HCO        + H$^-$       & $\rightarrow$ & H$_2$CO   + e$^-$                 \\
C          + O$^-$       & $\rightarrow$ & CO        + e$^-$                 \\
H          + O$^-$       & $\rightarrow$ & OH        + e$^-$                 \\
O          + O$^-$       & $\rightarrow$ & O$_2$     + e$^-$                 \\
CO         + O$^-$       & $\rightarrow$ & CO$_2$    + e$^-$                 \\
H$_2$      + O$^-$       & $\rightarrow$ & H$_2$O    + e$^-$                 \\
C          + e$^-$       & $\rightarrow$ & C$^-$                             \\
H          + e$^-$       & $\rightarrow$ & H$^-$                             \\
O          + e$^-$       & $\rightarrow$ & O$^-$                             \\
H$_2^+$    + e$^-$       & $\rightarrow$ & H$_2$     + Photon              \\
H$_2^+$    + e$^-$       & $\rightarrow$ & H         + H                   \\
OH$^+$     + e$^-$       & $\rightarrow$ & H         + O                   \\
H$_2$O$^+$ + e$^-$       & $\rightarrow$ & O         + H$_2$                  \\
H$_2$O$^+$ + e$^-$       & $\rightarrow$ & H         + OH                  \\
H$_2$O$^+$ + e$^-$       & $\rightarrow$ & H         + H         + O       \\
H$_3^+$    + e$^-$       & $\rightarrow$ & H         + H         + H       \\
H$_3^+$    + e$^-$       & $\rightarrow$ & H         + H$_2$                  \\
HCO$^+$    + e$^-$       & $\rightarrow$ & H         + CO                  \\
C$^+$      + e$^-$       & $\rightarrow$ & C         + Photon              \\
H$^+$      + e$^-$       & $\rightarrow$ & H         + Photon              \\
O$^+$      + e$^-$       & $\rightarrow$ & O         + Photon              \\
H$_3$O$^+$ + e$^-$       & $\rightarrow$ & H         + H         + OH      \\
H$_3$O$^+$ + e$^-$       & $\rightarrow$ & H         + H$_2$O                 \\
H$_3$O$^+$ + e$^-$       & $\rightarrow$ & H$_2$     + OH                  \\
H$_3$O$^+$ + e$^-$       & $\rightarrow$ & H         + O         + H$_2$      \\
&&\\
\end{tabular}
\tabularnewline\hline
\end{tabular} \hfill
\label{tab:appendixG2}
\end{small}
\end{table*}

\begin{table*}
\begin{small}
\caption{Surface reactions (41).}
\begin{tabular}{@{}c@{}}
\hline
\hline
\begin{tabular}{lll|c|c}
\multicolumn{5}{c}{Products remain on the surface} \\
Reactants                       &               & Products                      & E$_a$/K & Reference $\dagger$ \\
\hline
$\bot$H + $\bot$H               & $\rightarrow$ & $\bot$H$_2$                   & 0             & \\
$\bot$H + $\bot$O               & $\rightarrow$ & $\bot$OH                      & 0             & \\
$\bot$H + $\bot$OH              & $\rightarrow$ & $\bot$H$_2$O                  & 0             & \\
$\bot$H + $\bot$O$_3$           & $\rightarrow$ & $\bot$OH + $\bot$O$_2$        & 450           & a, i, c(480), t(0) \\
$\bot$H + $\bot$H$_2$O$_2$      & $\rightarrow$ & $\bot$OH + $\bot$H$_2$O       & 1000          & o(800-1250) \\
$\bot$H + $\bot$O$_2$           & $\rightarrow$ & $\bot$HO$_2$                  & 200           & b(0-250), j(0-200), t(0) \\
$\bot$H + $\bot$H$_2$O          & $\rightarrow$ & $\bot$OH + $\bot$H$_2$        & 9600          & r \\
$\bot$H + $\bot$HO$_2$          & $\rightarrow$ & $\bot$OH + $\bot$OH           & 0             & \\
$\bot$H + $\bot$CO              & $\rightarrow$ & $\bot$HCO                     & 600           & h, k $\star$ \\
$\bot$H + $\bot$HCO             & $\rightarrow$ & $\bot$H$_2$CO                 & 0             & \\
$\bot$H + $\bot$H$_2$CO         & $\rightarrow$ & $\bot$CH$_3$O                 & 400           & h, k $\star$ \\
$\bot$H + $\bot$CH$_3$O         & $\rightarrow$ & $\bot$CH$_3$OH                & 0             & \\
$\bot$H + $\bot$HCO             & $\rightarrow$ & $\bot$CO + $\bot$H$_2$        & 400           & s \\
$\bot$H + $\bot$H$_2$CO         & $\rightarrow$ & $\bot$HCO $\bot$H$_2$         & 400           & s \\
$\bot$H + $\bot$CH$_3$O         & $\rightarrow$ & $\bot$H$_2$CO + $\bot$H$_2$   & 150           & n \\
$\bot$H + $\bot$CH$_3$OH        & $\rightarrow$ & $\bot$CH$_3$O + $\bot$H$_2$   & 3000          & n \\
$\bot$H + $\bot$CO$_2$          & $\rightarrow$ & $\bot$CO + $\bot$OH           & 10000         & n \\
$\bot$O + $\bot$O               & $\rightarrow$ & $\bot$O$_2$                   & 0             & \\
$\bot$O + $\bot$O$_2$           & $\rightarrow$ & $\bot$O$_3$                   & 0             & \\
$\bot$O + $\bot$O$_3$           & $\rightarrow$ & $\bot$O$_2$ + $\bot$O$_2$     & 2300          & p($\geq$2300), r(2000) \\
$\bot$O + $\bot$HO$_2$          & $\rightarrow$ & $\bot$O$_2$ + $\bot$OH        & 0             & \\
$\bot$O + $\bot$H$_2$           & $\rightarrow$ & $\bot$H + $\bot$OH            & 4640          & l, o(3165) \\
$\bot$O + $\bot$OH              & $\rightarrow$ & $\bot$O$_2$ + $\bot$H         & 0             & \\
$\bot$O + $\bot$CO              & $\rightarrow$ & $\bot$CO$_2$                  & 600           & p, e(290) \\
$\bot$O + $\bot$HCO             & $\rightarrow$ & $\bot$CO$_2$ + $\bot$H        & 0             & \\
$\bot$O + $\bot$H$_2$CO         & $\rightarrow$ & $\bot$CO$_2$ + $\bot$H$_2$    & 335           & q \\
$\bot$OH + $\bot$OH             & $\rightarrow$ & $\bot$H$_2$O$_2$              & 0             & \\
$\bot$OH + $\bot$H$_2$          & $\rightarrow$ & $\bot$H + $\bot$H$_2$O        & 2100          & g \\
$\bot$OH + $\bot$CO             & $\rightarrow$ & $\bot$CO$_2$ + $\bot$H        & 400           & m, d(554), f(519) \\
$\bot$OH + $\bot$HCO            & $\rightarrow$ & $\bot$CO$_2$ + $\bot$H$_2$    & 0             & \\
$\bot$HO$_2$ + $\bot$H$_2$      & $\rightarrow$ & $\bot$H + $\bot$H$_2$O$_2$    & 5000          & o \\
\end{tabular}
\begin{tabular}{lll|c|c}
\multicolumn{5}{c}{Products desorb into the gas phase} \\
Reactants                       &               & Products              & E$_a$/K         & Reference $\dagger$ \\
\hline
$\bot$H + $\bot$H               & $\rightarrow$ & H$_2$                 & 0               & \\
$\bot$H + $\bot$O               & $\rightarrow$ & OH                    & 0               & \\
$\bot$H + $\bot$OH              & $\rightarrow$ & H$_2$O                & 0               & \\
$\bot$H + $\bot$O$_3$           & $\rightarrow$ & OH + O$_2$            & 450             & a, i, c(480), t(0) \\
$\bot$H + $\bot$H$_2$O$_2$      & $\rightarrow$ & OH + H$_2$O           & 1000            & o(800-1250) \\
$\bot$H + $\bot$CO              & $\rightarrow$ & HCO                   & 600             & h, k $\star$ \\
$\bot$H + $\bot$HCO             & $\rightarrow$ & H$_2$CO               & 0               & \\
$\bot$H + $\bot$H$_2$CO         & $\rightarrow$ & CH$_3$O               & 400             & h, k $\star$ \\
$\bot$H + $\bot$CH$_3$O         & $\rightarrow$ & CH$_3$OH              & 0               & \\
$\bot$O + $\bot$O               & $\rightarrow$ & O$_2$                 & 0               & \\
%$\bot$O + $\bot$O$_3$           & $\rightarrow$ & O$_2$ + O$_2$         & 0 & \\ %added by mistake, but makes no difference
%$\bot$O + $\bot$HO$_2$          & $\rightarrow$ & O$_2$ + OH            & 0 & \\ %added by mistake, but makes no difference
&&&&\\
&&&&\\
&&&&\\
&&&&\\
&&&&\\
&&&&\\
&&&&\\
&&&&\\
&&&&\\
&&&&\\
&&&&\\
&&&&\\
&&&&\\
&&&&\\
&&&&\\
&&&&\\
&&&&\\
&&&&\\
&&&&\\
&&&&\\
&&&&\\
\end{tabular}
\tabularnewline\hline
\end{tabular} \hfill
\label{tab:appendix1} \\
\raggedright
$\dagger$ The numbers in parentheses give alternative barriers (in K) from the respective studies. \\
$\star$ These are `effective' barriers deduced from the original provided rates.  \\
\centering
References: \\
\raggedright
$^{\textbf{a}}$ \cite{1978JChPh..69..350L}, 
$^{\textbf{b}}$ \cite{1988JChPh..88.6273W}, 
$^{\textbf{c}}$ \cite{1989JPCRD..18..881A}, 
$^{\textbf{d}}$ \cite{2000JChPh.113.5138D}, 
$^{\textbf{e}}$ \cite{2001ApJ...555L..61R}, 
$^{\textbf{f}}$ \cite{2001CPL...349..547Y}, 
$^{\textbf{g}}$ \cite{2004ACP.....4.1461A}, 
$^{\textbf{h}}$ \cite{2005ApJ...626..262A}, 
$^{\textbf{i}}$ \cite{2007ApJ...668..294C}, 
$^{\textbf{j}}$ \cite{2008CPL...456...27M}, 
$^{\textbf{k}}$ \cite{2009A&A...505..629F}, 
$^{\textbf{l}}$ \cite{2010ApJ...713..662A}, 
$^{\textbf{m}}$ \cite{2011ApJ...735..121N}, 
$^{\textbf{n}}$ \textit{NIST} database mean value; http://kinetics.nist.gov, 
$^{\textbf{o}}$ \cite{2013PCCP...15.8287L}, 
$^{\textbf{p}}$ \cite{2013A&A...559A..49M}, 
$^{\textbf{q}}$ \cite{2015A&A...577A...2M}, 
$^{\textbf{r}}$ \cite{2010A&A...522A..74C}, 
$^{\textbf{s}}$ Best estimate (priv. comm. Minissale), 
$^{\textbf{t}}$ \cite{2009ApJ...705L.195M} \\
\end{small}
\end{table*}

\begin{table*}
\begin{small}
\caption{Chemisorption reactions (5).}
\begin{tabular}{lll|c}
\hline
\hline
Reactants               &               & Products      & E$_a$/K \\
\hline
~~~H                    & $\rightarrow$ & $\bot$H$_c$   & 1000 \\
~~~H + $\bot$H$_c$      & $\rightarrow$ & ~~~H$_2$      & 1000 \\
$\bot$H$_c$             & $\rightarrow$ & ~~~H          & no E$_a$, E$_{\rm binding}=10000$\,K \\
$\bot$H                 & $\rightarrow$ & $\bot$H$_c$   & see \cite{2010ApJ...715..698C} \\
$\bot$H + $\bot$H$_c$   & $\rightarrow$ & ~~~H$_2$      & see \cite{2010ApJ...715..698C} \\
\hline
\end{tabular} \hfill
\label{tab:appendix3}
\end{small}
\end{table*}

\end{document}